\documentclass{article}

\usepackage{amsfonts, amsmath, amssymb, amsthm}

\usepackage[margin=2cm]{geometry}

\newcommand{\dd}{{\rm d}}

\newcommand{\bs}[1]{{\boldsymbol #1}}

\newcommand{\R}{\mathbb{R}}

\usepackage{graphicx}

\newcommand{\figwidth}{0.9\textwidth}

\usepackage{url}

\begin{document}

\title{Noisy bounded confidence models for opinion dynamics: the effect of 
boundary conditions on phase transitions}

\author{B. D. Goddard$^{1}$, B. Gooding$^{1}$, G. A. Pavliotis$^{2}$ and H. Short$^{1}$}
\date{%
$^{1}$School of Mathematics and Maxwell Institute for Mathematical Sciences, University of Edinburgh,
Edinburgh EH9 3FD, United Kingdom\\
$^{2}$Department of Mathematics, Imperial College London, London SW7 2AZ, United Kingdom
\\[2ex]%
\today
}

\maketitle

\begin{abstract}
{We study SDE and PDE models for opinion dynamics under bounded confidence,
for a range of different boundary conditions, with and without the inclusion of 
a radical population.  We perform
exhaustive numerical studies with pseudospectral methods to determine the effects
of the boundary conditions, suggesting that the no-flux case most faithfully
reproduces the underlying mechanisms in the associated deterministic models of Hegselmann and Krause.  
We also compare the SDE and PDE models, and use tools from analysis to study phase transitions,
including a systematic description of an appropriate order parameter.}
{bounded confidence, opinion dynamics, no-flux boundary conditions, phase transitions.} 
\\
2000 Math Subject Classification: 82C22, 82C26, 65M70
\end{abstract}

\section{Introduction and Previous Work}~\label{S:Introduction}
This work focuses on bounded confidence models for opinion dynamics under social influence,~\cite{CFL09,PT17,PT18},
which is part of the larger field of mathematical modelling in the social sciences~\cite{CFL09,T16,T18}.
In contrast to graph-based models in which communication occurs only between connected individuals,
here communication is instead limited by the difference in their opinions, which is treated as a 
continuous variable. 
A typical example of a continuous opinion is an individual's political orientation, which is not
restricted to a few discrete choices, such as extreme left or right, but rather can vary
across a spectrum.  
The motivation for bounded confidence models
comes from `biased assimilation'~\cite{LR79}, which postulates that 
individuals  are more strongly influenced
by others with similar opinions to their own.  A related consideration is that genuine discussion generally
only occurs between individuals who already share some common ground, i.e., they already have
sufficiently close opinions.   This can be modelled by the coupling between
individuals increasing as their opinions become more similar.

Here we consider models with a hard cut-off in opinion space; individuals with opinions differing by
more than the confidence bound $R$ do not influence each other. 
Such models were originally proposed as deterministic, discrete-time
processes~\cite{DNAW00,WDAN02,HK02}, which have also been extended to include noise or 
randomness~\cite{PTHG09,PTHG11,PTHG13,SCH17}, modelling uncertainty in observations
or external influences.  Related models include continuous time
ODEs~\cite{BHT10,YDH14}, SDEs, and (in the limit of many individuals) 
PDEs~\cite{WLEC17,KPE19}.
See~\cite{CFL09,PT17,PT18} for comprehensive reviews.  
In the below, due to the inclusion of noise in the SDE and (implicitly in) the PDE models, 
we refer to them as non-deterministic.  
More general formulations arise in mathematical biology as the Keller-Segel model~\cite{KS70} for slime mold,
and other similar models~\cite{MEK99,MEKBS03,TBL06}.  
Since the individuals which interact changes according to the dynamics, such models are often
described through `co-evolutionary networks'~\cite{PT18}, for example the Vicsek model of phase transitions~\cite{VCBJCS95}, 
the Cucker-Smale model for flocking~\cite{CS07-1,CS07-2,MT11},
and robotics~\cite{JLM03,BCM09,CFT12}.
We also highlight recent
mathematical work concerning the well-posedness and long-time behaviour
of related mean-field models~\cite{CJLW17,CGPS20,GMP20}.

A typical question one asks about such models is how a uniform, or disordered, initial condition
evolves under the interactions of individuals; typically the system is driven to a more ordered state~\cite{Y92,CFL09,KTH12}.
This has clear analogues with order--disorder phase transistions~\cite{Y92}.
Two popular  models are due to Hegselmann and Krausse (HK)~\cite{HK02} 
and Deffaunt and Weisbuch (DW)~\cite{DNAW00,WDAN02}.  See~\cite{PT18} for a recent review.
Both are discrete in time and rely on the idea of repeated averaging under bounded confidence.  
In DW agents interact 
in randomly chosen pairs who then either do or do not compromise (depending on the separation
of their opinions). In HK an individual moves to the average opinion of all agents within a distance $R$ of themselves
in opinion space.
In such models, an initially homogeneous or uniform distribution of opinions is unstable.  
This is due to boundary effects, in which those with extreme opinions can only be influenced 
by those with less extreme opinions, and hence tend to move towards more moderate
values, causing clusters.
Three possible outcomes are: (i) consensus (a single cluster); 
(ii) polarization (two distinct clusters);
(iii) fragmentation (more than two distinct clusters, or no distinct clusters).
 It has been noted that the number of clusters is typically $1/(2R)$, which is related to the $2R$-conjecture~\cite{BHT07,CFL09}.
This conjecture, which originally referred to the HK model, states that a uniform distribution of agents
converges to clusters separated by distances of roughly $2R$.  This was studied in~\cite{WLEC17} for
a PDE model with periodic boundary conditions; it would be interesting to investigate this further for the
different boundary conditions proposed below.  This is a topic of future work.

This convergence to a steady state is not normally observed in real-world systems~\cite{BN05,YOASS13,CTSM13}.  
Two possible factors missing from these models are (i) influence of external
agents or information, such as `radicals', leaders, or advertising; (ii) uncertainty in the dynamics.  
For (i), it is usually assumed that such influences are constant in time~\cite{DMPW09,YOASS13,WCB15,HK15,M15, KPE19}, 
or weakly susceptible to other opinions~\cite{ZZTK16}.  
Numerical simulations have demonstrated counterintuitive effects concerning the introduction of radials, such as 
increasing radical numbers decreasing the number of individuals sharing their opinion after long times.
We will demonstrate similar effects in our non-deterministic models; see Section~\ref{s:Radicals}.
For (ii), it has been argued~\cite{BN05} that diffusion is an essential
element of opinion dynamics, allowing the modelling of realistic political systems with disorder-order
transitions and complex lifecycles.  This has close links to the use of environmental noise in statistical physics
models, which can lead to phase transitions~\cite{CBV99,BN05,PTHG09,SWTT11,PTHG11,GJ12,PTHG13,WLEC17,KPE19}.
Additionally, in many applications, it is not only the equilibrium that is interesting,
but also the dynamical path to that equilibrium~\cite{CFL09}.  For example, if the opinion measures
political persuasion, then one would typically be interested in the distribution at a particular time (e.g.,
on an election date) than in the long-term equilibrium.  We note here that existing approaches fail to correctly model
the appropriate boundary conditions for such applications; we discuss this in detail later.
In Sections~\ref{s:NoRadicals} and~\ref{s:Radicals} we systematically investigate both the effects of the
noise strength and the dynamical paths to equilibrium.

Due to the non-linear, non-local nature of many opinion dynamics models, and the resulting challenges of analytical investigation,
the use of careful and systematic computer simulations has been important in social dynamics for decades, see e.g.,
\cite{HF98,HKS01,S01,ZZTK16,WLEC17}.  For $N$ individuals, the agent-based, ODE, and SDE models typically 
have computational costs that scale as $N^2$, which is the cost of determining the pairwise distances.
An interesting regime, which is amenable to both mathematical analysis
and decreased computational cost, is the mean-field limit where the number of individuals
becomes very large, $N\to \infty$.  Then the high-dimensional descriptions of individual opinions
reduces to a 1+1 (space-time) dimensional non-local, non-linear PDE, for which there exist a range of accurate
and efficient numerical approaches~\cite{WLEC17,NGYSK17,GPV19,KPE19}.
Such PDE models have been referred to by
a variety of names, including density-based~\cite{L07}, continuum~\cite{BHT10,HO16}, Eulerian~\cite{CFT12,MJB14},
hydrodynamic~\cite{MT11,MT14}, kinetic~\cite{DMPW09,BT15,BS16}, or mean-field~\cite{GPY17,WLEC17,NTCL18,KPE19}.
We note, in particular, that kinetic models provide an intermediate level of coarse-graining, between the discrete, `particle',
models and the continuum, "hydrodynamics-type" PDEs.  In kinetic models, interactions are modelled through instantaneous, local `collisions', or interactions,
between individuals.  This contrasts with the `soft' interactions included in the PDE models presented here.

Returning to the SDE and PDE models, there can be crucial differences between the two models: 
Firstly, order-disorder phase transitions are rigorously 
defined only in the thermodynamic limit as only then can there be
non-uniqueness of invariant measures.  However, such terms 
are still used for similar behaviour in the finite-$N$ models~\cite{TT07}.  Secondly, there are questions about how large 
the number of agents needs to be in order for the SDE and PDE models to be in good agreement.  
Many techniques used to study PDE models arise from statistical mechanics, but care must be taken
when directly transferring them to the social sciences.  This is due to the different notions of a `large'
number of individuals/particles; in molecular systems this could be of the order $10^{23}$, whereas in
social systems, it is more likely to be in the hundreds or thousands.  In such cases, finite size
effects are likely to be important~\cite{TT07}; we demonstrate this in Supplementary Material Section SM3.
However, these differences must be balanced by the relative computational complexities: as above, the 
computational cost of the SDE, agent-based model scales as $N^2$, whereas that of the PDE is 
independent of $N$, depending instead on the discretization.

The mean-field limit of a system of weakly interacting agents in the presence of noise
is given by a non-linear, non-local PDE of Fokker-Planck type. 
For such PDES (as well as the corresponding SDEs) it is necessary to specify boundary conditions.  Previous work has
been restricted to periodic domains, presumably for mathematical and computational ease, 
see e.g., \cite{GPY17,WLEC17,KPE19}.
However, care must be taken when comparing such implementations to the original deterministic models.
For example, in the original models, if all initial opinions lie in an interval $[a,b]$, then
this holds for all time~\cite{PT18}. 
To aid comparison, it would be natural to treat the non-deterministic models in the same way, requiring
the individual opinions to lie in some prescribed interval, which, without loss of generality, may be chosen
to be $[0,1]$.  
One criticism of periodic boundary conditions is that they
conflate the two extreme opinions at 0 and 1; this implies that two opposite extremes 
(for example on the political spectrum) are close.  This is clearly not a realistic assumption.  
Whilst the even 2-periodic choice in~\cite{KPE19} 
overcomes this conflation, it does so at the expense or introducing a mirror system, 
which can strongly influence the dynamics and has (potentially) undesirable
effects on quantities such as the order parameter; see Section~\ref{s:OrderParameter}.  
Here we introduce an additional choice of boundary
conditions, namely no-flux.  This can be regarded as a mixed (Robin) boundary condition for the mean-field Fokker-Planck-McKean-Vlasov PDE.  
As in the existing two cases, this preserves the mass conservation property
of the model, but does not conflate extreme opinions nor require the introduction of an auxiliary system.
\emph{We propose that no-flux boundary conditions are much more compatible with the original, discrete HK models in terms 
of the mechanism of cluster formation from a uniform initial condition}.  For example, as we will see in Section~\ref{s:NoRadicals},
large $R$ (confidence bound) and small $\sigma$ (noise) causes an initially uniform distribution to develop a cluster in the middle of the interval, 
rather than remaining in a uniform steady state, which is predicted when periodic boundary conditions are chosen.
In particular, we note that previous results, such as the numerical experiments and resultant phase diagrams in~\cite{WLEC17}  and the rigorous analysis reported in~\cite{GPY17} depend crucially on the assumption of periodic boundary conditions.
Given that the choice of periodic boundary conditions is clearly not optimal from a modelling perspective, 
it is not obvious whether the numerical experiments reported in these papers are relevant to the study of
opinion dynamics.

The challenge with no-flux boundary conditions is both analytical and numerical.  
The additional analytical challenges arise principally from the fact that, as already mentioned in~\cite{GMP20}, the no-flux boundary conditions for the McKean-Vlasov PDE are nonlinear and nonlocal. This means, in particular, that the uniform distribution is no longer a stationary state. Furthermore, doing linear stability analysis, in the form of~\cite{GPY17}[Sec. 4.1] is more involved, since the perturbation from the stationary state has to be chosen in such a way that it satisfies the boundary conditions and that it has zero mass.  We note that, at least formally and in the absence of radical groups, the mean field PDE is a Wasserstein gradient flow for the free energy~\cite{V03}. We reiterate that care has to be taken when dealing with the nonlinear and nonlocal boundary conditions that we consider in this paper. See~\cite{GMP20} for details.
In terms of numerical methods for SDE models, reflecting boundary
conditions are known to be challenging~\cite{S61,P14}, and the corresponding PDE models can no longer use efficient Fourier methods.  
Here we implement efficient and robust Fourier and Chebyshev pseudospectral methods, based on~\cite{NGYSK17,2DChebClass}.
We emphasise the fact that, due to their greater realism and compatibility with the original deterministic models, we 
favour the use of no-flux boundary conditions in future studies; we include the periodic cases for comparison
with the existing literature.

For completeness, we also briefly discuss the possible choice of zero Dirichlet boundary conditions.  The key disadvantage in
such as choice is that the system no longer preserves mass, i.e., the number of agents can change over time.  This may,
however, be an appropriate boundary condition in some situations, for example, if people with extreme political views stop 
interacting with others.  Of course, if one considers extreme opinions to be rare, and the vast majority of the population
has opinions localised in the centre of opinion space, then Dirichlet boundary conditions will give essentially identical
results to any other choice.  This could be achieved, for example, in a low-noise system where the initial condition
is centred towards the middle of the opinion space. 
However, when considering models with radicals who may aim to drive opinions to extremes,
this is clearly not the case.  An important example of such a system is a binary referendum, where the extremes of the
system correspond to a strong likelihood of voting one way or the other.  We do not consider such Dirichlet boundary conditions in this work, primarily because the lack of mass conservation prevents a meaningful comparison to the other choices. We mention that whether boundaries for mean-field SDEs modelling opinion formation are accessible or not  is an interesting question. In particular, it is possible that, for appropriate choices of the interaction between agents, the dynamics never reaches the boundaries of opinion space. This is an interesting and intriguing question -- in particular, the development of a Feller-type classification for mean field SDEs in a one dimensional bounded domain -- that we plan to return to in future work.  

Our main contributions are:
\begin{itemize}
\item a unification of existing and novel SDE and PDE models for bounded confidence opinion dynamics;
\item a systematic numerical study of the transient and long-time dynamics of these models, under three different
boundary conditions and a wide range of parameter regimes, with and without a `radical' population;
\item a careful discussion of an `order parameter';
\item the insight that the no-flux boundary conditions most faithfully reproduce the underlying mechanisms
of the original deterministic models.
\end{itemize}

The remainder of the paper is organised as follows:  In Section~\ref{s:Model} we introduce the models, as well
as the associated order parameter, and convergence to equilibrium.  Section~\ref{s:Numerics}
describes the details of the numerical methods for both the PDE and SDE models.  In Sections~\ref{s:NoRadicals}
and~\ref{s:Radicals} we present the numerical experiments for systems without and with radicals, respectively.
Section~\ref{s:Conclusions} contains our conclusions and a description of some open problems.  The Supplementary
Material contains more discussion on the order parameter,
validation of the numerical methods against existing results from the literature, a thorough comparison of the SDE and PDE models,
as well as some further examples.


\section{Model} \label{s:Model}

\subsection{Dynamics}

The original model of Hegselmann and Krause (HK) is discrete in time and space.  It considers a set of $N$
agents, with agent $i$ having opinion $x_i$ with $x = (x_1, \ldots, x_N)$.  
Bounded confidence is introduced by defining a confidence level $R$ and a set for each agent 
$I(i,x) = \{ 1\leq j \leq N : |x_i-x_j| \leq R\}$, i.e. the set of all individuals whose opinion is within $R$ of that of
individual $i$.  
At each time the opinion of individual $i$ is updated through
\[
	x_i(t+1) = \big| I\big(i,x(t)\big) \big|^{-1} \sum_{j \in I(i,x(t))} x_j(t),
\]
where here $|I|$ denotes the number of elements of $I$.  In words, an individual's opinion at the next time step
is given by the mean of the opinions of individuals within their confidence interval. It is clear to see here that 
if two groups are separated by a distance of $R$ or more then they will form decoupled subsystems, which then 
never interact.  Note that there exist alternative models in which the attraction increases with separation~\cite{MT14}.

In the SDE models~\cite{WLEC17,KPE19}, which originate in statistical physics, the mean is taken not over
the set $I$, but over all individuals, with zero weight on those outside $I$, i.e. $|I|$ is replaced by $N$ in the normalisation.
This leads to a simpler mean-field PDE but it is also possible to retain the original normalisation~\cite{GPY16}.
The original choice of normalisation is perhaps more physically relevant, especially in swarming/flocking
models, in which the bounded confidence is based on physical, rather than opinion, distance~\cite{CS07-1,CS07-2}.  
In such cases, it is plausible to assume that an individual is completely unaware of those outside its confidence bound.
In opinion dynamics models, it is perhaps more reasonable to assume that an individual polls the opinion of all
other individuals (as if in a completely connected network) and simply ignores the opinion of those individuals
outside their confidence bound.
Introducing the $1/N$ scaling results in dynamics which are slowed down by the presence of individuals who do
not interact, and are, in fact, unaware of each other~\cite{MT11}; 
other works have claimed that the results are insensitive to this choice~\cite{CFT12,GPY16}.

For a system of $N$ individuals with opinions $x_i$, and a confidence bound $R$,  the dynamics that we
consider in this paper, given 
a suitable initial condition, are described by~\cite{WLEC17}
\begin{equation} \label{eq:SDE}
	\dd x_i = - \frac{1}{N} \sum_{j: |x_i-x_j|\leq R} (x_i-x_j) \dd t + \sigma \dd W_t^{(i)},
\end{equation}
where $W_t^{(i)}$ are independent Wiener processes. As described in Section~\ref{S:Introduction},
the motivation for including noise comes from
agents' `free will', or uncertainty in measurement and communication.
Taking the mean-field limit of $N\to \infty$ results in a Fokker-Planck PDE for the density of 
opinions $\rho(x,t)$~\cite{GPY17,WLEC17}:
\begin{equation} \label{eq:PDE}
	\partial_t \rho(x,t) = \partial_x \Big( \rho(x,t) \int (x-y) \rho(y,t) \bs{1}_{|x-y|\leq R} (y)\,\dd y \Big) 
	                               + \frac{\sigma^2}{2} \partial_{xx} \rho(x,t),
\end{equation}
where $\rho$ corresponds to the empirical measure 
$\rho^N(x,t) = N^{-1} \sum_j \delta_{x_j}(\dd x)$ as $N \to \infty$.  

\subsection{Boundary Conditions}

It remains to discuss the boundary conditions (BC) imposed on \eqref{eq:SDE} and \eqref{eq:PDE}.  As discussed above,
we consider three separate cases: (i) Periodic; (ii) No-Flux; (iii) Even 2-Periodic, all on $[0,1]$, representing opinion space.
For \textbf{BC (i)} [Periodic]
the natural interpretation is that $\rho(x,t)$ is extended periodically to the whole of $\R$ with the periodicity condition
$\rho(x+1,t) = \rho(x,t)$, $\forall x \in \R$.  We note that $\rho(\cdot,t)$ is not a probability density on $\R$,
but does serve as a (normalised) probability density on [0,1].  If [0,1] represents the whole of opinion space then
this is clearly not a good modelling assumption; in particular it conflates the two extreme opinions.
For \textbf{BC (ii)} [No-Flux], it is helpful to rewrite \eqref{eq:PDE} in terms of the flux, $j$:
\begin{equation} \label{eq:PDEFlux}
	\partial_t \rho(x,t) = - \partial_x j(x,t), \qquad
	j(x,t)                     =  - \rho(x,t) \int (x-y) \rho(y,t) \bs{1}_{|x-y|\leq R} (y) \,\dd y - \frac{\sigma^2}{2} \partial_x \rho(x,t).
\end{equation}
We then impose no-flux boundary conditions, i.e., $j(0,t) = j(1,t) = 0$.  We emphasise that these boundary conditions
are non-local and non-linear due to the nature of the flux in the PDE~\eqref{eq:PDE}. 

The mean field PDE~\eqref{eq:PDEFlux} can be written as a gradient flow in the sense of Jordan-Kinderlehrer-Otto~\cite{CGPS20}[Sec. 6.2]:
\begin{equation}\label{e:gradient}
\partial_t \rho(x,t) = \partial_x \left(\rho \partial_x \frac{\delta \mathcal{F}}{\delta \rho} \right),
\end{equation}
where the free energy for the Hegselmann-Krause model is
\begin{equation}\label{e:free_energy}
\mathcal{F}[\rho] = \frac{\sigma^2}{2} \int \rho \log (\rho) \, dx + \int \int W(x-y) \rho(x) \rho(y) \, dx dy,
\end{equation}
with 
\begin{equation}\label{e:interaction}
W(x) = -\left( \left( |x| - \frac{R}{2} \right)_{-} \right)^2
\end{equation}
where $\left( \cdot \right)_{-}$ denotes the negative part of a function. In~\cite{CGPS20}[Sec. 6.2] it was shown that, for periodic boundary conditions and at sufficiently small $R$, an appropriately rescaled Hegselmann-Krause model exhibits a discontinuous phase transition. Similar results for the nonlinear, nonlocal no-flux boundary conditions that we consider in this paper will be presented elsewhere~\cite{GMOP21}.

For \textbf{BC (iii)} [Even 2-Periodic] we follow~\cite{KPE19}.  
Rather than considering $\rho(x,t)$ on $x \in [0,1]$, we consider its unique even, 2-periodic extension which satisfies
$\rho(-x,t) = \rho(x,t)$ and $\rho(x+2,t) = \rho(x,t)$, $\forall x \in \R$.  Again, $\rho(\cdot,t)$ is a probability density on $[0,1]$.
For clarity, in this case the sum over $j$ in \eqref{eq:SDE} is over the full, even 2-periodic system with $2N$ 
individuals.  \emph{We will show that the no-flux boundary conditions most faithfully reproduce the behaviour of the original HK models},
for example, the instability of a uniform initial condition, and the physical interpretation of the domain.

\subsection{Radicals}

As described in~\cite{HK15,KPE19}, a natural extension of bounded confidence models is to include $N_r > 0$
\emph{radicals}, or extreme groups.  These should be thought of as individuals with fixed opinions who, nevertheless, affect the opinions
of those who interact with them.  As discussed in~\cite{HK15}, such a formalism can also be used to model advertising, 
charismatic leaders, and other external effects. 
In the language of statistical mechanics, radicals act as an external potential. Radicals (or stubborn agents) have also been included as
a generalisation of the voter model, see, e.g.,~\cite{YOASS11,YOASS13}.

For the SDE, we retain the indexing of the `normal' individuals as $1, \ldots, N$, and add in radicals indexed by 
$N+1, \ldots, N+N_r$.  The dynamics are then governed by
\begin{align*}
	\dd x_i &= - \frac{1}{N} \sum_{j: |x_i-x_j|\leq R} (x_i-x_j) \dd t + \sigma \dd W_t^{(i)}, \quad i = 1, \ldots N \\
	\dd x_i &= 0,\quad  i = N+1, \ldots, N+N_r.
\end{align*}
Note that the sum in the first equation now runs over $j=1, \ldots, N+N_r$.  
The corresponding Fokker-Planck PDE is 
\begin{align*}
	\partial_t \rho(x,t) &= - \partial_x j(x,t), \quad \\
	j(x,t)                     &=  - \rho(x,t) \int (x-y) \big( \rho(y,t) +M \rho_r(y) \big) \bs{1}_{|x-y|\leq R}(y) \, \dd y - \frac{\sigma^2}{2} \partial_x \rho(x,t),
\end{align*}
where $\rho_r$ determines the (fixed) distribution of the radicals.
We find it convenient
to fix $\rho_r$ as a probability distribution (in the senses described above for the periodic boundary conditions) and scale the mass with a parameter
$M$.  For physical reasons, $\rho_r$ should be non-negative (although it is interesting to consider
negative/repulsive opinions), and also $M \ll 1$, 
otherwise the interpretation as $\rho_r$ as the density of `radicals' is lost.  However, these 
restrictions are not intrinsic to the model itself.  Note that, in the case of even 2-periodic
boundary conditions, $x \in [-1,1]$, the full space, and the convolution is also taken over this whole domain.

\subsection{Order Parameter} \label{s:OrderParameter}

To enable quantification of the resulting opinion densities, \cite{WLEC17} introduced the order parameter, which
has both a discrete and continuum definition:
\[
	\tilde{Q}_d(t) = \frac{1}{N^2} \sum_{i,j=1}^N \bs{1}_{|x_i(t) - x_j(t)|\leq R}, \qquad
	\tilde{Q}_c(t) = \int \int \rho(x,t) \rho(y,t) \bs{1}_{|x-y|\leq R} (y) \, \dd x \dd y.
\]
In the sequel, we denote both quantities by $\tilde{Q}$, as their use is unambiguously defined by the data 
to which they refer.  We will shortly explain the seemingly extraneous tilde notation.

The order parameter measures the order (or disorder) of the opinions.  
One interpretation is that it measures the proportion of pairs of individuals within a radius $R$ of each other
in opinion space.  Naturally, disorder is described by a uniform distribution, whereas clustering (localised states)
corresponds to consensus.
For a uniform distribution $x_i = i/N$, for each $i$ we find $\sum_j |x_i-x_j| = 2RN$, 
and so $\tilde{Q} = 2R$, which provides an $R$-dependent lower bound for $\tilde{Q}$; the same holds for 
 $\rho(x,t) = 1$, $\forall x \in [0,1]$ in the continuum case. For a single cluster of individuals of width less than $R$, we find $\tilde{Q}=1$,
which is the maximum value of $\tilde{Q}$ for both the periodic and no-flux boundary conditions.
For $n$ equal clusters, of width less than $R$, and separated by at least $R$, we find that $\tilde{Q}=1/n$,
suggesting that the order parameter is essentially the inverse of the number of well-separated clusters.
Note that there is not an injective mapping between densities and order parameters.  
The systematic derivation of order parameters for models of opinion dynamics will be studied in
future work.  We emphasize that the identification of the order parameter for models of opinion dynamics
is not a straightforward matter, since it is not clear what the `physical' significance of such an order
parameter is.  This contrasts with other situations such as the magnetization in Ising-type models.

We now note that the order parameter for the even 2-periodic case is somewhat different.  We now have
(at least) two choices: compute the order parameter for $N$ individuals on $[0,1]$, or for $2N$ individuals
on $[-1,1]$.  In the first case, it is unclear how one should define the distance
between two individuals on $[0,1]$, since the dynamics are defined on $[-1,1]$ with even/periodic boundary conditions.  
It is also desirable to use the same definition of $I$ in both the dynamics and the order parameter; this is only possible
when defining $\tilde{Q}$ using $\rho$ on the whole domain $[-1,1]$.

Consider a uniform distribution of $2N$ individuals on $[-1,1]$ with periodic boundary conditions, which
results in $\sum_{i,j=1}^{2N} \bs{1}_{|x_i - x_j|\leq R} = 2N^2R$.  Hence, to obtain the same value
of $\tilde{Q}$ as in the periodic case, one requires a prefactor of $1/(2N^2)$.  The same normalisation
gives the corresponding results for single and multiple clusters which are well-separated from each other and
the domain boundaries.
Motivated by this, we redefine the order parameters, removing tildes:
\[
	Q_d(t) = c_{\rm BC} \frac{1}{N^2} \sum_{i,j=1}^{N_{\rm BC}} \bs{1}_{|x_i(t) - x_j(t)|\leq R}, \quad
	Q_c(t) = c_{\rm BC}\int \int \rho(x,t) \rho(y,t) \bs{1}_{|x-y|\leq R}(y) \, \dd x \dd y.
\]
where $c_{\rm BC} = 1$, $N_{\rm BC} = N$ for periodic and no-flux boundary conditions and $c_{\rm{BC}} = 1/2$,
$N_{\rm BC} = 2N$ for even 2-periodic boundary conditions. Note that, as above,
if $R$ is small, or the opinion distribution is located away from the boundaries of the domain, 
then our definition is equivalent to that used previously~\cite{KPE19}.
However, this leads to some non-standard results for $Q$ in the even 2-periodic case.  Consider a 
single cluster of particles in $[1-R/2,1]$, and the corresponding mirror cluster in $[-1,-1+R/2]$; this leads
to a value of $Q=2$.  
We use  $Q>1$ as a signature
that the even 2-periodic boundary conditions have had a significant effect on the dynamics.
We give some examples of the behaviour of the order parameter in Supplementary Material Section 
SM1.

\subsubsection{Stationary States and Convergence to Equilibrium} \label{s:tEq}

First, we remind the reader that stationary states for the Hegselmann-Krause dynamics are defined as stationary solutions of the mean field PDE, either in the absence or the presence of radicals. When writing the mean field PDE as a conservation law, Equation~\eqref{eq:PDEFlux}, then the stationary Fokker-Planck equation becomes

\begin{equation}
\label{e:stationary}
	 - \partial_x j(x) = 0, \qquad	j(x)  =  - \rho(x) \int (x-y) \rho(y) \bs{1}_{|x-y|\leq R}(y) \, \dd y 
	 - \frac{\sigma^2}{2} \partial_x \rho(x).
\end{equation}
In view of the no-flux boundary conditions $j(0) = j(1) = 0$, the stationary Fokker-Planck equation becomes
$$
j(x) = 0.
$$
Since the Hegselmann-Krause PDE admits a gradient flow formulation~\eqref{e:gradient}, it is possible to give a variational characterisation of stationary states in terms of critical points of the free energy functional~\eqref{e:free_energy}. The equivalence between the different characterisations of stationary states of the mean field PDE with periodic boundary conditions, as (a) stationary solutions of the Fokker-Planck equation, (b) critical points of the free energy functional, (c) global minimizers of the entropy dissipation functional and (d)  solutions of the Kirkwood-Monroe integral equation is presented in~\cite{CGPS20}[Prop. 2.4]. Similar results for the case of no-flux boundary conditions will be presented elsewhere~\cite{GMOP21}.

Away from the phase transition, we expect that we have exponentially fast convergence to equilibrium, measured either in relative entropy or in an appropriately weighted $L^2$ norm. It is useful to have a measure of how quickly the solution of~\eqref{eq:PDE} converges
to equilibrium, i.e. of the exponent in the exponential estimate.  We have found that
a robust measure of being close to equilibrium is that the maximum value of $\partial_t \rho$
is lower than a given tolerance $\epsilon_{\rm{eq}}$.  When this first
happens defines an equilibrium time $t_{\rm{eq}}$ for the system.  We note that this time
clearly depends on the tolerance chosen, and so should be regarded as a measure of \emph{relative}
time to equilibrium for different parameter regimes, rather than a firm statement that we have
reached equilibrium.  Unless otherwise stated, we choose $\epsilon_{\rm{eq}}=10^{-4}$.



\section{Numerical Methods} \label{s:Numerics}

In this section, we give some details of the numerical methods employed for the SDE and PDE models,
which were described in the previous section.

\subsection{Continuum Model} 

There are two principal methods that have been used to obtain accurate solutions
of the continuum equations (2.2).  The first, used by~\cite{WLEC17}, is a semi-implicit
pseudo-spectral method, which is a standard approach used in fluids dynamics and
statistical mechanics.  Their method approximates the solution at a finite number
of points in real space, with a corresponding Fourier transform.  The derivatives are straightforward
to approximate in Fourier space, by multiplication, whereas the convolutions can be
approximated by inverse Fourier transformation to real space.  
For smooth solutions, such methods are
exponentially accurate in the number of collocation points, and automatically
satisfy the periodic boundary conditions and mass conservation.  Time-stepping
is performed with a semi-implicit update.

A related method is used in~\cite{KPE19} where, rather than approximating the solution
on a spatial grid, it is expanded in a basis of the corresponding Hilbert space, with the
expansion truncated on a finite set.  This leads to a system of ODEs for the coefficients
of the basis functions, which then produce an approximation to the solution of the PDE.
Once again, this is particularly suited to periodic geometries, as there is a clear basis
set to use (Fourier or complex exponentials), and the periodic boundary conditions
are applied through the use of periodic functions.

We note now that the no-flux case is more challenging in this regard.  The boundary
condition is now non-local, since the flux (see~\eqref{eq:PDEFlux}) contains the interaction term.  Our
approach is based on the Matlab package 2DChebClass~\cite{2DChebClass}, 
which utilises a Chebyshev pseudospectral
method in which the convolution integrals are computed in real space.  This has been
shown to give highly accurate and efficient solutions for related non-local, non-linear
PDEs in one and two dimensions.  The main challenges over standard implementations~\cite{Trefethen00,Boyd01}
are: (i) how to compute the convolutions with a finite support
($[x-R,x+R]$); (ii) how to enforce the boundary conditions.  

\subsubsection{Computation of Convolutions}

For (i), we note that the problem has already been solved in 2DChebClass~\cite{2DChebClass}, in particular for the 
computation of hard-rod free energies in (dynamical) density functional theory.
In short, for a given computational point $y_n$, the interval of interest is
$\mathcal{I}_{R,n} = [ \max(0,y_n-R), \min(1,y_n+R)]$.  
For a pseudospectral method, given a vector of $M$ spatial collocation points $\bs{y} = \{y_m\}$, 
a function $f$ is approximated by a vector of values at each of these points, which we denote by
$\bs{f}[\bs{y}]$, such that $\bs{f}_m = f(y_m)$.
Suppose we wish to evaluate the convolution
\[
	C_R(y) = \int_{ \mathcal{I}_R } K(y-y') \rho(y') \dd y'
\]
for some explicit function $K$, an interval $\mathcal{I}_R =  [ \max(0,y'-R), \min(1,y'+R)]$, and a general $\rho$.
This can be computed straightforwardly through
matrix multiplication, as follows:
\begin{itemize}
	\item Discretise the whole domain $[0,1]$ with a vector of 
		$N$ Chebyshev collocation points, $\bs{y} = \{y_n\}$;
	\item Choose a discretisation point $y_i$ and determine the interval
		$\mathcal{I}_{R,i} = [ \max(0,y_n-R), \min(1,y_n+R)]$;
	\item Discretise the interval $I_n$ with $\hat{N}$ Chebyshev collocation points, 
		$\hat{\bs{y}}_i = \{ \hat{y}_{i,\hat{n}} \}$;
	\item Compute the corresponding integration weight (row) vector corresponding to 
		$\hat{\bs{y}}_i$, denoted $\hat{\bs{w}}_i$, such that,
		$\hat{\bs{w}}_i  \times \bs{f}[\hat{\bs{y}}_i] \approx \int_{\mathcal{I}_{R,i}} f(y) \dd y$;
	\item Compute the $\hat{N}\times N$ (barycentric) interpolation matrix, $B_i$, from $\bs{y}$ to $\hat{\bs{y}}_i$,
		such that $B_i \times \bs{f}[\bs{y}] \approx \bs{f}[\hat{\bs{y}}_i]$.
	\item Evaluate the (explicit) kernel $K$ at $y_i - \hat{\bs{y}}_i$ with values placed on the diagonal 
		of a $\hat{N} \times \hat{N}$ matrix, with all other entries zero. We denote this matrix $K_i$;
	\item The $i$-th row of the $N \times N$ convolution matrix $C_R$ is then given by
		$(C_{R,i}) =  \hat{\bs{w}}_i \times K_i \times B_i$.
\end{itemize}
Then, for any length $N$ vector $\bs{\rho}$ with $\bs{\rho}_n = \rho(y_n)$, we have
 $C \times \bs{\rho} \approx \int_{  \mathcal{I}_R } K(y-y') \rho(y') \dd y'$.

In words, for each collocation point, we first interpolate $\rho$ from the full interval $[0,1]$
onto the convolution domain $\mathcal{I}_R$, giving $\rho(y')$.
We then multiply by the kernel evaluated at $y-y'$, giving $K(y-y')$.  Taking the product of these
two results and multiplying by $\hat{w}_i$ then results in an approximation to the convolution.
The main advantage of this procedure is that the convolution matrix depends only on $R$ and $K$,
and thus needs only to be computed once for each such pair.  It can then be reused at each timestep
within a single computation and, indeed, for any computations involving the same convolution.
Additionally, exactly the same procedure may be applied to the periodic domains, but with the 
simplification that the length of the convolution domain is the same for each collocation point.

\subsubsection{Enforcing Boundary Conditions}

It remains to determine how to accurately enforce the no-flux boundary conditions, i.e.\ point (ii) above.
After discretising space with a set $\{y_n\}$ of $N$ collocation points, a PDE is reduced to a set of $N$
(coupled) ODEs, describing the value of $\rho$ at each of the $y_n$.  In particular, there are $N-2$
collocation points in the interior of the domain, and 2 collocation points on the boundary (here, the points
at 0 and 1). Two standard approaches to tackling boundary conditions in pseudospectral methods 
are interpolant restriction and boundary bordering~\cite{Trefethen00,Boyd01}.  

For interpolant restriction, one would typically solve a problem with zero Dirichlet (or other simple) 
boundary conditions and then add on a particular solution which satisfies the desired boundary conditions.
The challenge when applying this method lies primarily in determining a particular solution
of the (non-local, non-linear) equations which satisfies the boundary conditions; this is far from straightforward in general,
and also leads to a problem-specific numerical method.

Suppose the system of ODEs is
written as $\dd \rho / \dd t = L \rho$, for some matrix $L$ corresponding to a discretised linear operator.  
Then the boundary conditions may be imposed by deleting rows and columns of the matrix $L$ and
replacing them by rows and columns which impose the desired boundary conditions.  Note that this is
again a problem-specific approach, and the substitution effectively has to be performed `by hand'.  It
is also still unclear how this could be performed straightforwardly for a non-local, non-linear problem,
such as those studied here.

Instead, we use a similar, but more general approach of differential-algebraic equations.  We 
replace the set of $N$ ODEs with the $N-2$ ODEs on the interior, and two
algebraic boundary conditions.  This set of equations may then be solved with a standard
numerical scheme, such as Matlab's ode15s.  The key benefits here are that it is
straightforward to implement for non-local, non-linear boundary conditions, and one does not
need to determine anything `by hand'.

\subsubsection{Even 2-Periodic Boundary Conditions}

We now briefly describe how to enforce the even 2-periodic boundary conditions.  This is done
by working on a periodic domain $[-1,1]$, using an analogous discretisation as in the standard
periodic case.  To enforce the evenness of the solution, we define it only on half of the domain,
$[0,1]$ and then, when operations such as differentiation, integration, or interpolation need to
be applied, we mirror it to the full domain $[-1,1]$.  Note that, for technical reasons, our numerical
implementation actually stores $\rho$ on $[-1,0]$, but the overall idea is identical.  This `mirroring'
operation automatically satisfies the periodic boundary conditions of the solution.  However,
we note that it can lead to numerical challenges since it does not enforce smoothness of the 
solution at 0 or $\pm 1$.  This can be overcome by choosing an initial condition which is smooth
on the whole domain as, at least morally, the diffusive part of the PDE ensures that this smoothness
is conserved.  However, in our examples, since we define the initial condition as a periodic
function on $[0,1]$, such smoothness is not guaranteed.

\subsubsection{Computational Details}

As described above, we use pseudospectral methods with Fourier (periodic, even 2-periodic)
or Chebyshev (no-flux) collocation points.  In each case we use 200 such points, distributed either on
$[0,1]$ (periodic, no-flux) or $[-1,1]$ (even 2-periodic), and interpolate to 400 evenly spaced points
for plotting.  For the convolution integrals we choose $\hat{N} = 100$.  We note that these numbers
may seem high for pseudospectral approaches, but here there are (at least) two motivating reasons.
Firstly, we require highly accurate solutions over very long times (e.g., up to time $10^4$ or $10^5$)
and, secondly, in many parameter regimes, the densities become strongly peaked with very
high derivatives.  We have found that these constraints lead to a requirement for a relatively
large number of points \emph{for certain parameter regimes}, notably those with small $\sigma$
and large $R$.  This is unexpected as such problems become highly advection-dominated.  
For ease of comparison, we  use the same number of points for all computations.  
We note that a typical dynamical computation
takes significantly less than a second on a standard laptop, whilst precomputing the convolution integral matrices
(which needs to be done only once for each set of $\{$function, boundary condition, $R\}$)
takes on the order of 20--30s for the periodic geometries, and 5s for the Chebyshev case.

For the resulting discretised system of ODEs, we use Matlab's ode15s routine, with 
relative and absolute tolerances of $10^{-6}$ and $\epsilon_{\rm{eq}} = 10^{-4}$.  We note that for
the vast majority of the computations shown here it is possible to reduce the tolerances to around
$10^{-10}$ or even $10^{-12}$, with correspondingly smaller  $\epsilon_{\rm{eq}}$ but, again for
consistency, we fix these values for all computations, unless otherwise stated. 

\subsubsection{Validation Against Existing Results}
We refer the reader to Section SM2 of the Supplementary Material for an extensive validation
of our numerical codes against existing results from~\cite{WLEC17} and~\cite{KPE19}.

\subsection{Discrete Dynamics}

For the numerical solution of the SDE, we use the standard Euler-Maruyama method.  We note that,
since the noise in the model is purely additive, in our case this is equivalent to the Milstein method,
and hence has both weak and strong orders of convergence equal to the time step, $\dd t$. 
Unless otherwise stated, we average our results over $10$ runs of $10^4$ particles with $\dd t = 10^{-2}$.

Since the method is standard, we only highlight some particular subtleties with the implementation
of the boundary conditions.  Here we consider a set of positions 
$\bs{x}(t) = \{x_1(t), \ldots, x_N(t), x_{N+1}, \ldots, x_{N+N_r}\}$  at time $t$, where the first $N$ positions
correspond to the normals, and the remaining $N_r$ to the radicals.
We define the (signed) separation between two positions $x_i$ and $x_j$ by $S(x_i,x_j)$; for the
no-flux case this is simply $x_i-x_j$, whilst for the periodic cases it is the minimum separation
of the two particles.  For example, for a periodic interval of length 1, $S(0.8,0.2)$ is -0.4, rather
than 0.6.  Note that the lengths of the periodic and even 2-periodic intervals are one and two, 
respectively.
We define the set of particles within the bounded confidence interval of particle $i$ as
$\mathcal{I}_i(\bs{x}) = \{ j : | S(x_i,x_j) \leq R \}$.

Given $\bs{x}(t)$, the proposed positions, ignoring the boundary conditions, 
at the next time step $t + \dd t$ are given by
\begin{align*}
	\tilde{x}_i(t+\dd t) &= x_i(t) - \frac{1}{N}  \sum_{j \in \mathcal{I}_i( \hat{ \bs{x}} )} S(x_i,\hat{x}_j)
			   + \sigma \sqrt{\dd t} * w_i(t), \quad i = 1, \dots, N\\
	\tilde{x}_i(t+\dd t) &= x_i(t), \quad i = N+1, \dots, N+N_r,	
\end{align*}
where $w_i(t)$ is a random variable drawn from a normal distribution with mean zero and variance one.
Here, for the periodic and no-flux boundary conditions, $\hat{\bs{x}} = \bs{x}$, whereas for the 
even 2-periodic case, $\hat{\bs{x}} = \bs{x} \cup -\bs{x}$, i.e.\ the union of the positions and their
reflections.

Given the proposed positions at time $t + \dd t$, it is necessary to apply the boundary conditions.
The periodic case is standard; after each time step we ensure that the positions lie in the interval $[0,1]$
by taking $x_i(t+\dd t) = \tilde{x}_i(t+\dd t) \mod 1$.  For the no-flux case, we use standard
reflective boundary conditions~\cite{P14}: 
If $\tilde{x}_i(t + \dd t) \in [0,1]$ then set $x_i(t+\dd t) = \tilde{x}_i(t+\dd t)$.
If $\tilde{x}_i(t + \dd t) \notin [0,1]$ then propose a new $\tilde{x}_i(t + \dd t)$
by reflecting at the closest boundary (i.e.\ 0 or 1); repeat until $\tilde{x}_i(t + \dd t) \in [0,1]$ 
and then set $x_i(t+\dd t) = \tilde{x}_i(t+\dd t)$.  Note that, in practice, for sufficiently small $\dd t$,
it is highly unlikely to require more than one reflection at a given time step to ensure that $\tilde{x}_i$
lies in the interval.  For the even 2-periodic case, by the symmetry of the problem any individual
moving out of the interval $[0,1]$ has a corresponding `mirror' individual that moves exactly the same 
distance into the interval across the same boundary.  Thus, one way of implementing the even 2-periodic
boundary conditions is to apply the same procedure as for the no-flux case.  Note that this \emph{does not}
result in the same dynamics due to the different choices of $\hat{\bs{x}}$.

Finally, we note that it is necessary to be able to sample from a given density of particles, both
for the initial condition and for the radical population.  For simple distributions, such as uniform
or Gaussian, this can be achieved using standard pseudorandom number generators.  However,
to provide a general approach, for example to sample from the triangular radical distribution used
in~\cite{KPE19}, we implemented a sampling algorithm using slice sampling, which is a Monte-Carlo
approach capable of sampling from any given density.  To compare such densities, and the
results of the SDE dynamics, to the PDE results, it is necessary to histogram the results and 
correctly normalise them; for a a system with $N$ particles, which is to be averaged over
$N_{\rm runs}$ runs, and binned into a histogram with bin width $\dd x$, with bin counts $\{b_i\}$,
the correct normalisation is  $b_i/(N N_{\rm runs} \dd x)$.

\subsection{Comparison Between the Discrete and Continuum Models}
We direct the reader to Sections SM3 and SM4 of the  Supplementary Material for a comparison
between the SDE and PDE models.  Of particular note is that a relatively large number of agents
(of the order $10^3$--$10^4$) are required to ensure good agreement between the two models.
This is a significantly larger number of agents than used in previous studies, e.g.,~\cite{WLEC17,KPE19}.


\section{Numerical Experiments: Opinion Dynamics in the Absence of Radicals}
\label{s:NoRadicals}

We begin our numerical experiments by studying noisy opinion dynamics models without radicals.  
For a given boundary condition, there are only three other properties of the model: 
the values of $R$ and $\sigma$, and the initial condition.  

\subsection{Uniform Initial Condition}

\begin{figure}
\centering
\resizebox{\figwidth}{!}{
\raisebox{2cm}{\includegraphics[height = 2.5cm]{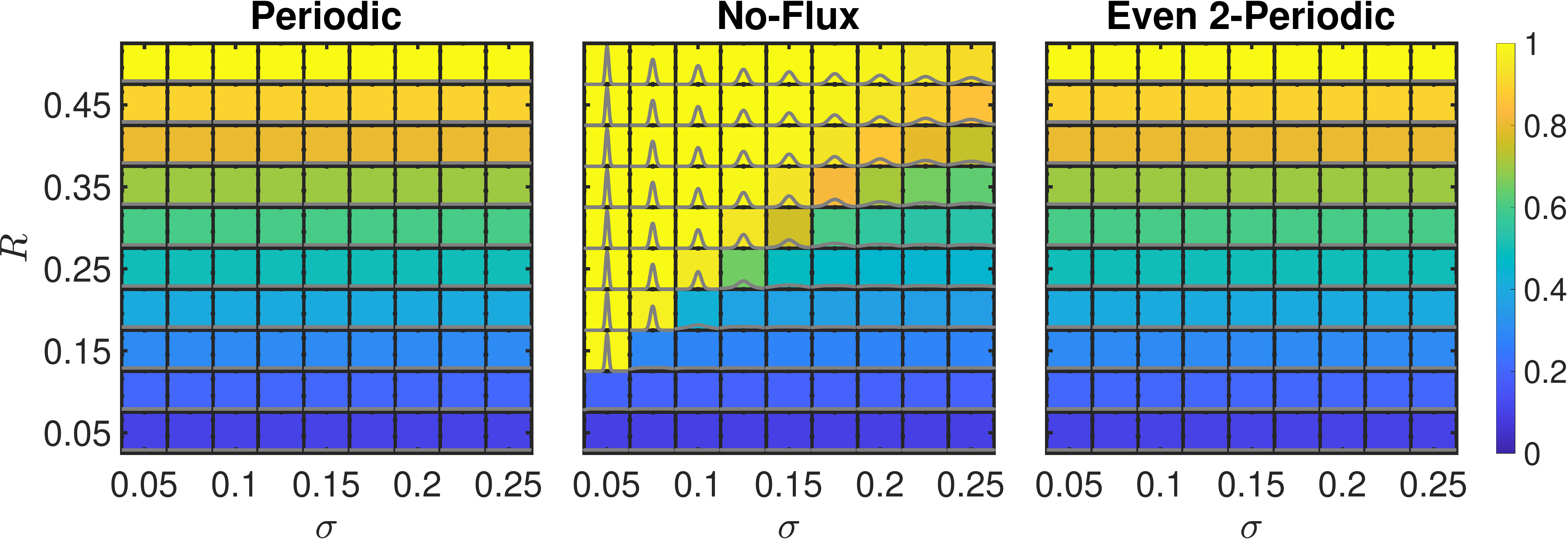}}
\includegraphics[height = 4.5cm]{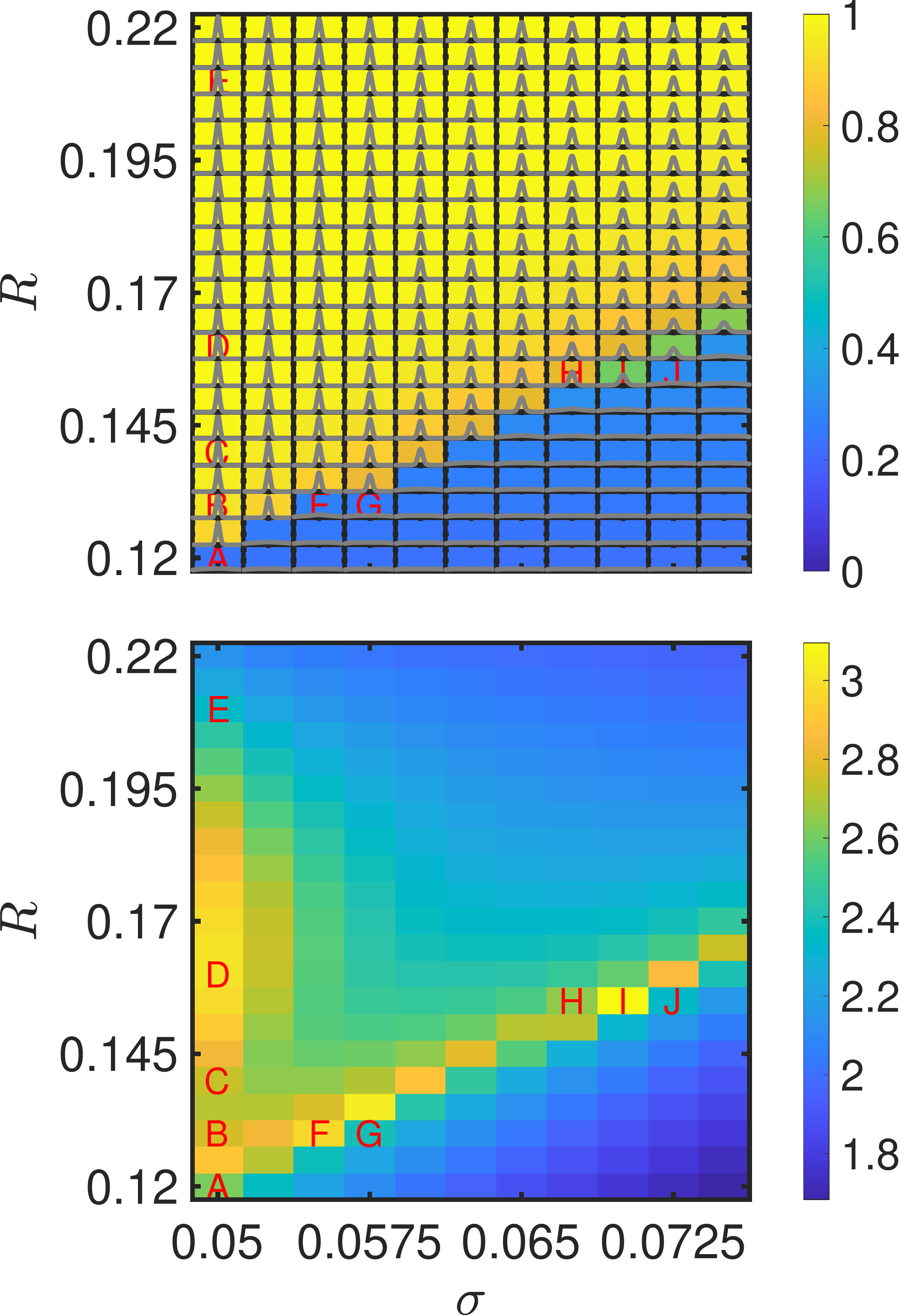}
\includegraphics[height = 4.5cm]{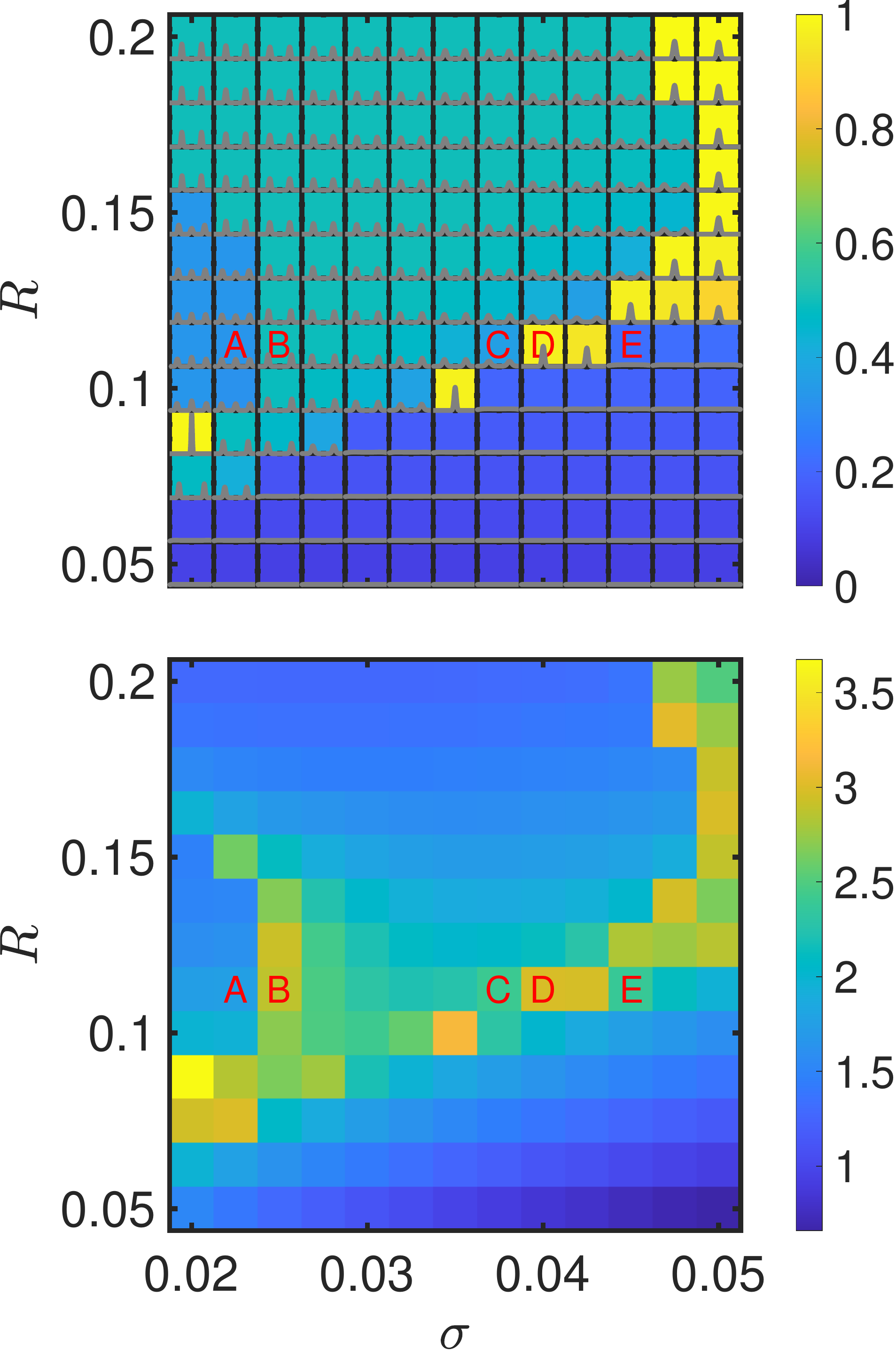}
}
\caption{Results for a uniform initial condition with varying $R$ and $\sigma$.
Colour denotes the order parameter at the final time (top) and equilibration time on a log scale (bottom); grey lines
are the final densities.  The middle and right plots are for no-flux boundary conditions and letters refer to
snapshots in Figures~\ref{fig:Uniform_Zoom_Snapshots} and~\ref{fig:Uniform_Small_Sigma_Snapshots}.}
\label{fig:Uniform}
\end{figure}

Our first system has a uniform initial condition, which is standard in many
opinion dynamics models, see, e.g. ~\cite{HK02,PLR06,MT14,HK15,HK17,WLEC17,KPE19}.
This initial condition is a steady state for both the periodic and even 2-periodic boundary conditions, which can
be seen in Figure~\ref{fig:Uniform}, and easily shown using the stationary PDE.  Here the different
values of the order parameter are due only to the intrinsic dependence of $Q$ on $R$.
However, this is the first demonstration of the significant effect of the boundary conditions
on the dynamics; the uniform distribution \emph{is not} a steady state for the no-flux case,
and the density tends to cluster in the centre of the interval. This is a result of the 
density being zero outside the domain $[0,1]$, so there tends to be an inwards net force on 
the individuals.  There is a clear trend as $R$ and $\sigma$ change: increasing $R$ tends
to enhance the cluster formation, whilst the opposite is true for increasing $\sigma$.
This has a simple explanation: for larger $R$ there is more interaction, and hence 
the individuals tend towards what is essentially the global mean, rather than the local
mean with a small confidence interval; increasing $\sigma$ adds more diffusion 
to the system, which tends to disperse clusters.

There is another striking feature of the middle plot in Figure~\ref{fig:Uniform}, namely
that there seems to be a sudden switch in behaviour as $R$ and/or $\sigma$ are varied,
with the long-time density being either strongly peaked or almost flat.  This is particularly
noticeable when $\sigma$ is small.  As such, in the middle panel of Figure~\ref{fig:Uniform} we zoom
in to a transition region (left panel) and also consider smaller values of $\sigma$ (right panel).
The bottom panels of Figure~\ref{fig:Uniform}
show the time to equilibrium, as defined in Section~\ref{s:tEq}, on a $\log$ scale.  
The dynamics take appreciably longer to reach equilibrium in regions where the long-time dynamics
is particularly sensitive to the choice of parameters.  Note that in the small-$\sigma$ case it 
was necessary to increase the equilibration tolerance to $\epsilon_{\rm eq} = 5 \times 10^{-4}$; there is overlap
between the two plots at $\sigma = 0.05$ to aid comparison.

\begin{figure}
\centering
\resizebox{\figwidth}{!}{
\includegraphics[width = 2.6cm]{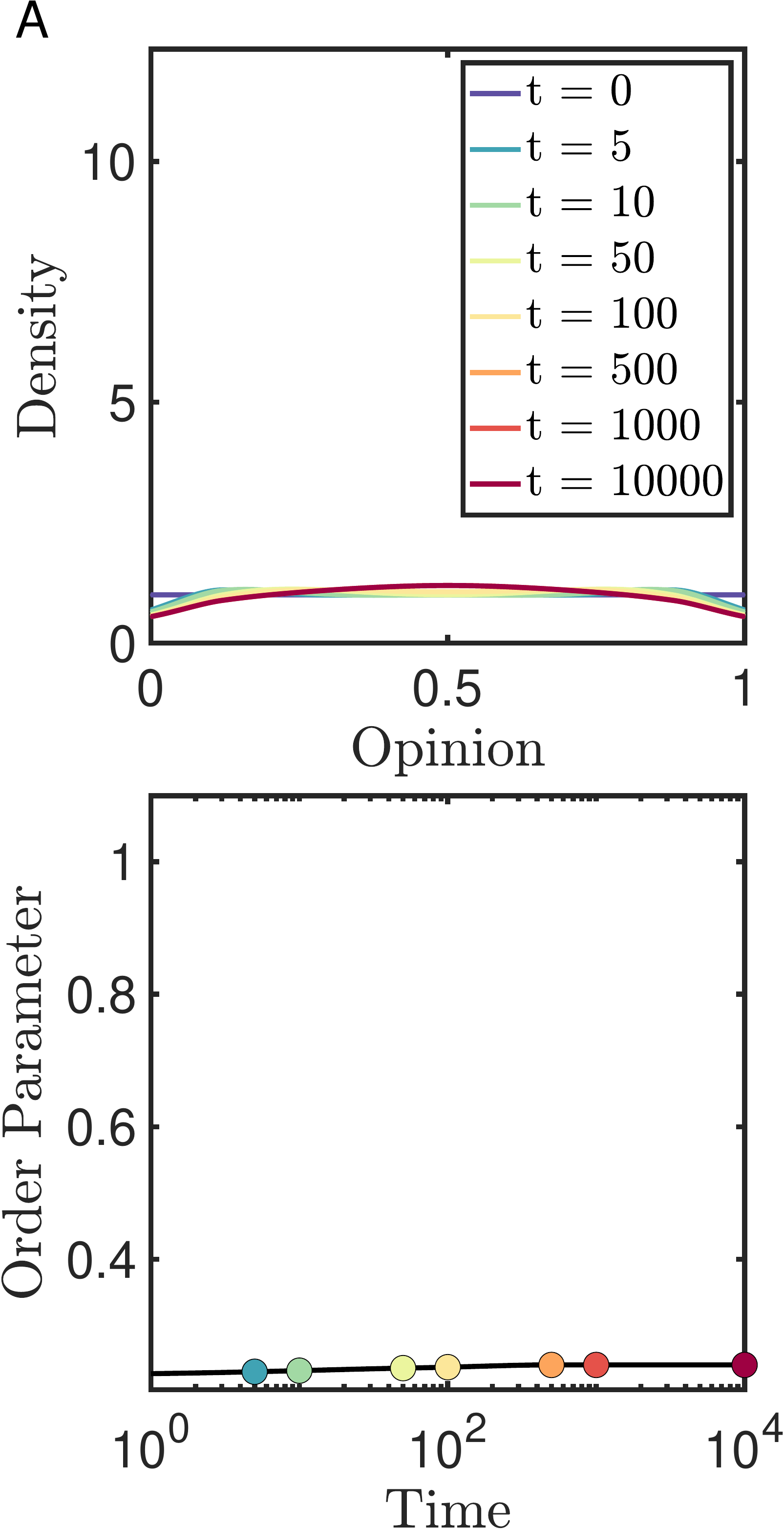}
\includegraphics[width = 2.6cm]{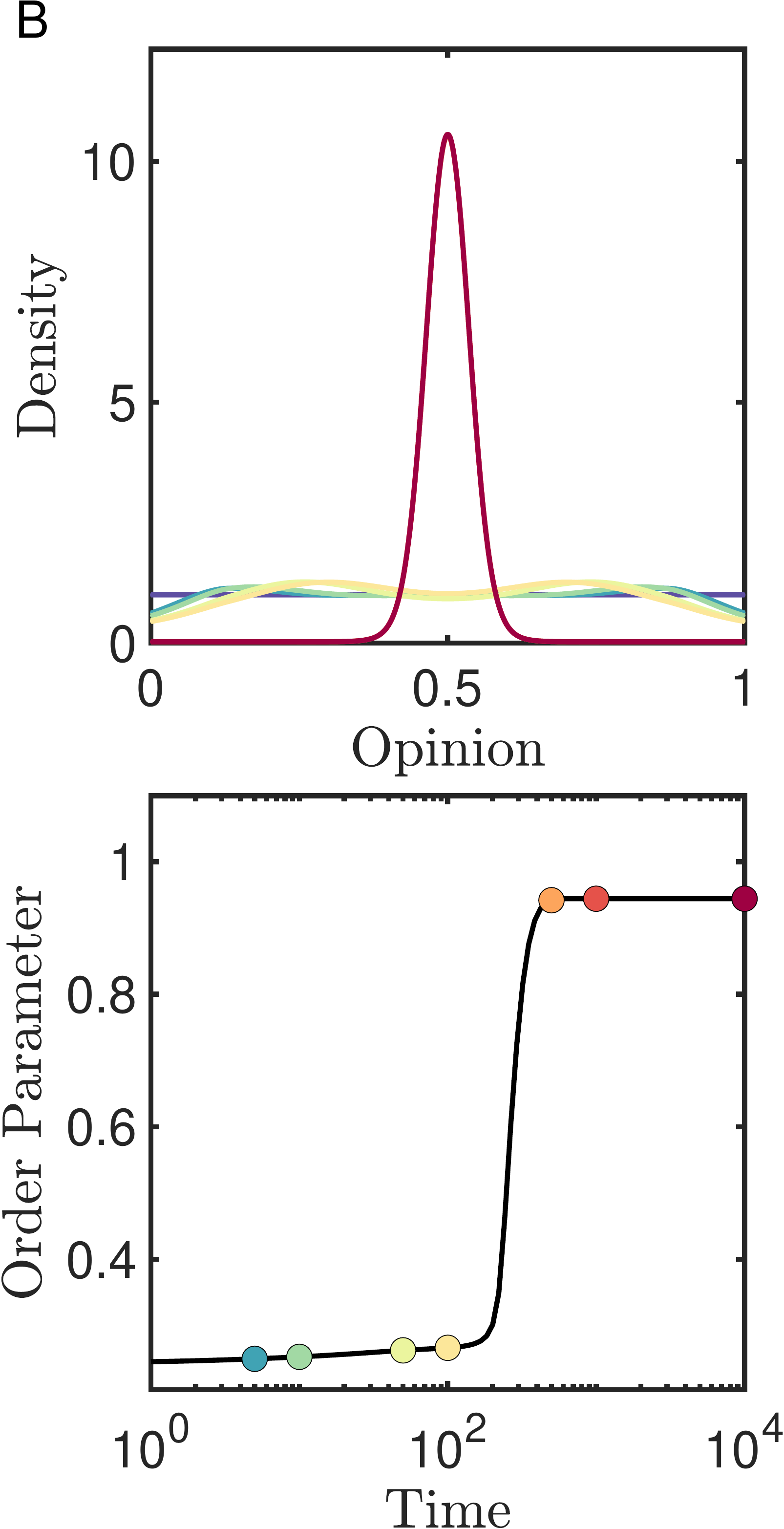}
\includegraphics[width = 2.6cm]{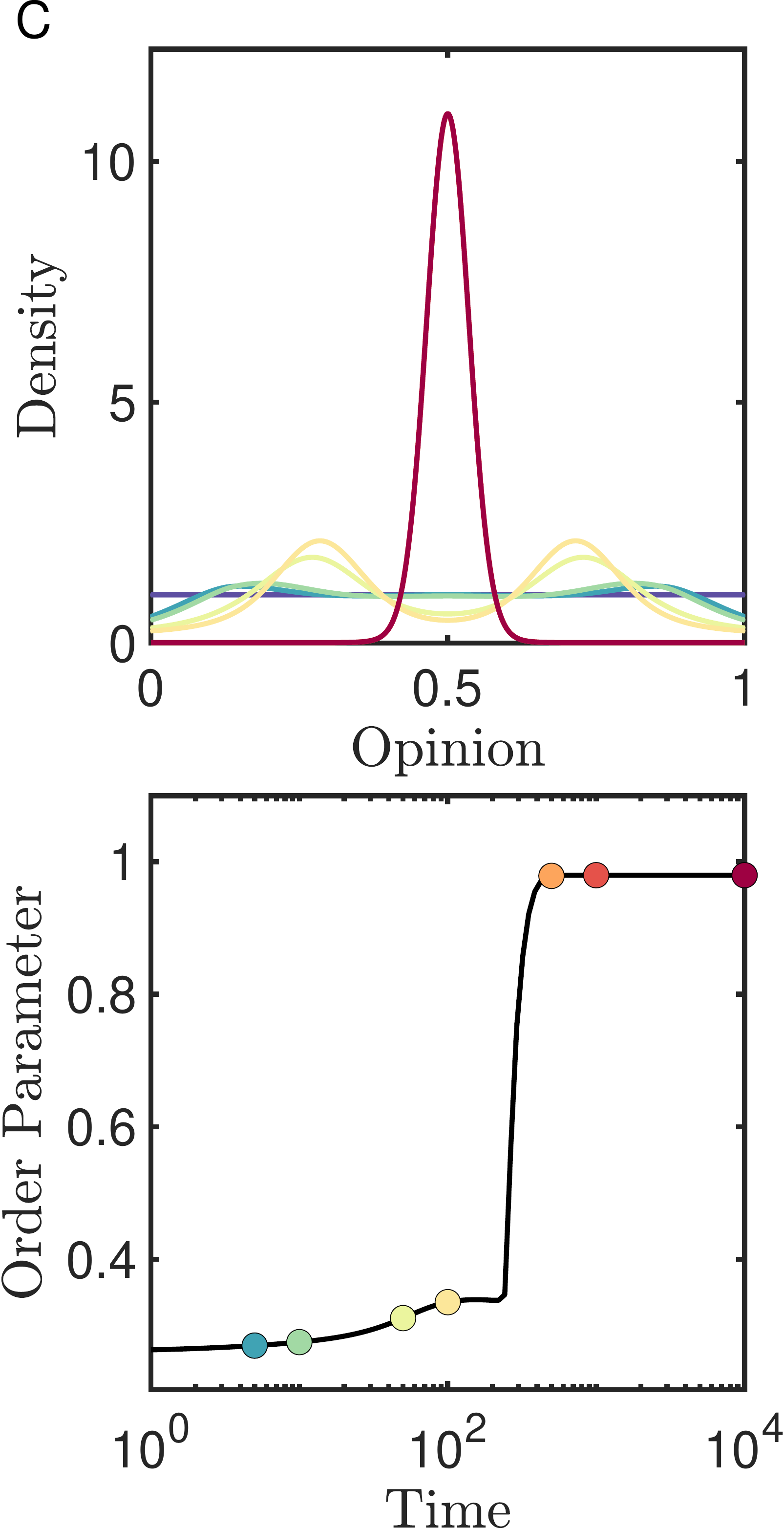}
\includegraphics[width = 2.6cm]{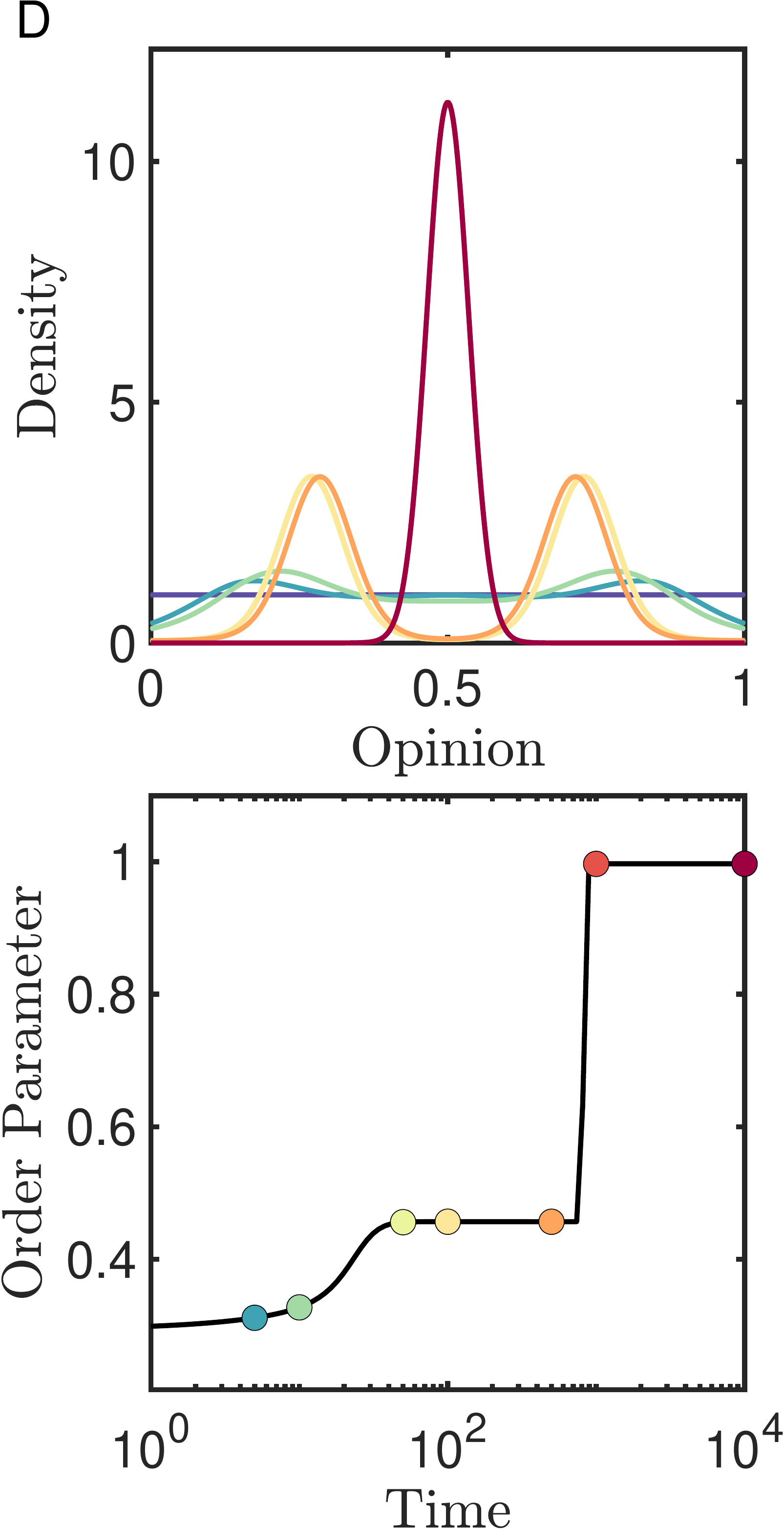}
\includegraphics[width = 2.6cm]{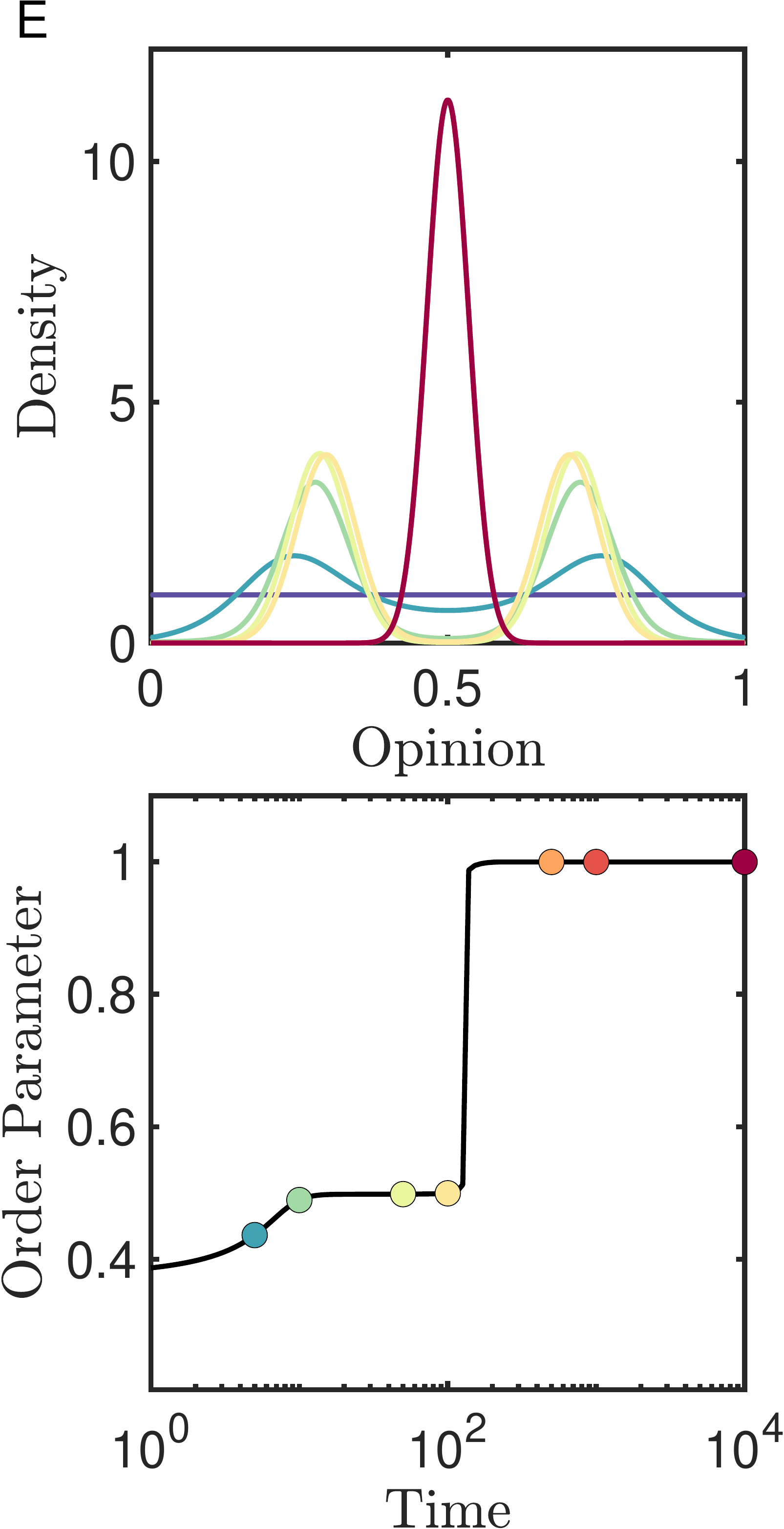}}\\
\resizebox{\figwidth}{!}{
\includegraphics[width = 2.6cm]{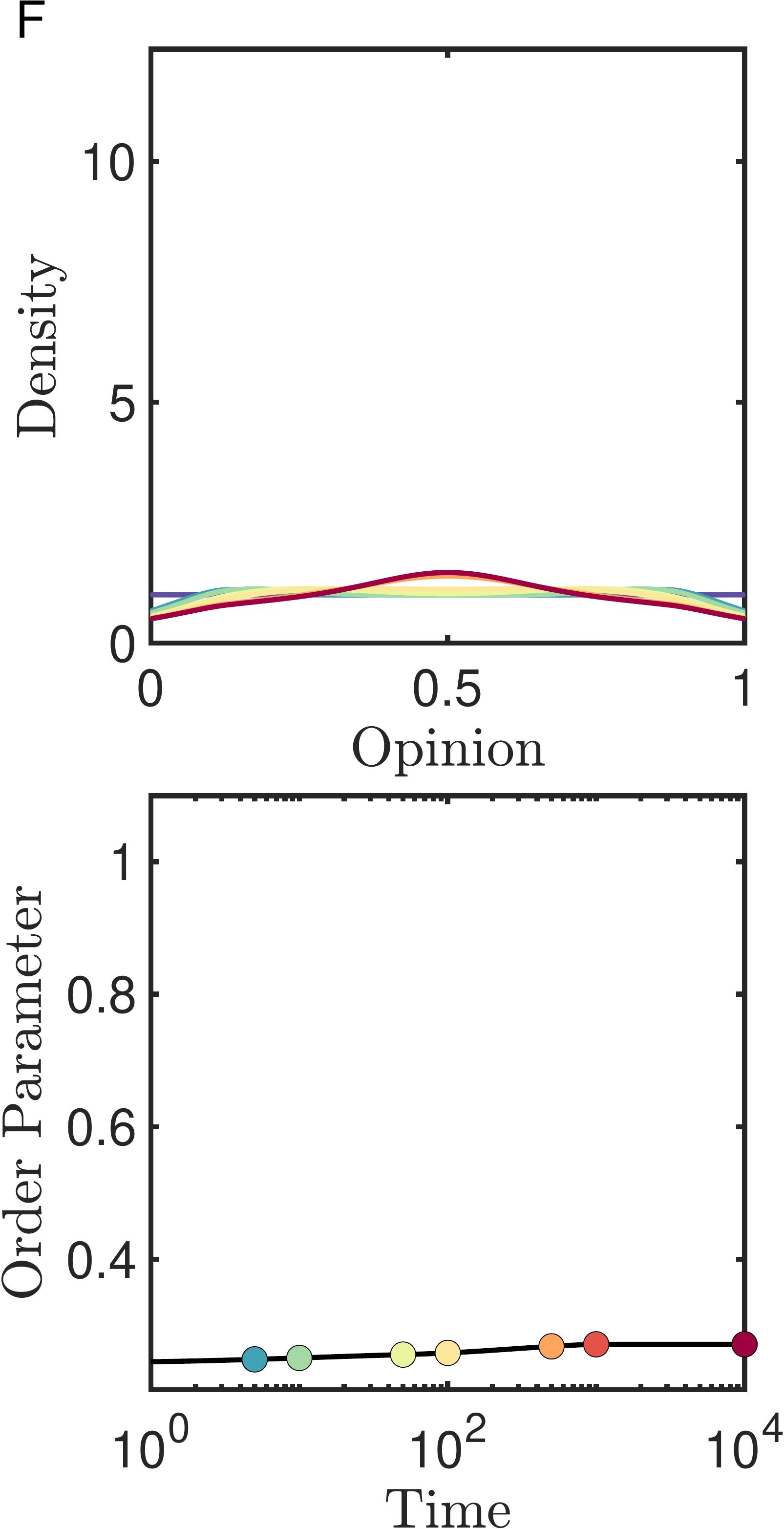}
\includegraphics[width = 2.6cm]{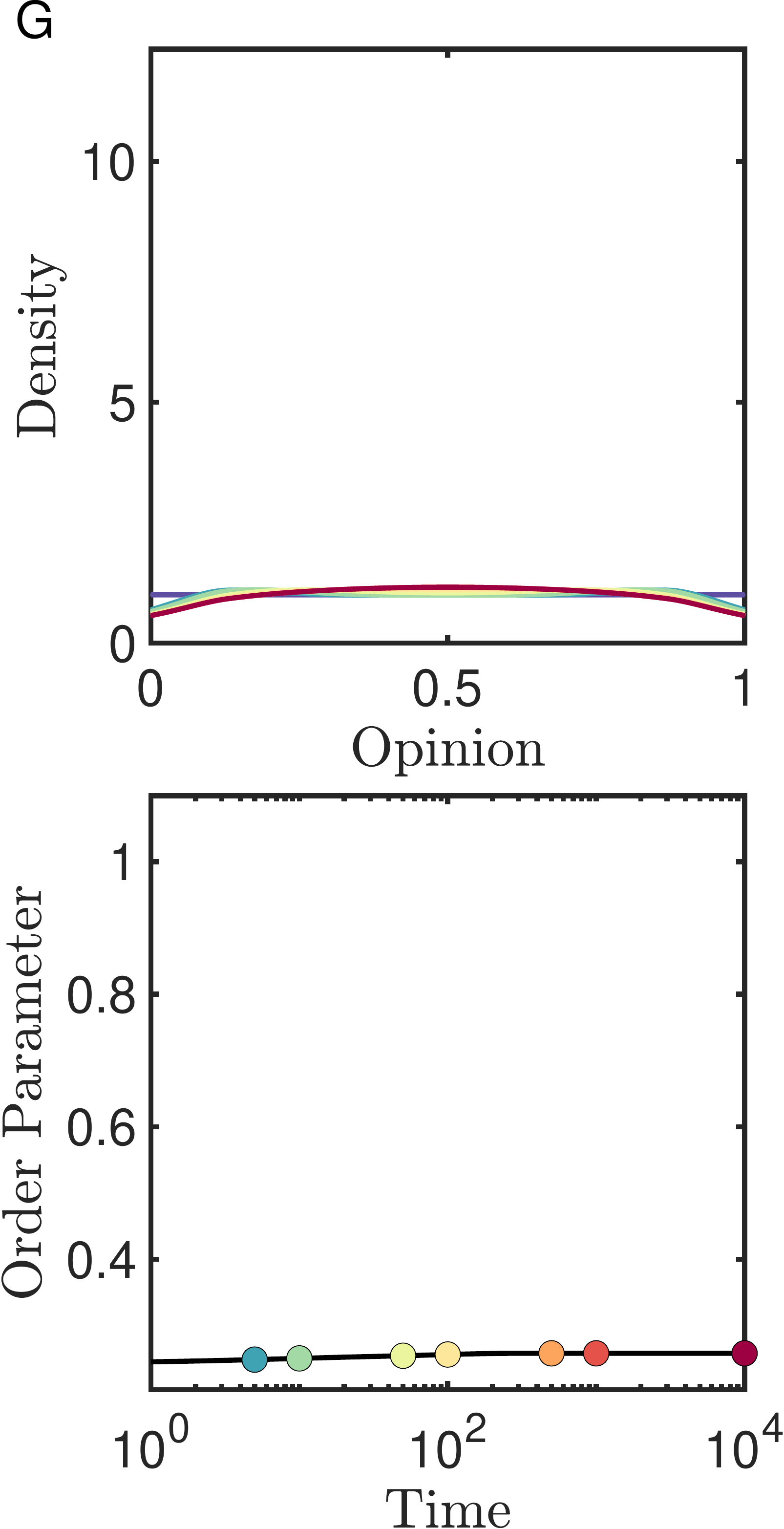}
\includegraphics[width = 2.6cm]{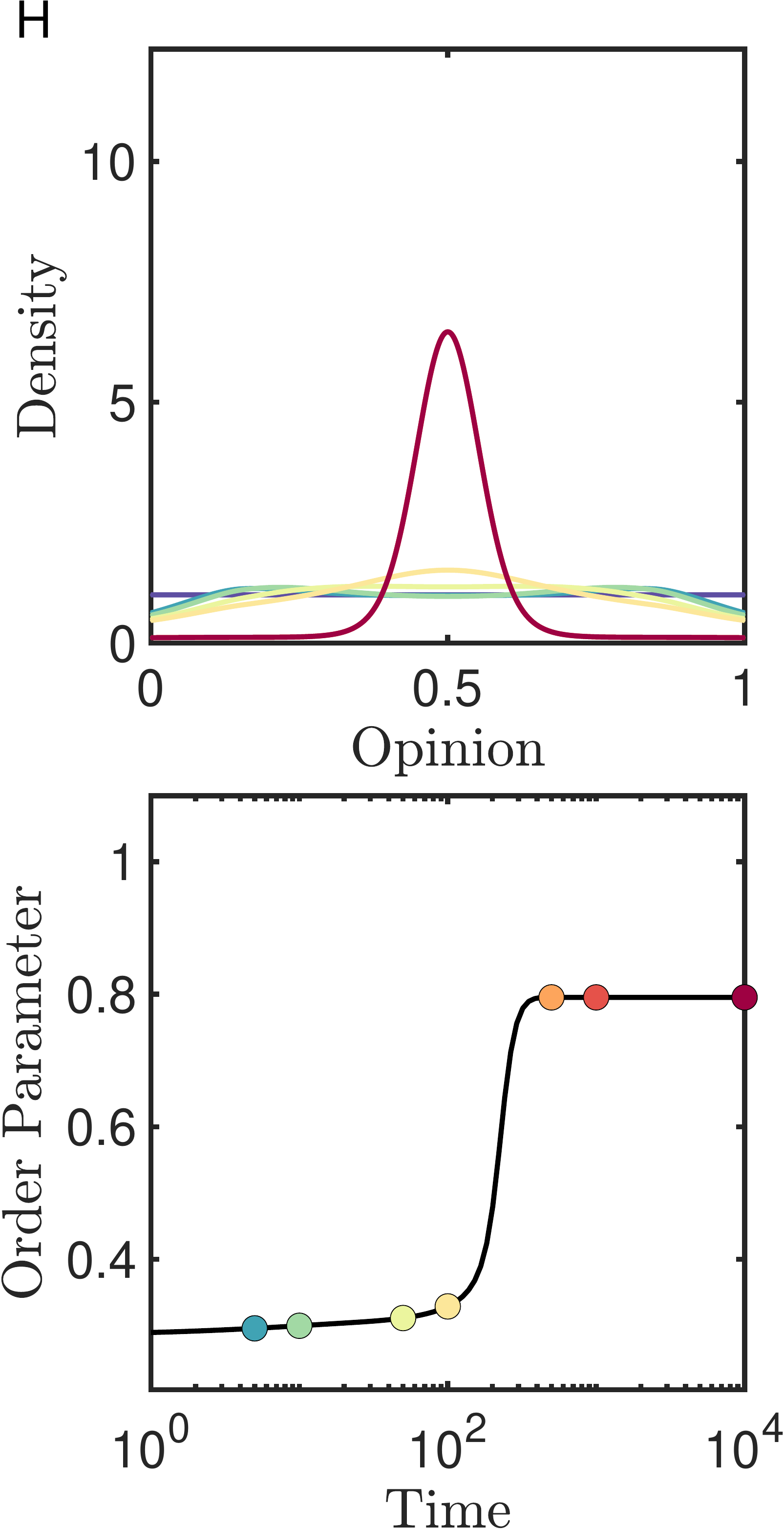}
\includegraphics[width = 2.6cm]{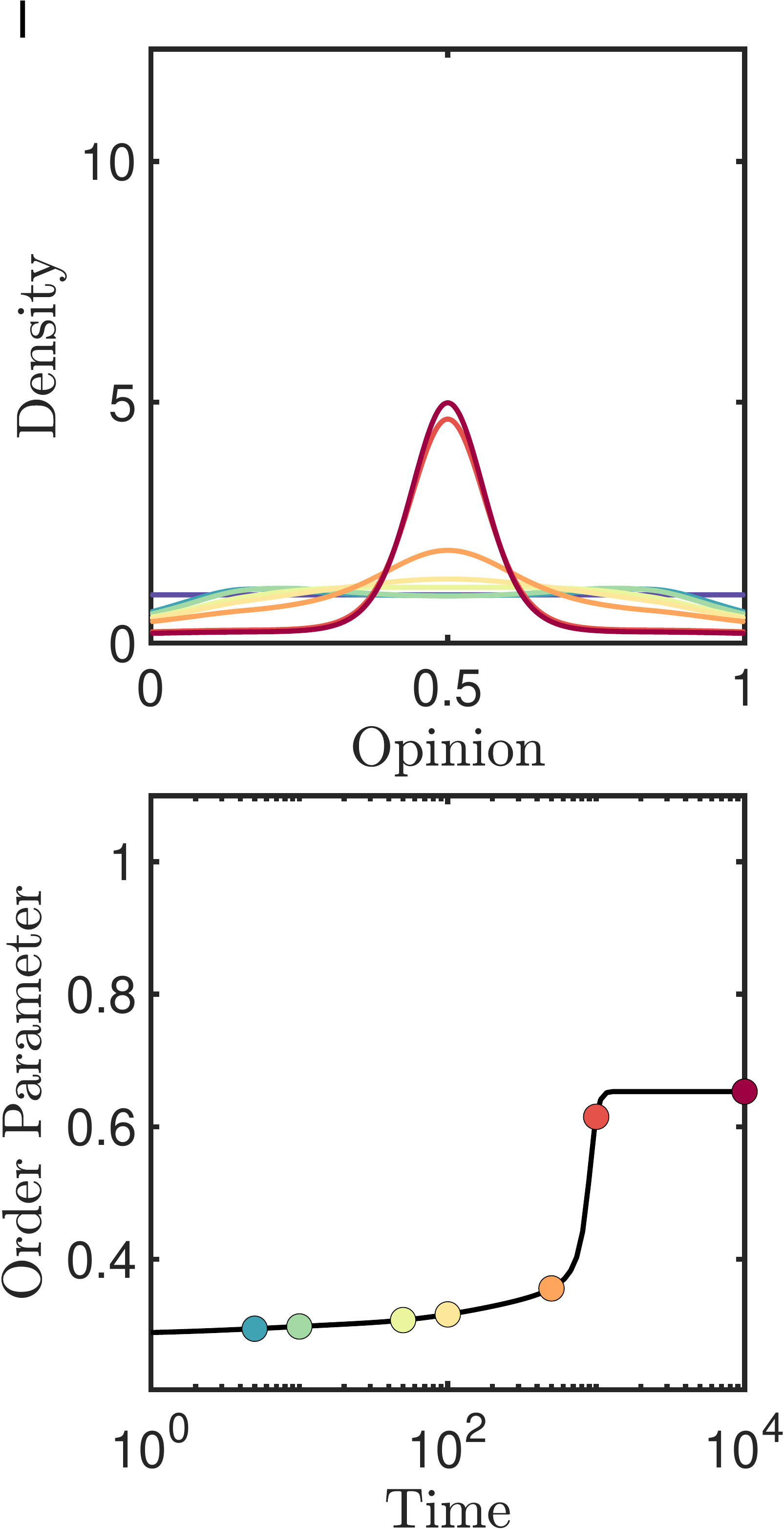}
\includegraphics[width = 2.6cm]{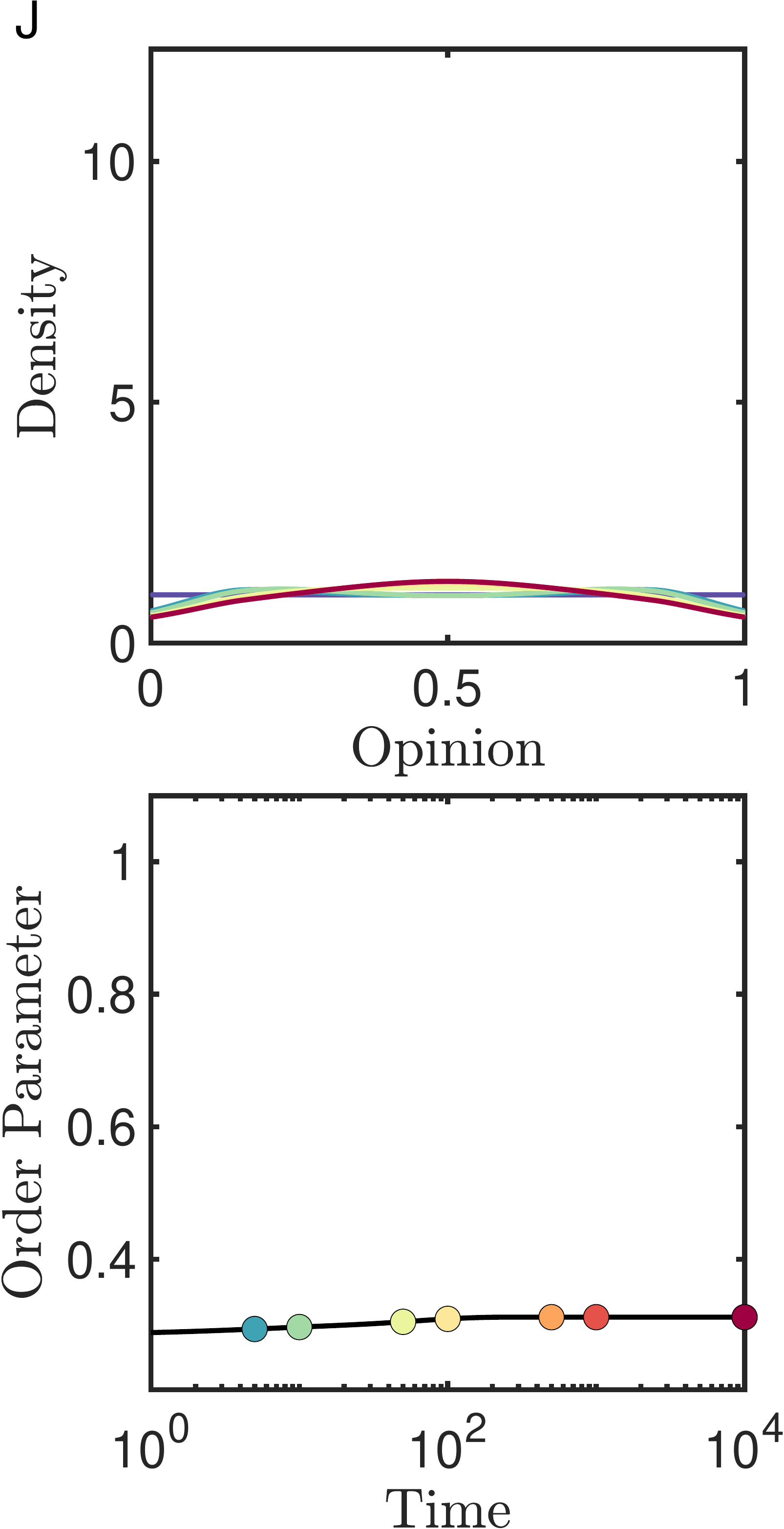}
}
\caption{Results for a uniform initial condition and no-flux
boundary conditions. Letters correspond to labelling in the middle panel of Figure~\ref{fig:Uniform}.  
In each case, we show snapshots of the densities at the indicated times (top) and 
the time evolution of the order parameter (bottom); coloured dots
show the order parameter at the times of the corresponding snapshots
in the top panel.}
\label{fig:Uniform_Zoom_Snapshots}
\end{figure}

\begin{figure}
\centering
\resizebox{\figwidth}{!}{
\includegraphics[width = 2.6cm]{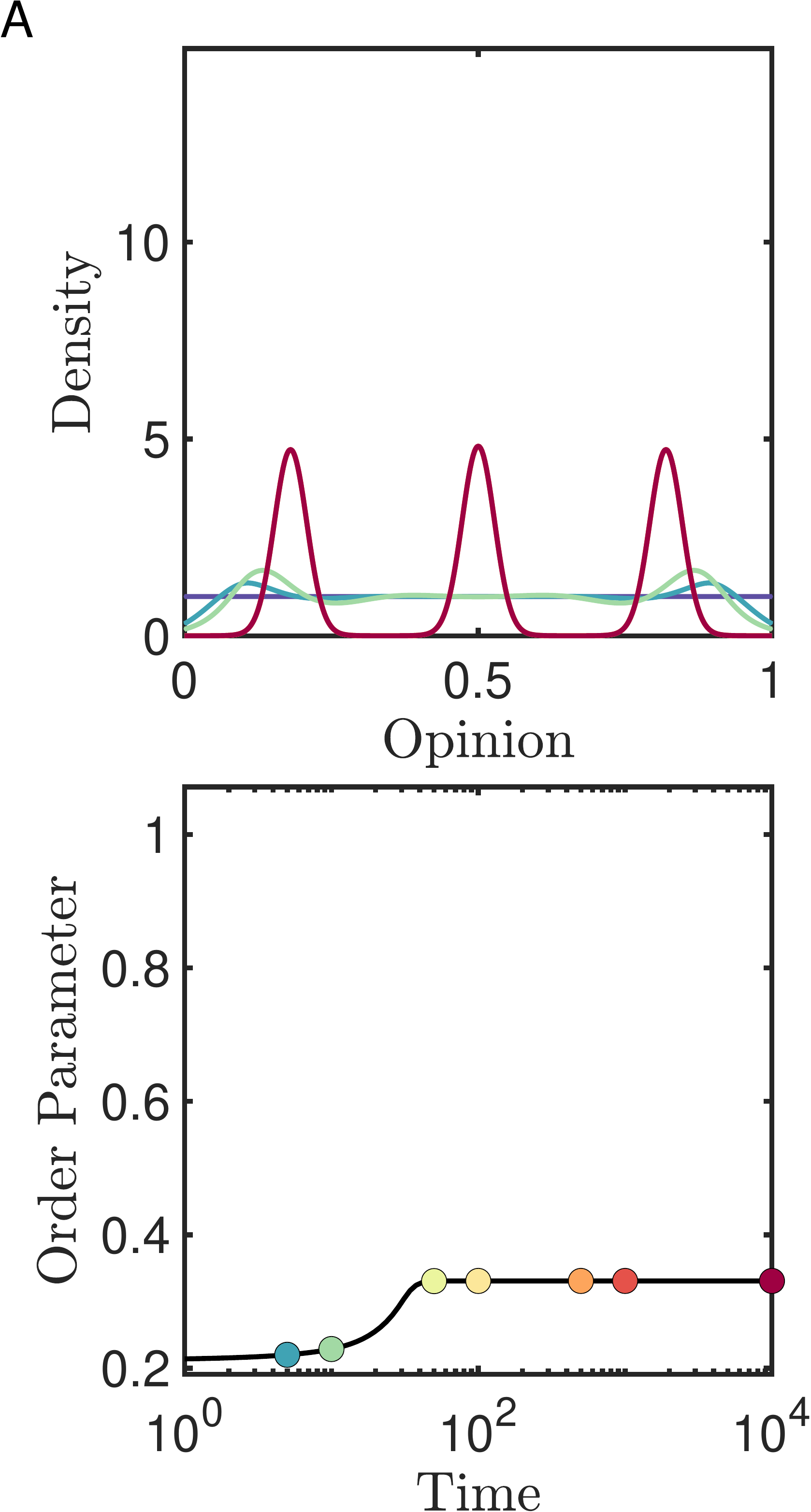}
\includegraphics[width = 2.6cm]{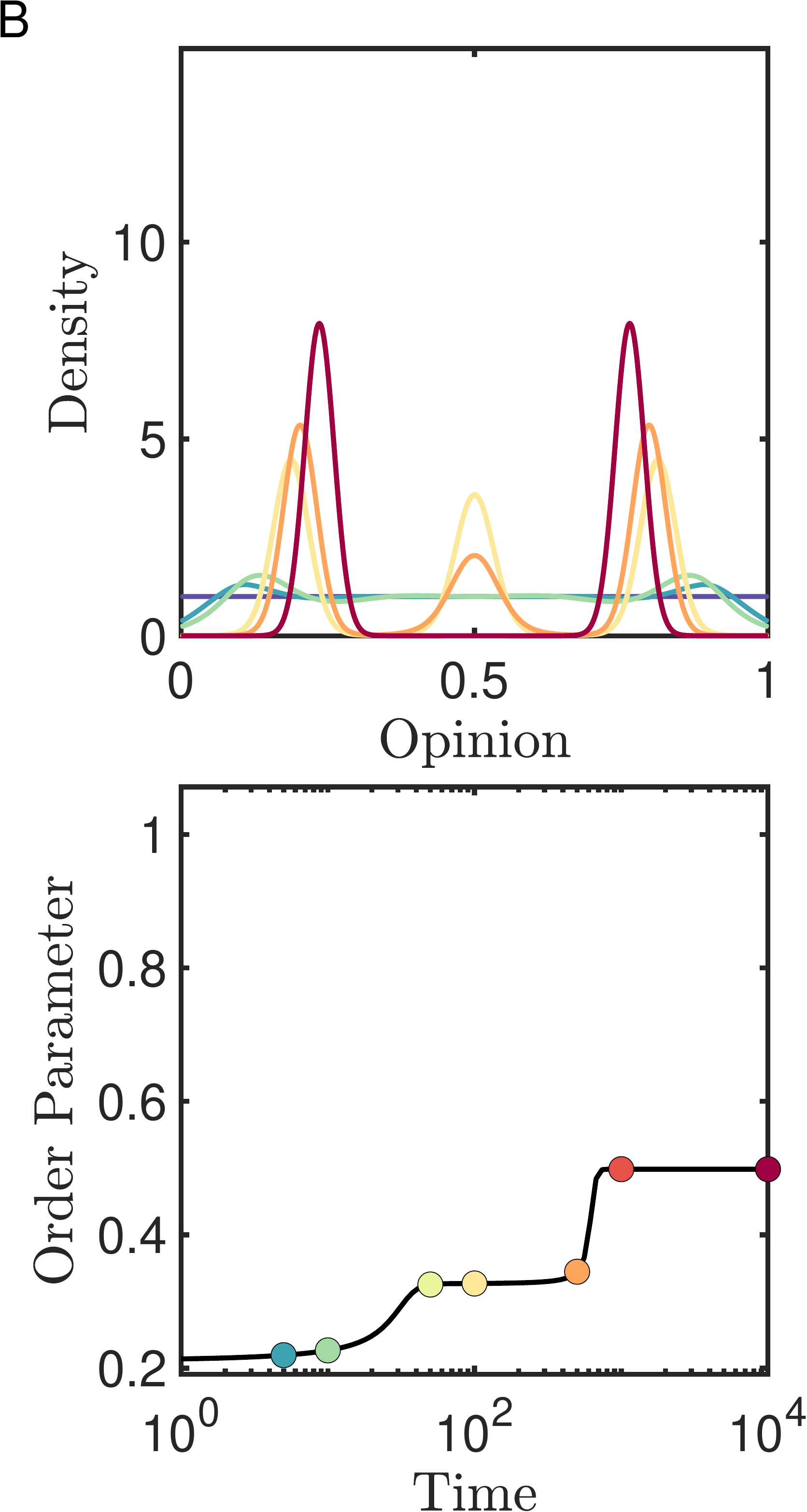}
\includegraphics[width = 2.6cm]{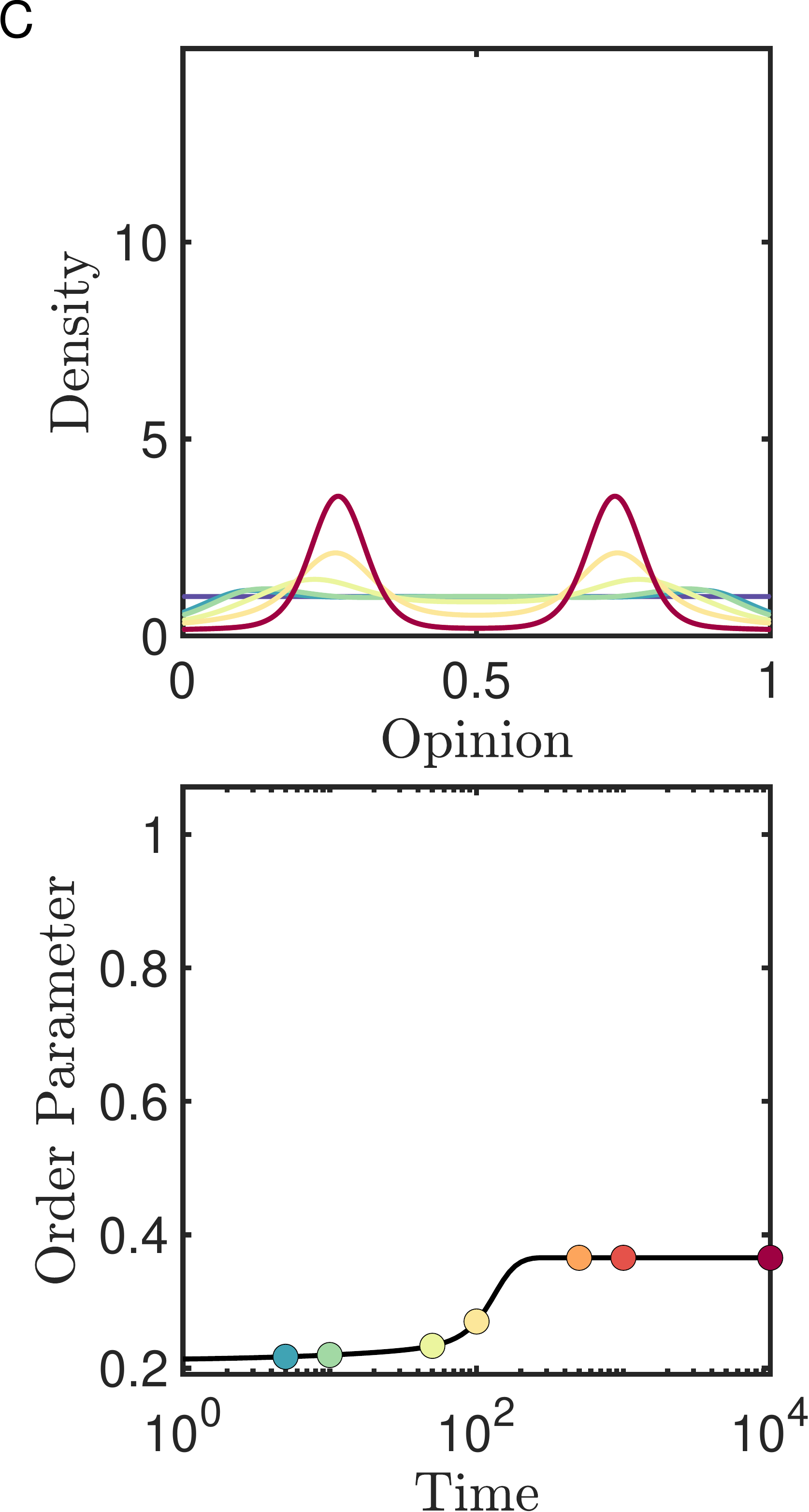}
\includegraphics[width = 2.6cm]{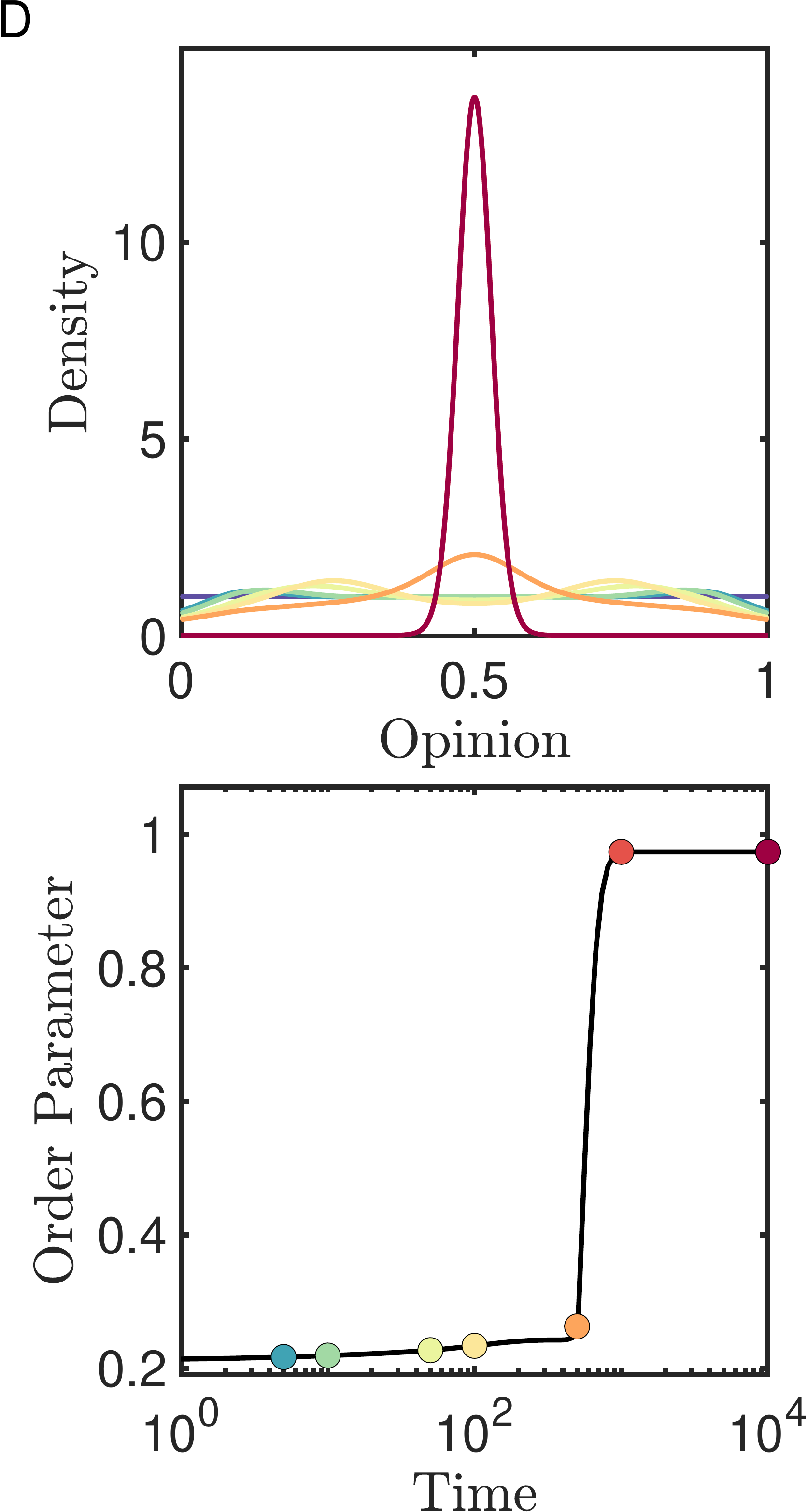}
\includegraphics[width = 2.6cm]{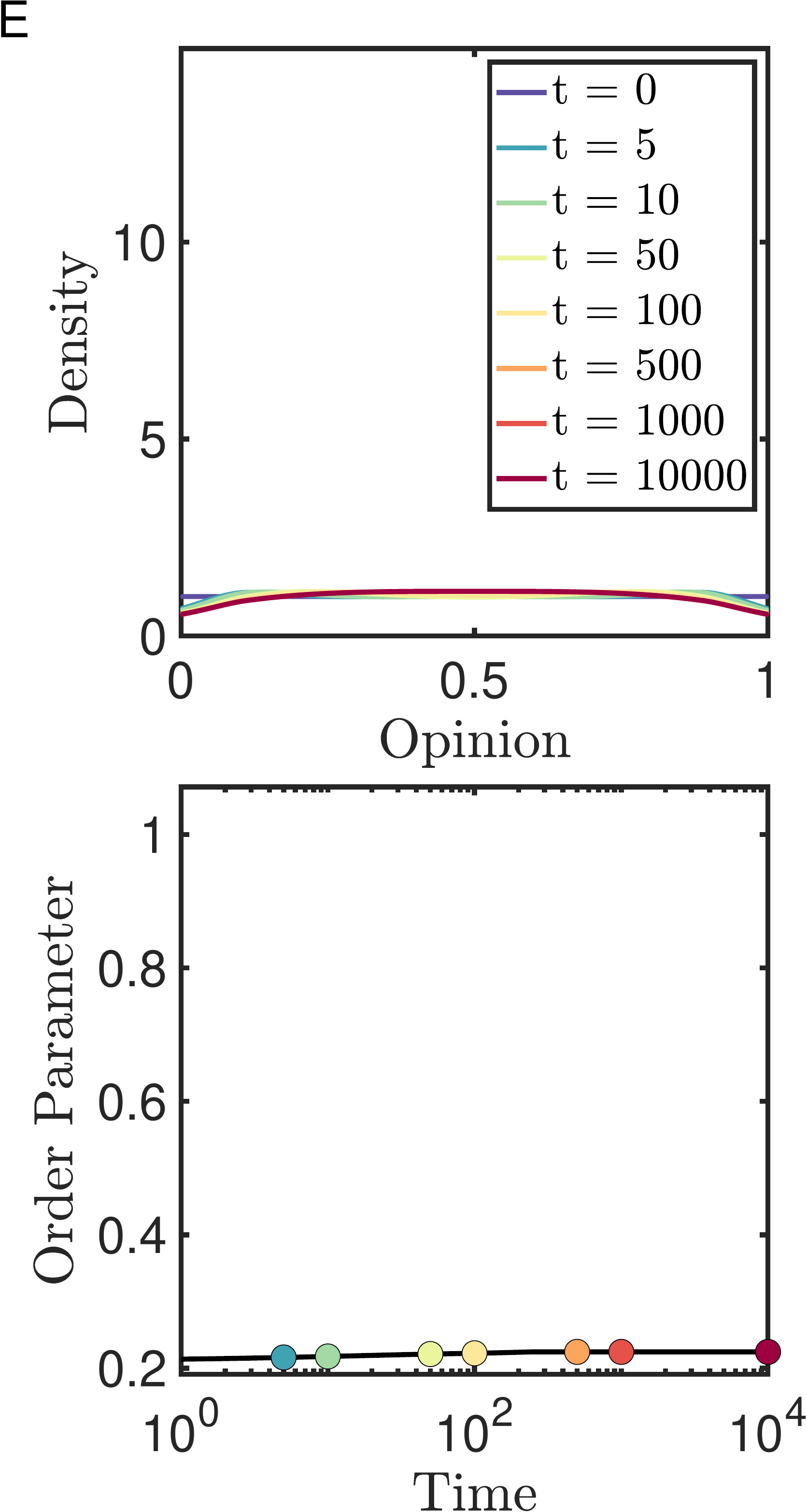}}\\
\resizebox{\figwidth}{!}{
\includegraphics[width = 2.6cm]{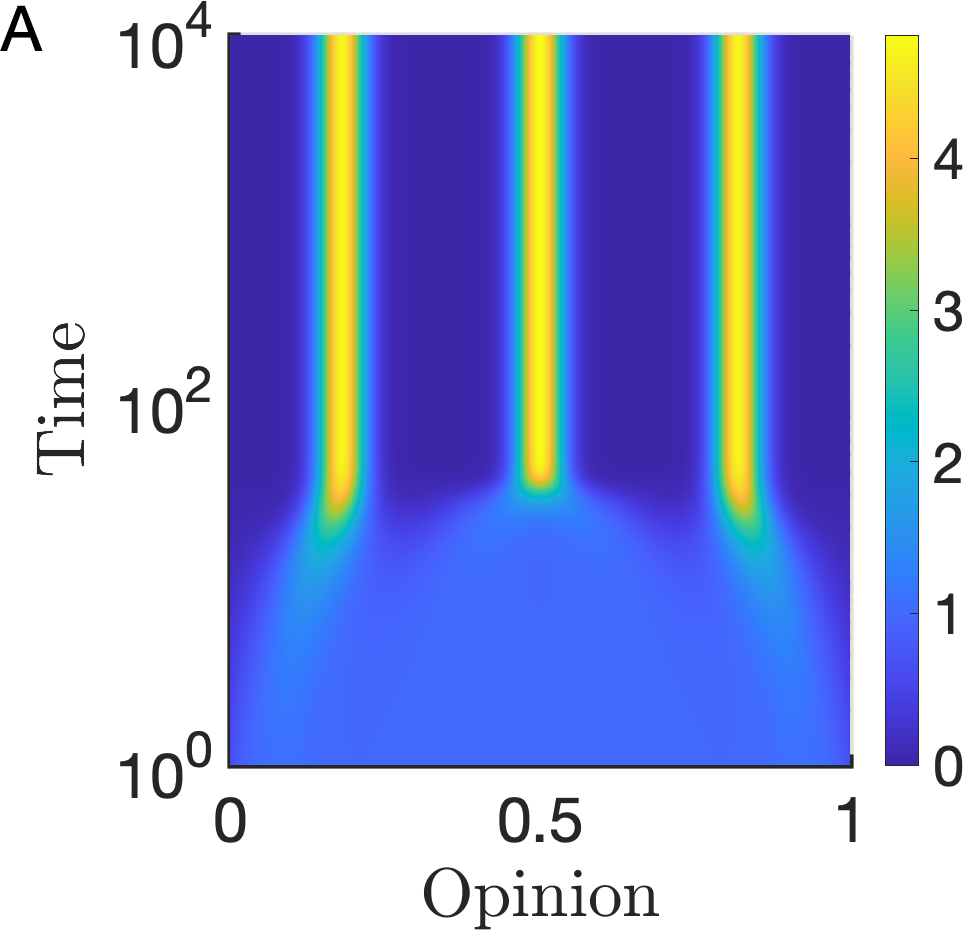}
\includegraphics[width = 2.6cm]{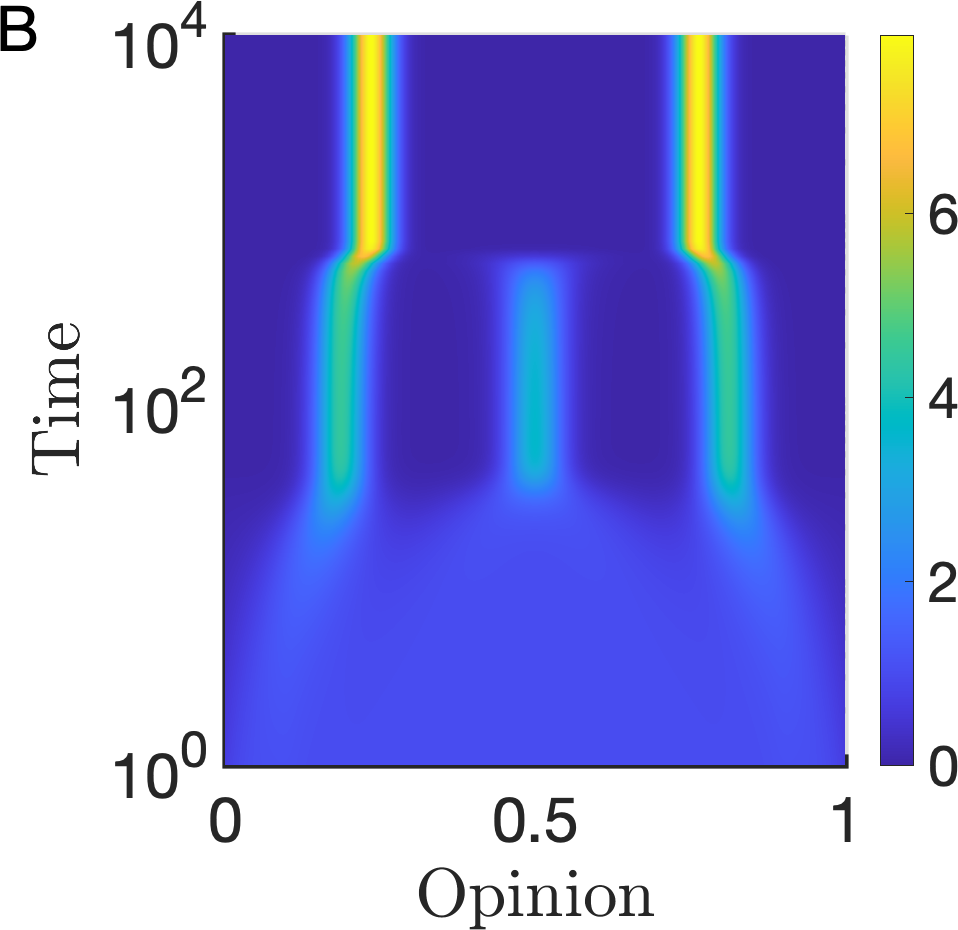}
\includegraphics[width = 2.6cm]{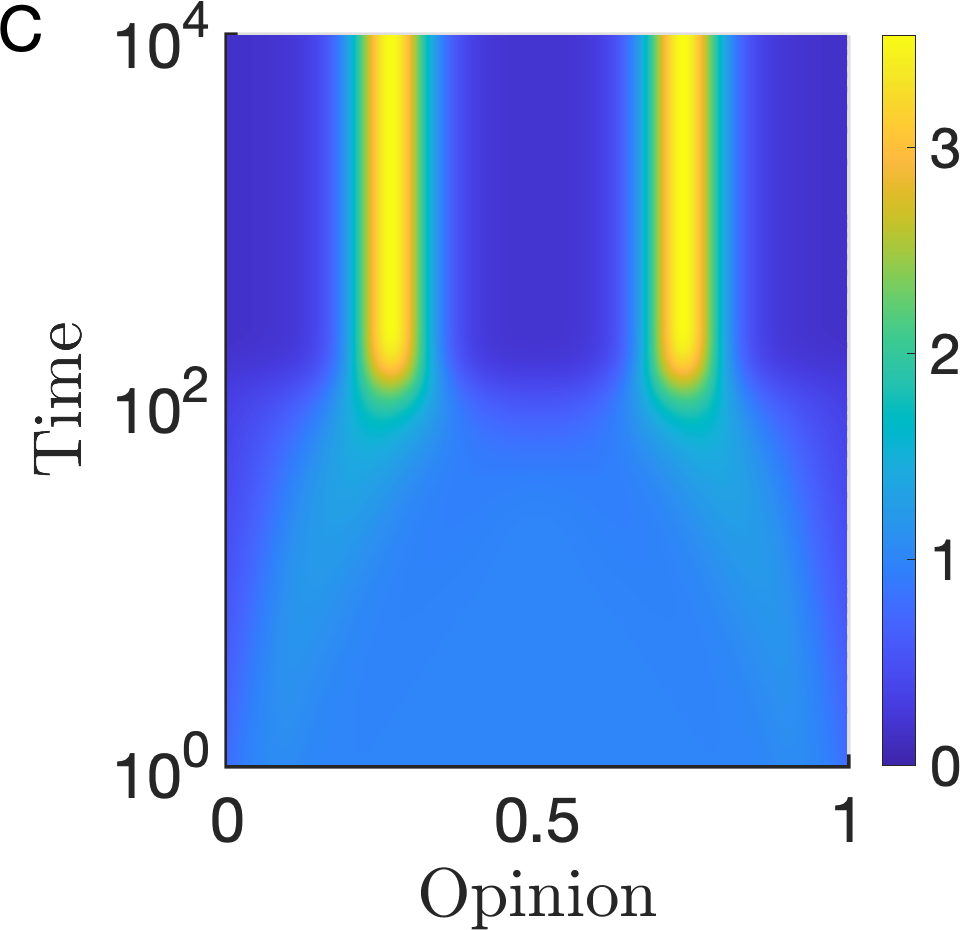}
\includegraphics[width = 2.6cm]{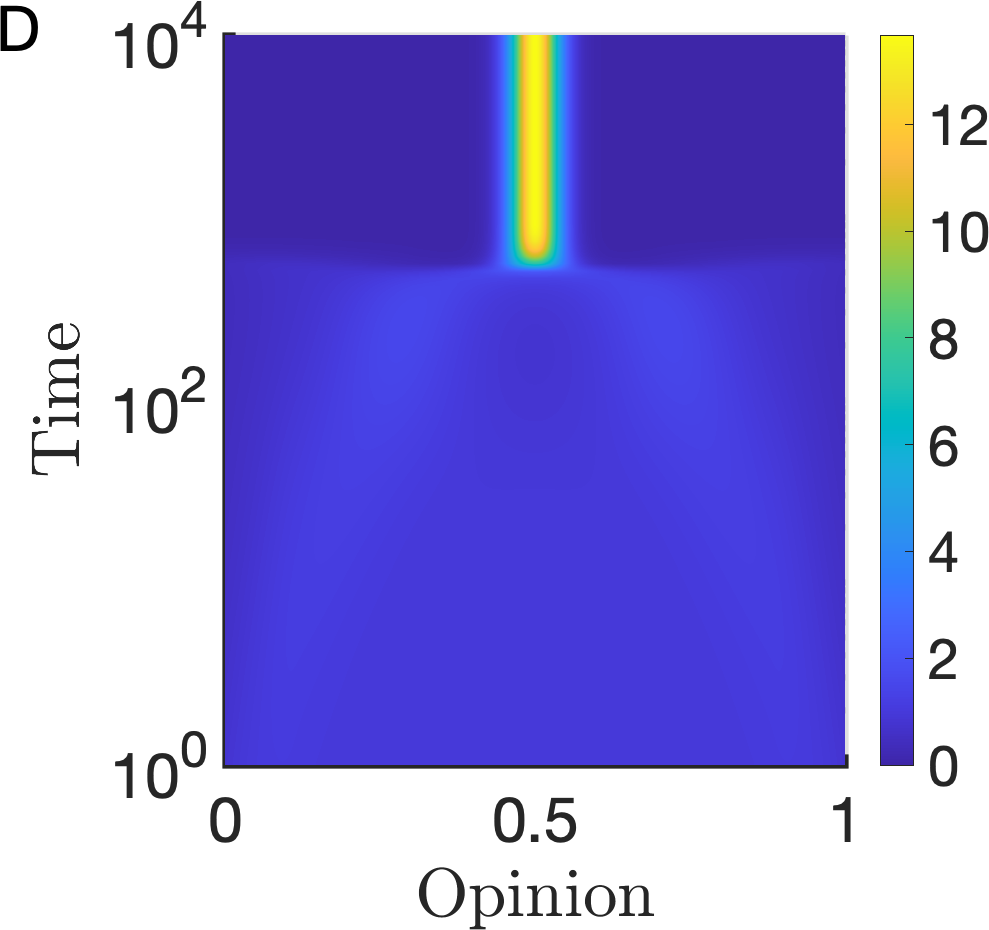}
\includegraphics[width = 2.6cm]{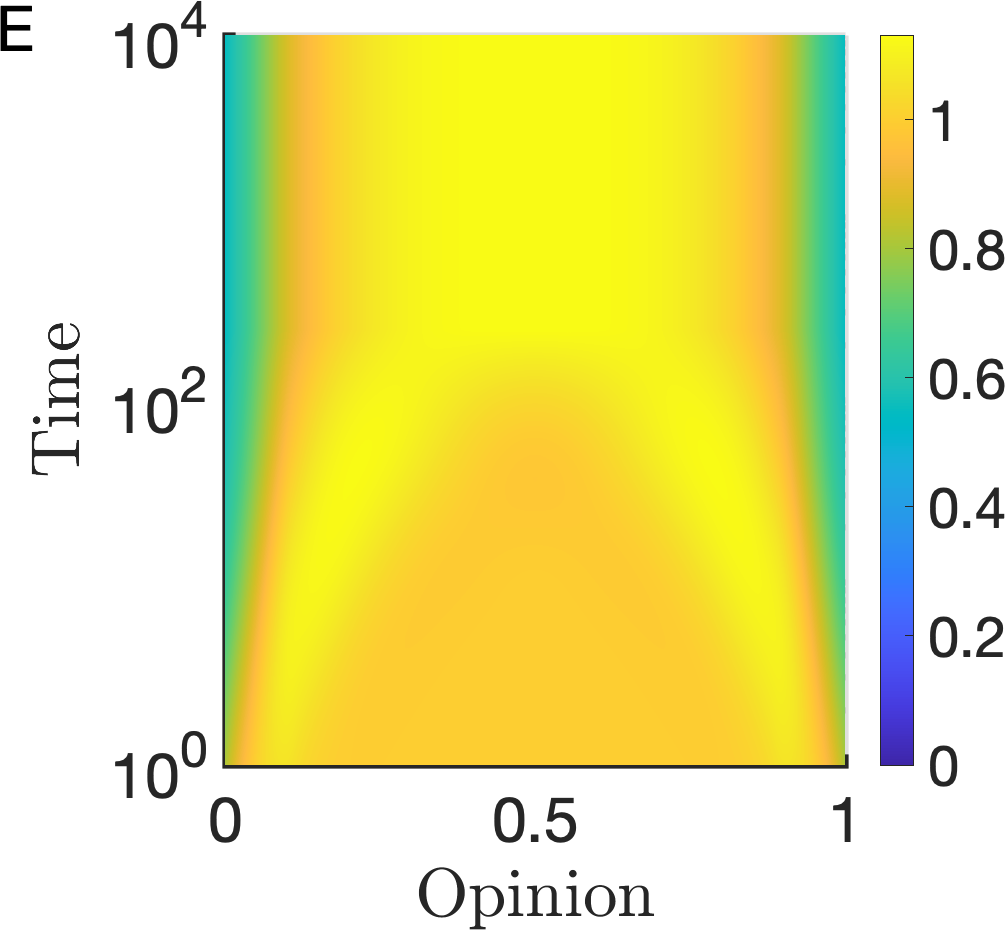}
}
\caption{As Figure~\ref{fig:Uniform_Zoom_Snapshots}, 
but for smaller values of $\sigma$; labels correspond to captions in the right panel of
Figure~\ref{fig:Uniform}. Additionally, we show time-space plots of the densities in the bottom row.}
\label{fig:Uniform_Small_Sigma_Snapshots}
\end{figure}

In Figures~\ref{fig:Uniform_Zoom_Snapshots} and~\ref{fig:Uniform_Small_Sigma_Snapshots} we show
the dynamics of parameter pairs labelled in the middle and right subplots of Figure~\ref{fig:Uniform},
respectively. We show snapshots
of the density at various times (top) and the order parameter as a function of time (bottom),
with coloured dots corresponding to the time--$Q$ values for the snapshots in the upper panels.
For the cases A--E in Figure~\ref{fig:Uniform_Zoom_Snapshots}, we see how the behaviour changes for
fixed $\sigma=0.05$ and increasing $R$.  For small $R$, a shallow, almost uniform, cluster 
slowly develops.  As $R$ increases the dynamics become richer.  At first, the density develops
two peaks, with the order parameter rising to a plateau. The two peaks then move together, 
before merging into a single cluster, indicated by a larger $Q$.
In particular, for case E, where $R=0.21$, the initial state has $Q\approx 0.4 \approx 2R$, as expected,
it then rises to $Q\approx 0.5$, indicating the presence of two clusters, before ending at
$Q\approx 1$, and a single cluster.  
Cases (B, F, G) and (H, I, J) demonstrate the effects of fixing $R$ and increasing $\sigma$.
In (B, F, G), we begin in the small-$\sigma$ regime (B), where a single,
steep cluster forms almost directly from the uniform state, 
As $\sigma$ increases (F, G), diffusion
dominates resulting in an approximately uniform distribution.
Similar behaviour is observed for larger $R$ (H, I, J).

Figure~\ref{fig:Uniform_Small_Sigma_Snapshots} shows the corresponding 
dynamics for the small-$\sigma$ case, fixing $R=0.125$ (see the right hand plot of 
Figure~\ref{fig:Uniform}). Here we observe a much richer collection
of possible long-time states and dynamics; to aid visualisation we plot the density
as a function of time and space in the bottom panels.
For small $\sigma=0.0225$ (A) we 
observe a final state with three clusters, and final order parameter
approximately $1/3$.  For slightly larger $\sigma = 0.025$ (B), the final
state has two clusters, with order parameter around $1/2$, but the dynamics
clearly pass through a transient state with three, non-equal clusters (see
the snapshots in panel B at times 100 and 500).  Increasing $\sigma$ further
next results in a direct transition to a two-cluster state (C), followed by
direct formation of a single cluster (D), and eventually an essentially
disordered/uniform long-time state (E).  We note that there are similar transitions
when fixing $\sigma$ and varying $R$ (not shown).

Here we find it informative to compare the right panel of 
Figure~\ref{fig:Uniform} with Figure 3 of
the original Hegselmann-Krause paper~\cite{HK02}, which shows the long-time equilibria (for noiseless dynamics)
as $R$ is increased.  
For small $R$, the state is homogeneous, whilst increasing $R$ results first in two clusters and then a 
single, central cluster, with rapid transitions as $R$ increases. As stated in~\cite{HK02}, as $R$ increases `we step from
fragmentation (plurality) over polarisation (polarity) to consensus (conformity)'.  This is a direct analogue of our
results just described.
In contrast, for the two periodic boundary conditions, the uniform initial condition is an equilibrium,
and we see no such $R$-induced transitions.  \emph{This suggests that no-flux boundary conditions more
faithfully reproduce the results, and underlying mechanisms, of the original models.}

We also compare to Figure 1 of~\cite{PT18}, which demonstrates the dependence on $R$ (their $d$) 
of the final number of clusters and equilibration time.  They note that both dependencies are 
non-monotonic, whereas it may be intuitively expected that increasing $R$ causes a reduction
in the number of clusters and a decrease in the equilibration time.  Such effects are also visible
when, e.g., fixing $\sigma = 0.2$ and varying $R$ in the right panel of Figure~\ref{fig:Uniform}.
This phenomenon of `abnormally' slow convergence has also been demonstrated in~\cite{L06}, who
described the resulting states as metastable.  Similar sensitivities have also been observed
in a noisy DW model~\cite{CTSM13}, which also noted the importance of the initial condition on determining
the long-time dynamics; we will now investigate the further choices of initial condition.  In Supplementary
Material SM4, we show a comparison with the SDE for short times; the agreement is very good.
The phenomenon of dynamical metastability for noisy opinion dynamics PDE models, also observed 
in~\cite{GPY16}, will be studied in future work.

\subsection{Single Gaussian Initial Condition}\label{s:Gaussian}

In this section we investigate the effects of a non-uniform initial condition
with a single, relatively broad cluster/consensus.  
Following~\cite{WLEC17}, we choose
\begin{equation}
	\rho_0(y) = Z^{-1} \exp\big( - C[d(y,y_0)]^2 \big),
	\label{eq:Gaussian_IC}
\end{equation}
where $C = 20$ and $Z$ is the normalisation constant.  Here $d(x,y)$ denotes the
1-periodic distance between two points. In~\cite{WLEC17}, $y_0$ was chosen to be $0.5$ but this is 
irrelevant beyond visualisation in the periodic case. 
However, for the no-flux and even 2-periodic cases, the choice
of $y_0$ can result in qualitatively different dynamics.  For example,
if $y_0=0.5$ then the initial condition is symmetric and the periodic and
even 2-periodic cases are identical; for other $y_0$ this is not the case.

\begin{figure}
\centering
\resizebox{\figwidth}{!}{
\includegraphics[width = 0.5\textwidth]{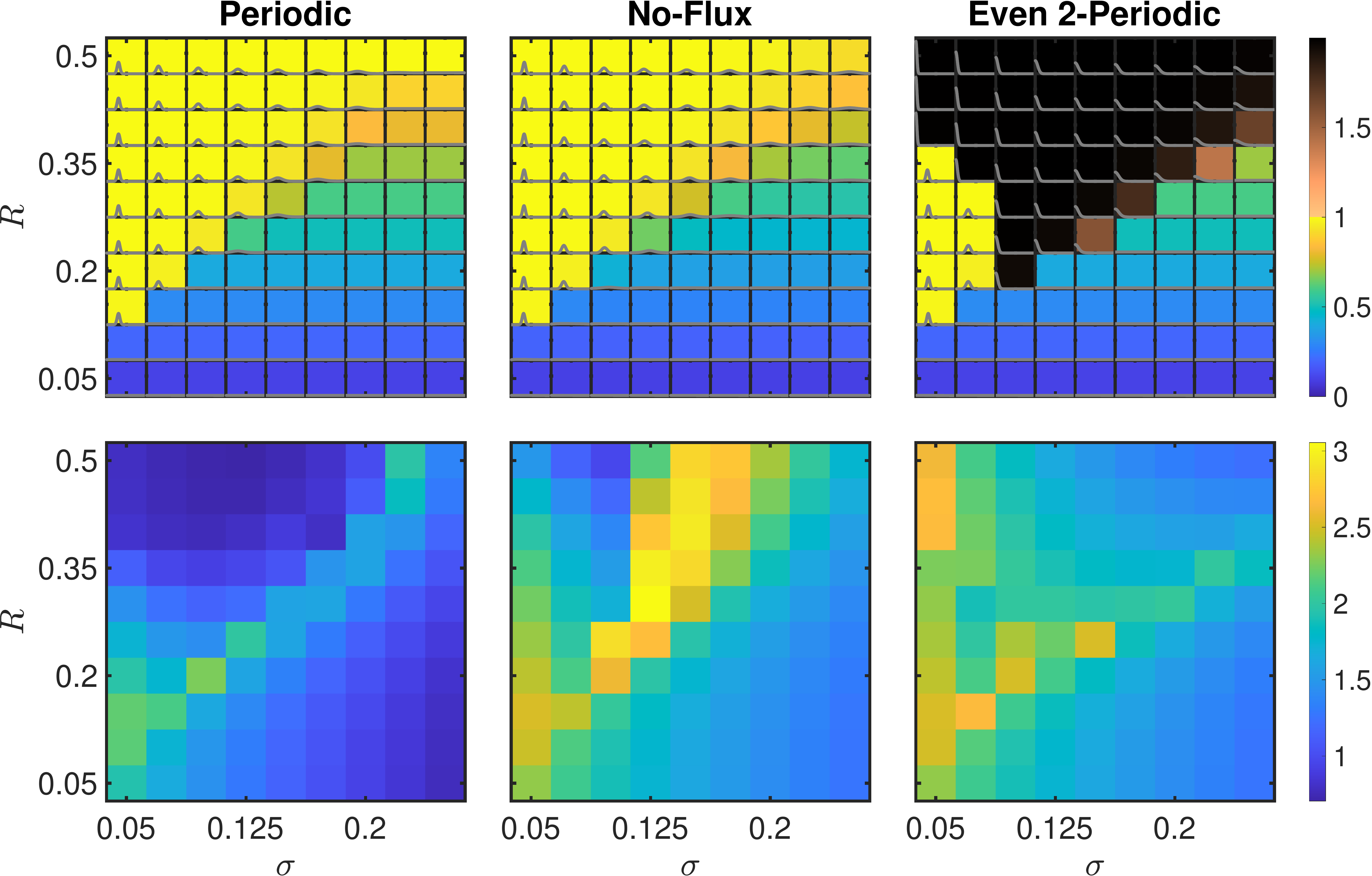}
\includegraphics[width = 0.5\textwidth]{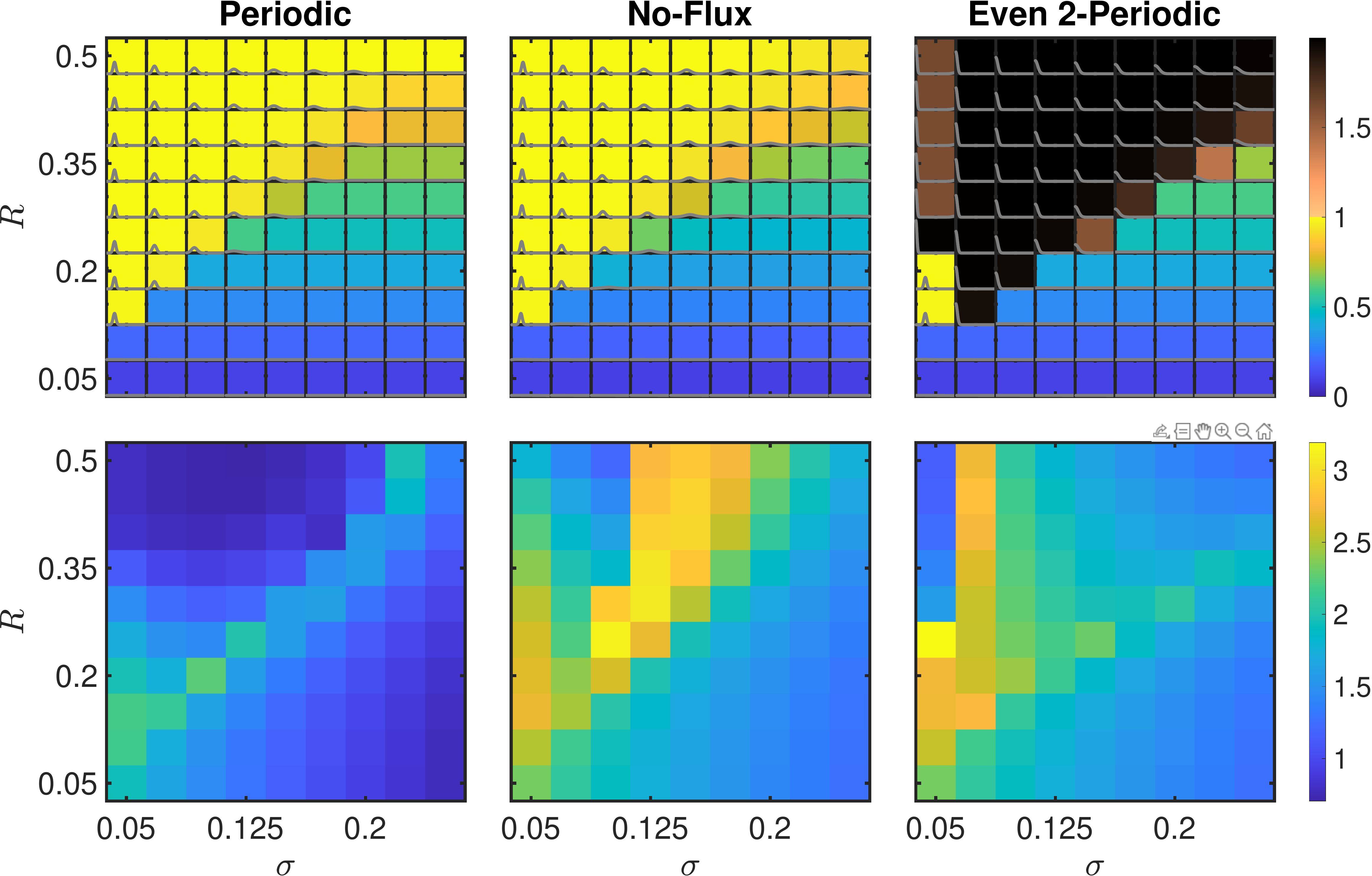}}\\
\resizebox{\figwidth}{!}{
 \includegraphics[width = 0.5\textwidth]{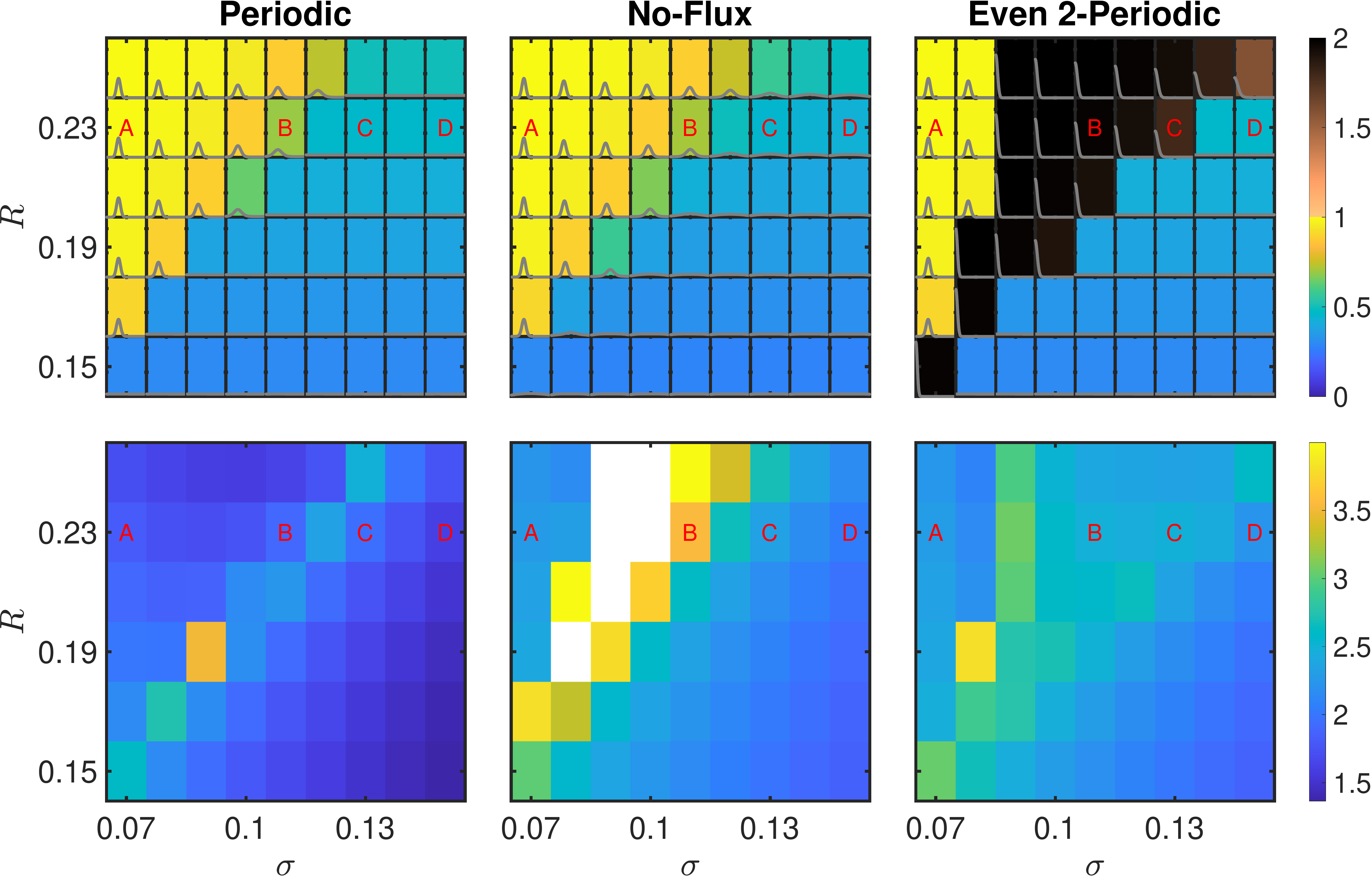}
\includegraphics[width = 0.5\textwidth]{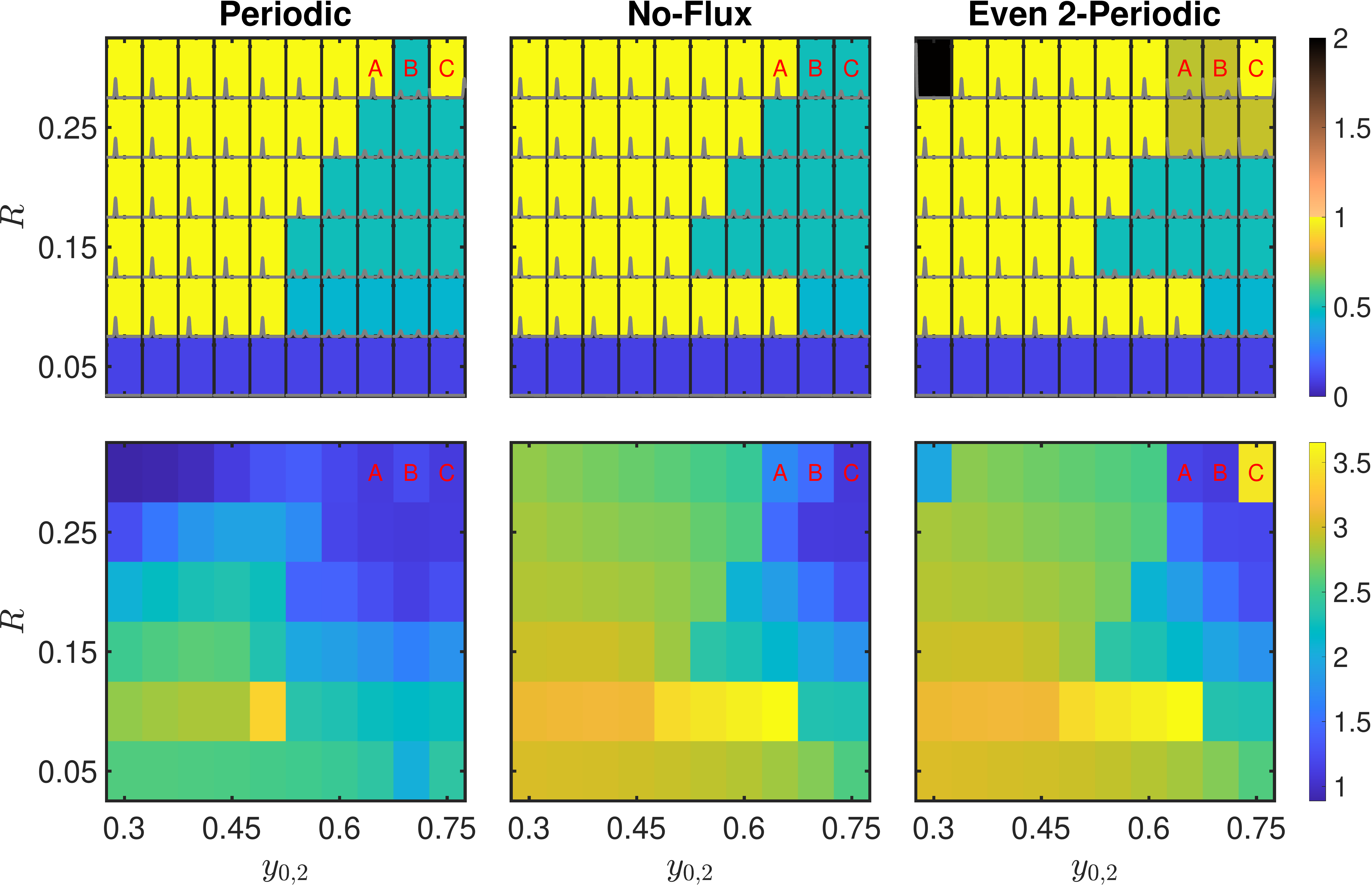}
}
\caption{
As Figure~\ref{fig:Uniform} but for a Gaussian initial condition \eqref{eq:Gaussian_IC}
with $y_0 = 0.3$ (top left, bottom left) and $y_0 = 0.2$ (top right), and an initial condition
which is a linear combination of Gaussians \eqref{eq:Two_Gaussians_IC} (bottom right)
for different boundary conditions. 
Note the additional colour bar scale from 1 to 2 for the even 2-periodic
case.  White denotes simulations which have not converged by the final time $(10^4)$.}
\label{fig:Gaussian}
\end{figure}

In the top panels of Figure~\ref{fig:Gaussian}
we display the final order parameters, densities, and equilibration times for a
range of values of $R$ and $\sigma$, for all three boundary conditions,
and $y_0$ equal to 0.3 (left), and 0.2 (right).  
We also show a zoom in parameter space for $y_0 = 0.3$ (bottom left),
and results for an initial condition which is a
Gaussian mixture (bottom right); see Section~\ref{s:TwoGaussians}.  White regions in the 
equilibration times denote simulations which have not converged.

We first discuss some general trends.  
With the exception of $y_0 = 0.2$ and even 2-periodic
boundary conditions (top right plot), the final result is either a single cluster, or an (almost)
uniform state.  This is to be expected as the only force which can break up the initial cluster is 
diffusion, which favours the uniform state. The formation of one large cluster on the 
left and one small cluster on the right for $y_0 = 0.2$, even 2-periodic boundary conditions,
small $\sigma$ and large $R$ is due to the periodicity of the initial condition. A
small amount of mass is originally concentrated at the right end of the domain, near
$y=1$, and gets trapped due to the attraction to its periodic image near $y=-1$.  
Relatedly, we note the additional
range of $Q$ for the even 2-periodic case, in particular for large $R$, where 
the density is significantly influenced by its periodic/even image.
As expected, increasing $\sigma$ also causes
a trend to disordered states, whereas increasing $R$
tends to increase the sharpness of the resulting cluster, with increased time-to-equilibrium
in regions separating qualitatively different final densities.

\begin{figure}
\centering
\resizebox{\figwidth}{!}{
\includegraphics[width = 0.5\textwidth]{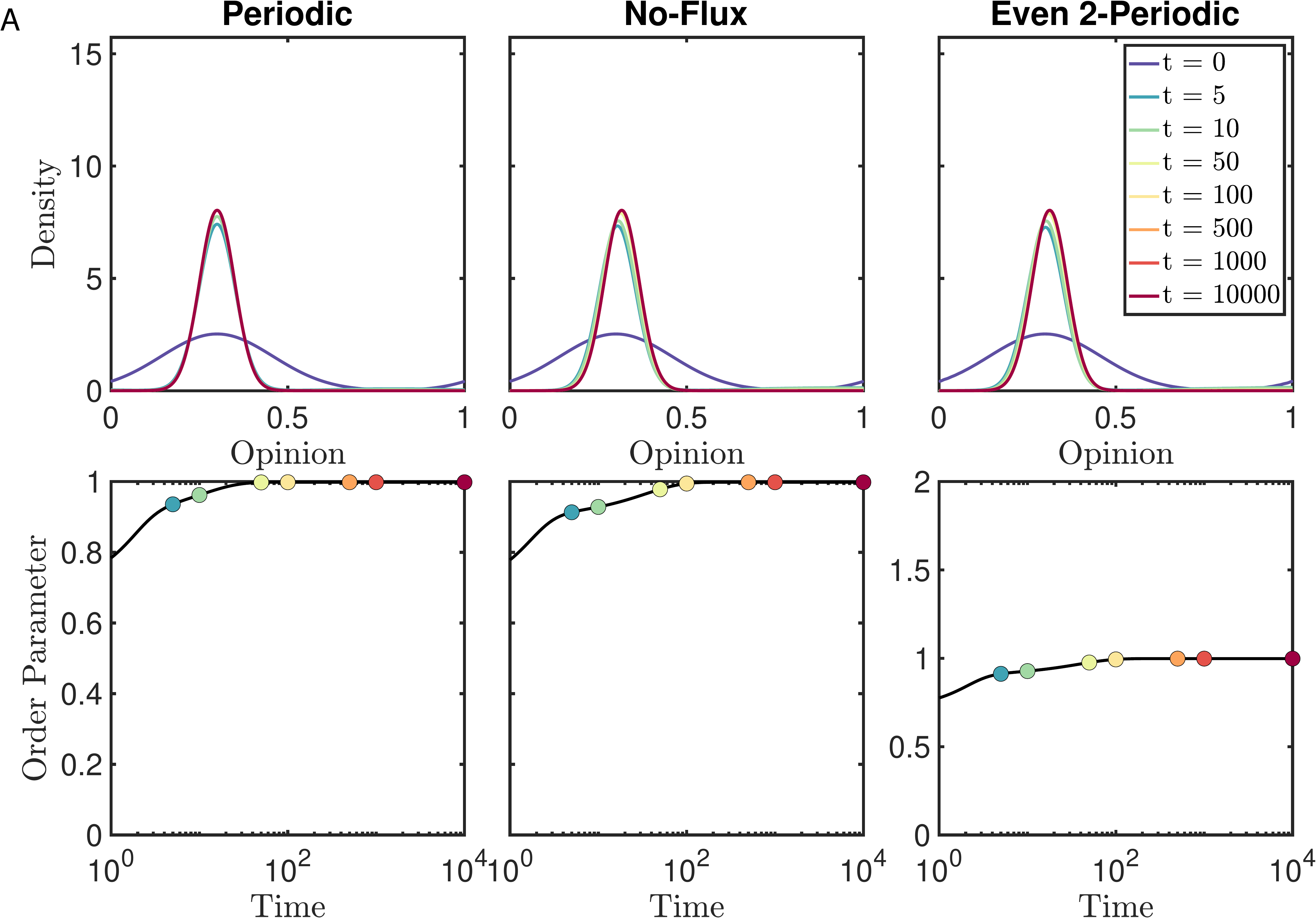}
\includegraphics[width = 0.5\textwidth]{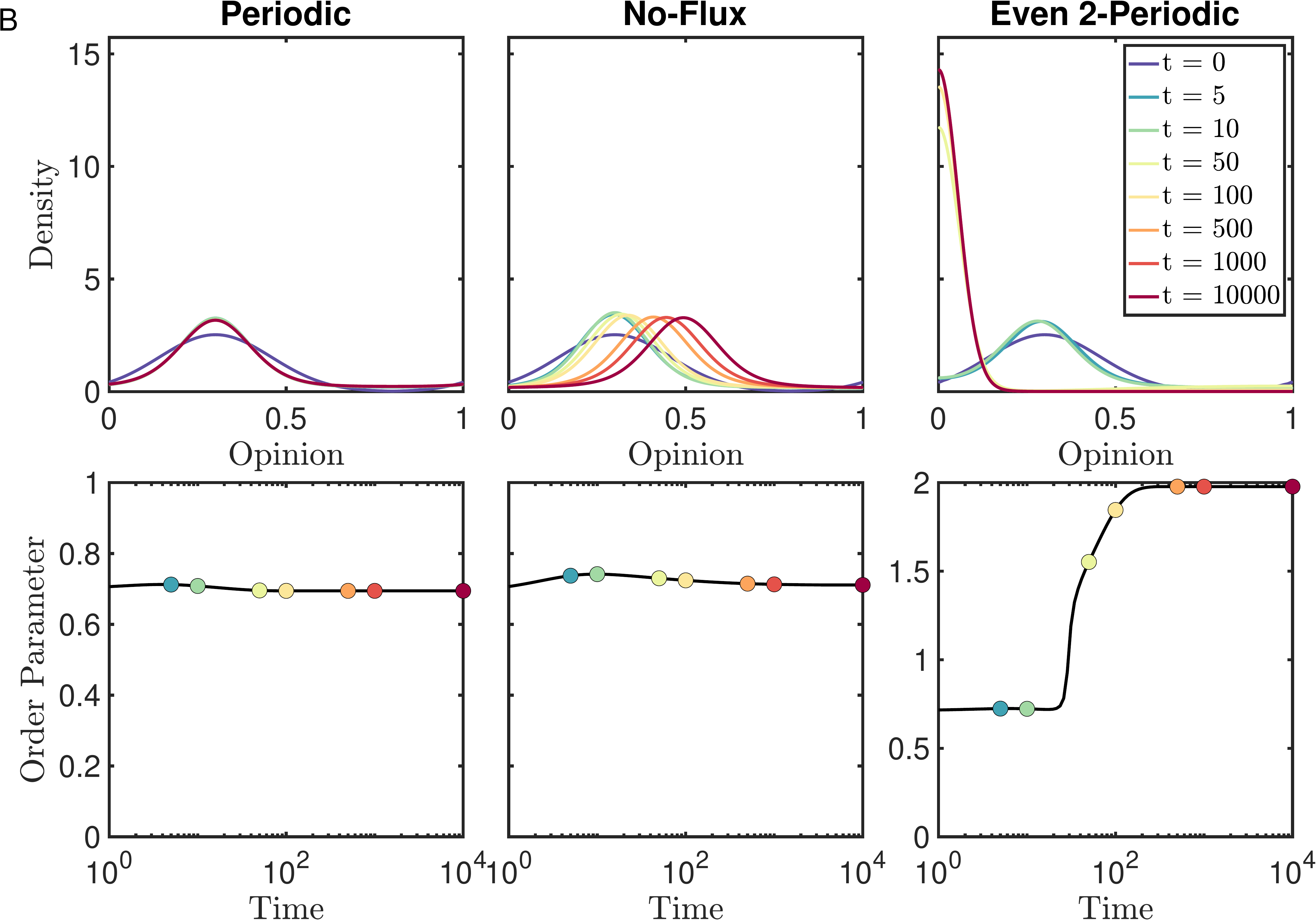}}\\[2mm]
\resizebox{\figwidth}{!}{
\includegraphics[width = 0.5\textwidth]{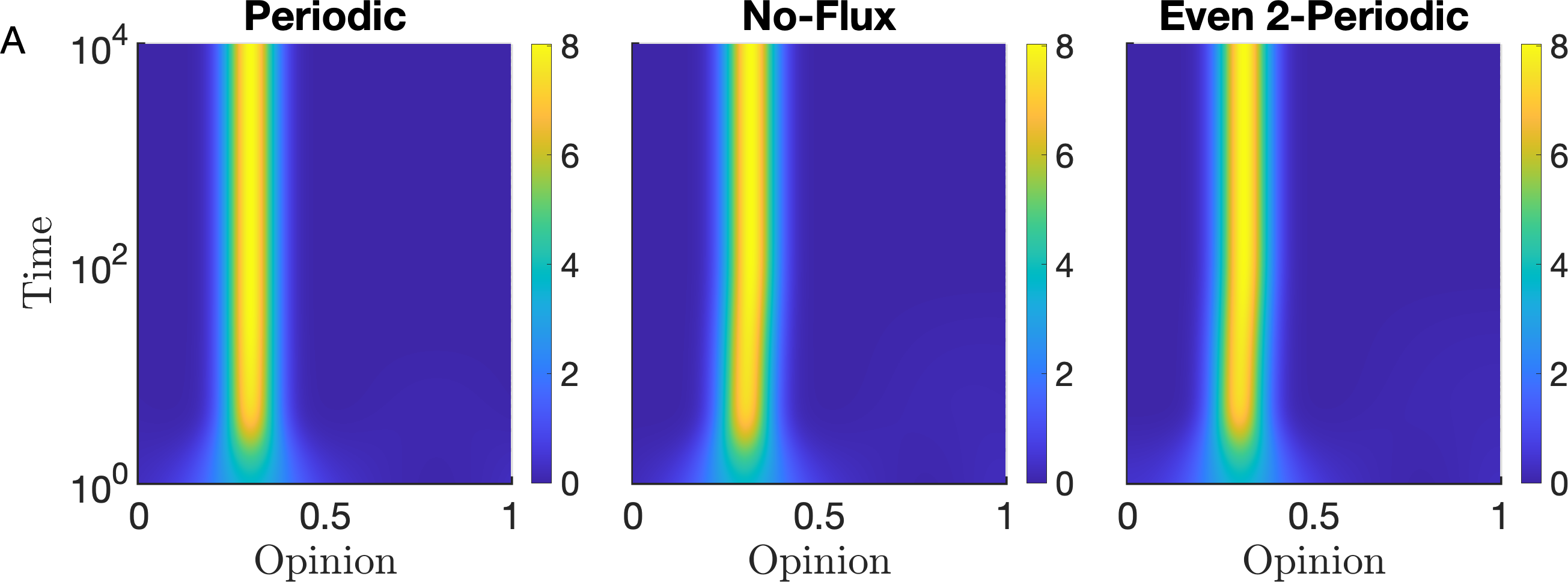}
\includegraphics[width = 0.5\textwidth]{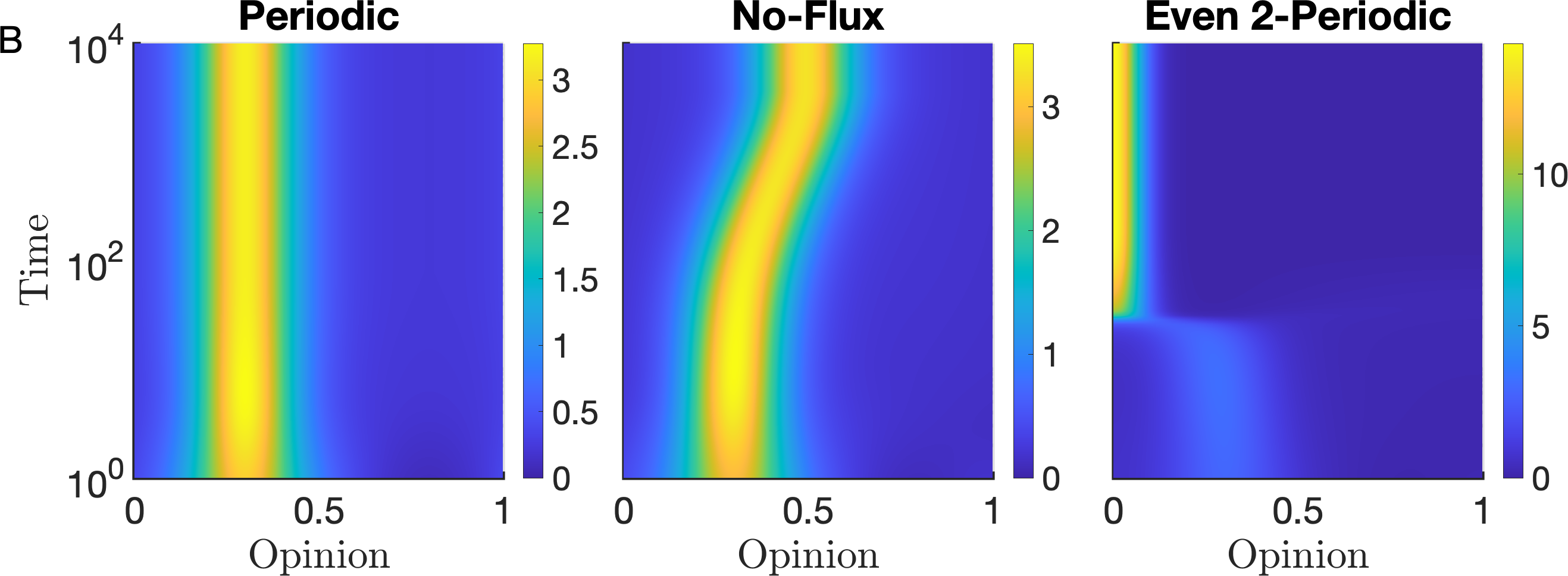}}\\[2mm]
\resizebox{\figwidth}{!}{
\includegraphics[width = 0.5\textwidth]{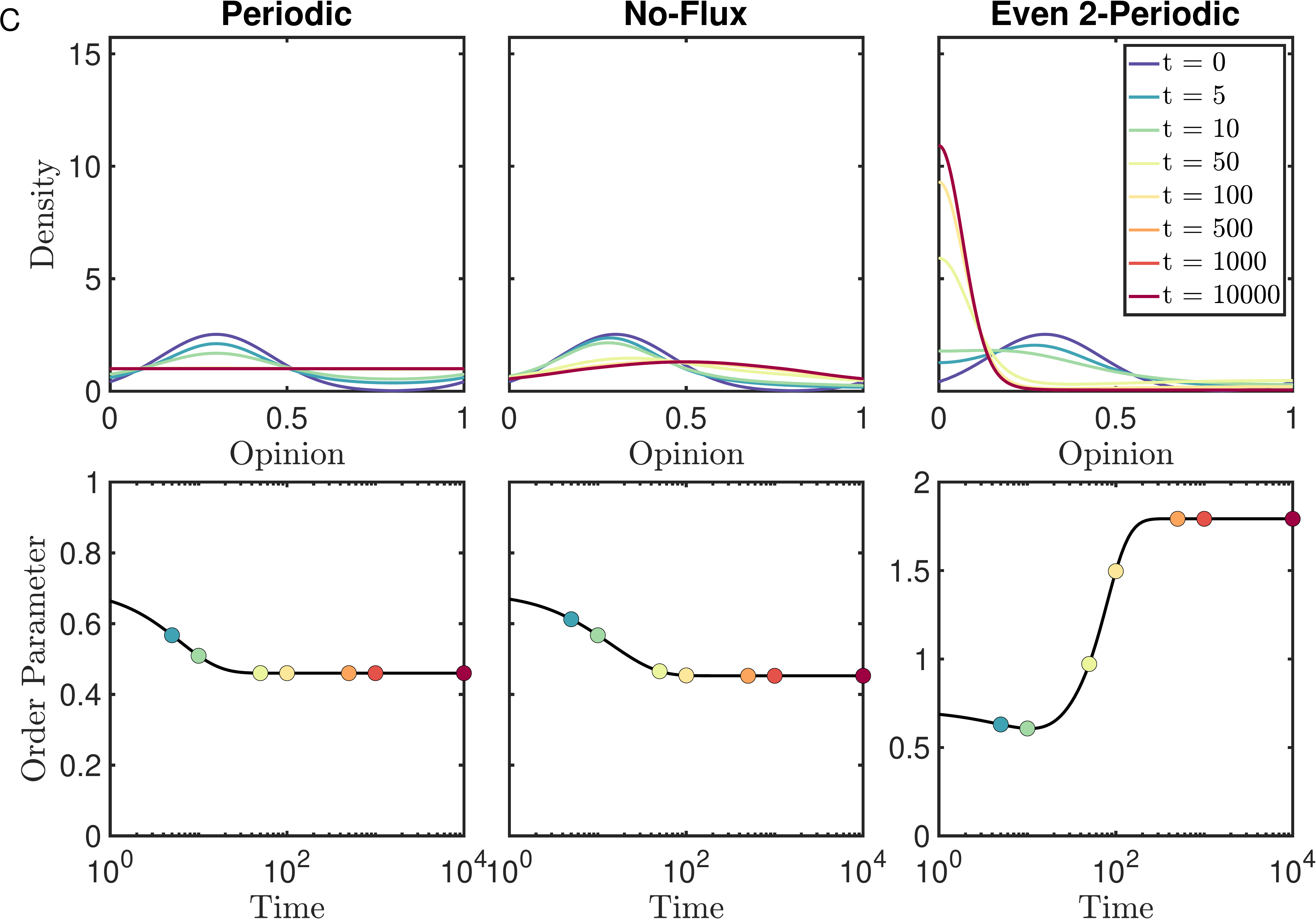}
\includegraphics[width = 0.5\textwidth]{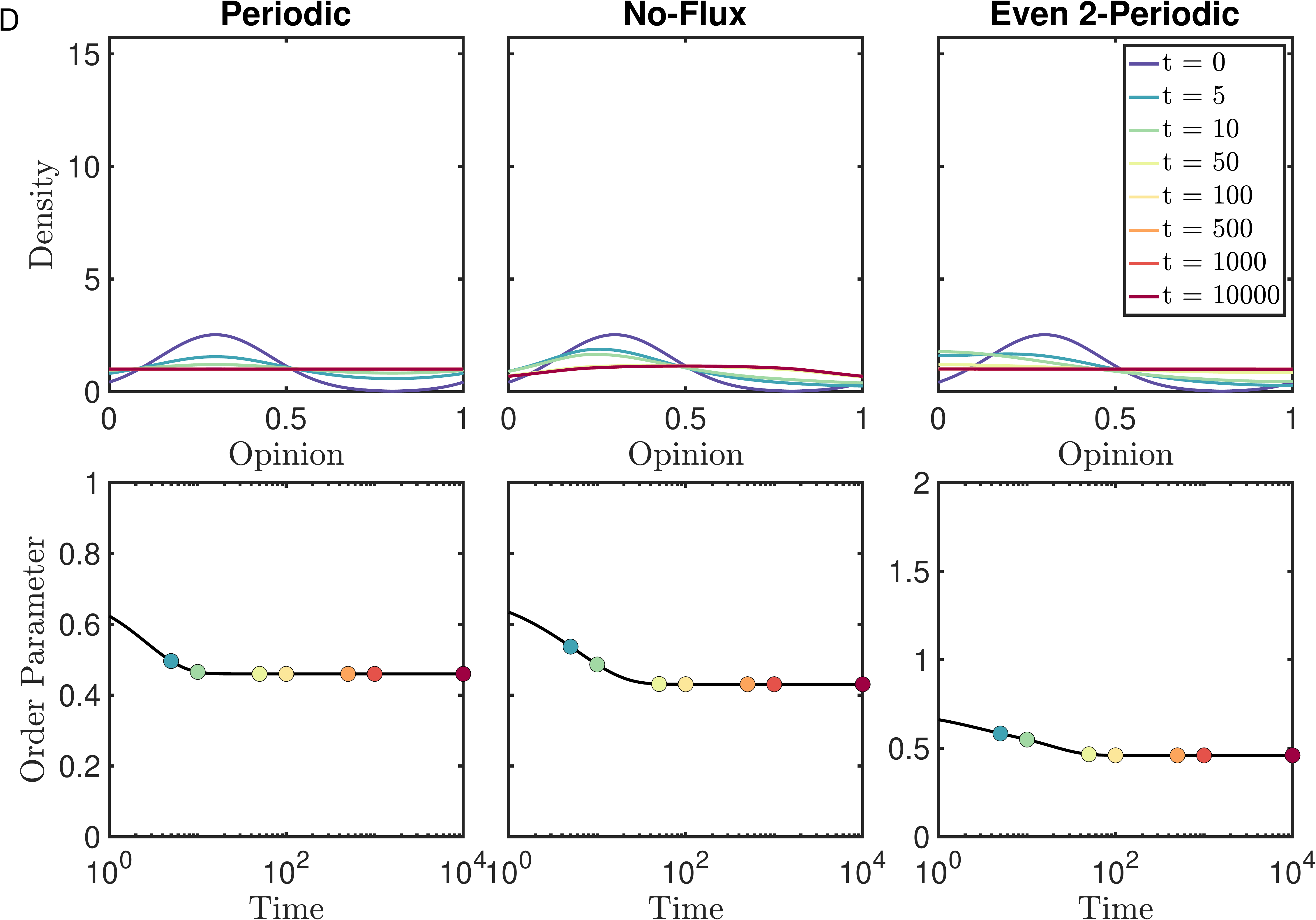}}\\[2mm]
\resizebox{\figwidth}{!}{
\includegraphics[width = 0.5\textwidth]{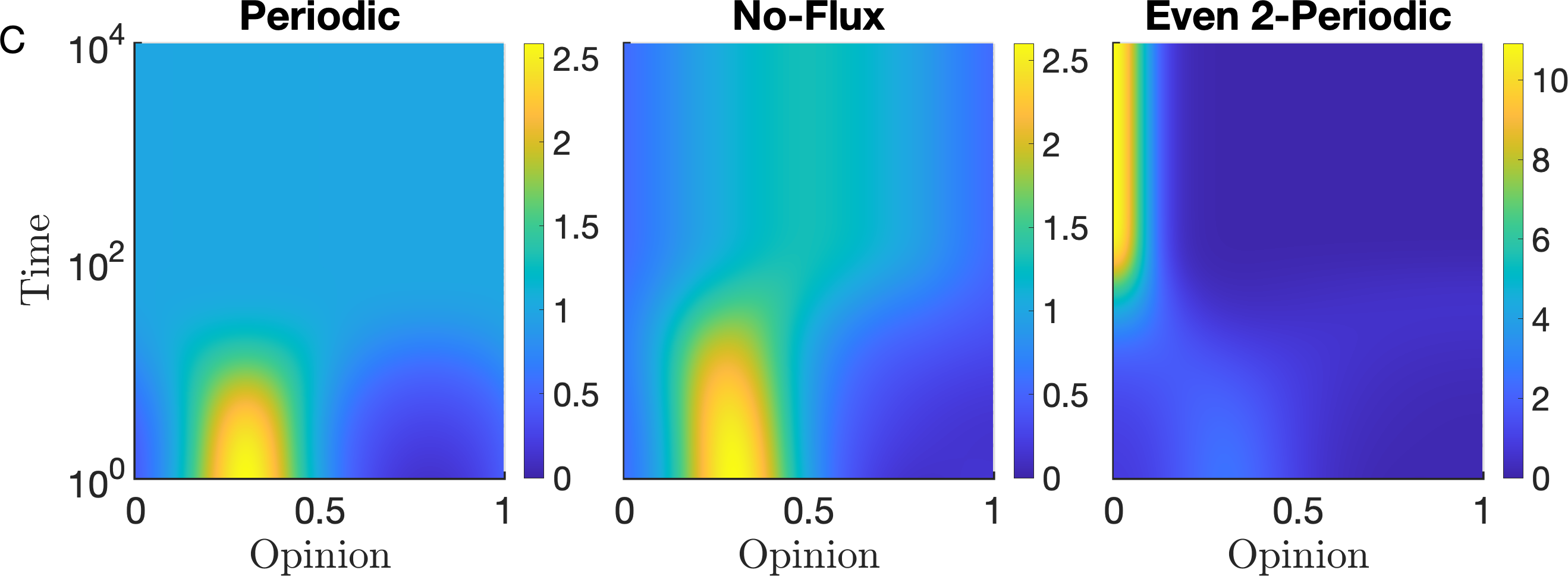}
\includegraphics[width = 0.5\textwidth]{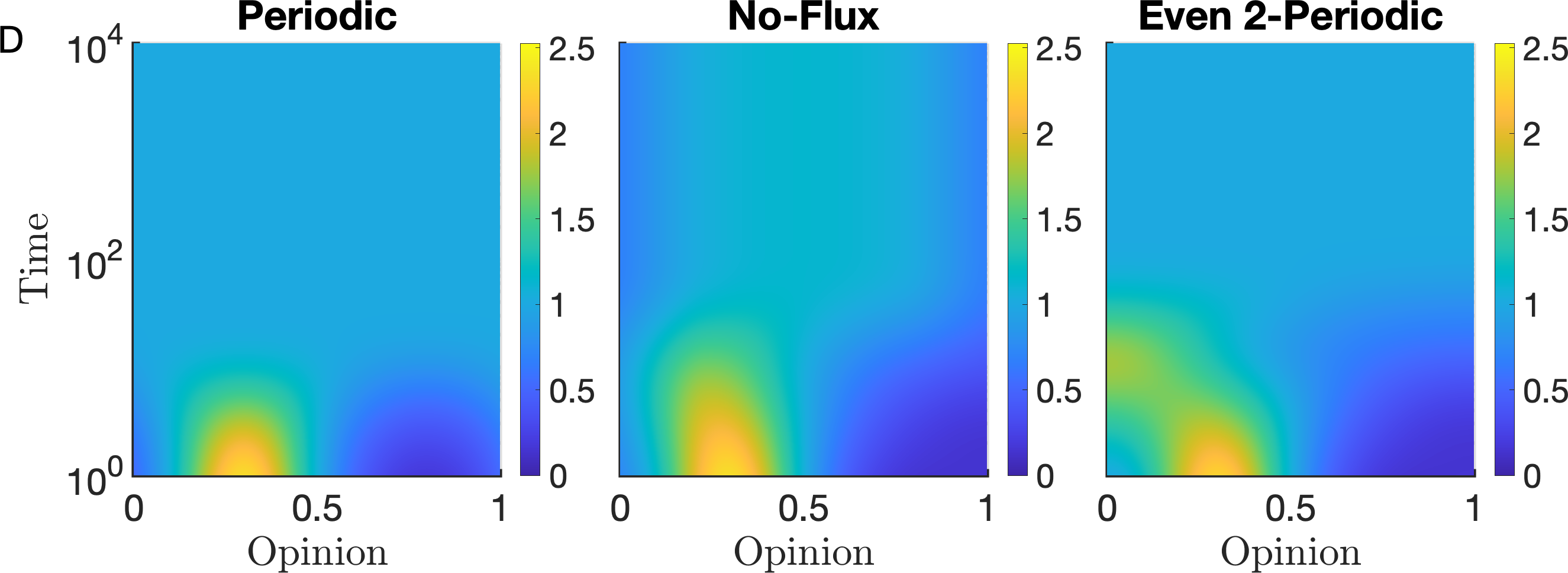}
}
\caption{
As Figure~\ref{fig:Uniform_Small_Sigma_Snapshots} but for a Gaussian initial condition.  
Labels correspond to those in the bottom left panel of Figure~\ref{fig:Gaussian}, 
for A ($R=0.23$, $\sigma = 0.07$), B ($R=0.23$, $\sigma = 0.11$),
C ($R=0.23$, $\sigma = 0.13$), and D ($R=0.23$, $\sigma = 0.15$).
}
\label{fig:Gaussian_Zoom_Snapshots}
\end{figure}

There are also some unexpected observations.
A feature of the no-flux boundary conditions is the movement towards $0.5$ of
the final maximum opinion for fixed $R$, as $\sigma$ increases.
This is likely a consequence of the competition between diffusion and attraction;
for smaller values of $\sigma$, the noise is not sufficiently strong to disperse the original cluster,
whereas larger noise can cause the cluster to move.  Examples of this dynamics
can be seen in Figures~\ref{fig:Gaussian_Zoom_Snapshots}; case B ($R=0.23$, $\sigma = 0.11$) 
shows a clear drift in the mean opinion over time for the no-flux boundary conditions.
Parameter sets B  ($R=0.23$, $\sigma = 0.11$) and C ($R=0.23$, $\sigma = 0.13$)
for the even 2-periodic case demonstrate the significant effect that this
choice of boundary condition has on the dynamics, with a strong cluster
forming at the left hand edge.  
It is also interesting to note that the
behaviour of the order parameter is not monotonic in a number of cases.
We have found excellent agreement with the 
associated SDE agent-based model, at least for short times for which the computational cost of the SDEs
is reasonable; see Supplementary Material Section SM4.

\subsection{Two Gaussians Initial Condition} \label{s:TwoGaussians}

To demonstrate the effect of multi-modal initial conditions, we study a
linear combination of two Gaussians, or a Gaussian mixture:
\begin{equation}
	\rho_0(y) = Z^{-1} \Big[ \exp\big( - C[d(y,y_{0,1})]^2 \big)
	                            + \exp\big( - C[d(y,y_{0,2}))]^2 \big) \Big],
	\label{eq:Two_Gaussians_IC}
\end{equation}
where $Z$ is the normalisation constant.
We note that there are many
parameters in this setup, and presumably also a correspondingly large
number of interesting transitions between regimes as the parameters are
varied, but for conciseness, we fix $C = 80$, $y_{0,1} = 0.2$, and
$\sigma = 0.03$.
In the bottom right plot of Figure~\ref{fig:Gaussian} we show the final-time density,
corresponding order parameter, and equilibration time as $R$ and $y_{0,2}$
are varied.
We note a general trend, as in previous cases, that (with a few
notable exceptions) increasing $R$ for fixed $y_{0,2}$ increases the tendency
for consensus to form, and also for consensus to be closer to the centre
of the interval. For fixed $R$, increasing $y_{0,2}$
(i.e., separating the initial clusters) tends to favour the formation
of two clusters, rather than a single consensus. 

\begin{figure}
\centering
\resizebox{\figwidth}{!}{
\includegraphics[width = 0.5\textwidth]{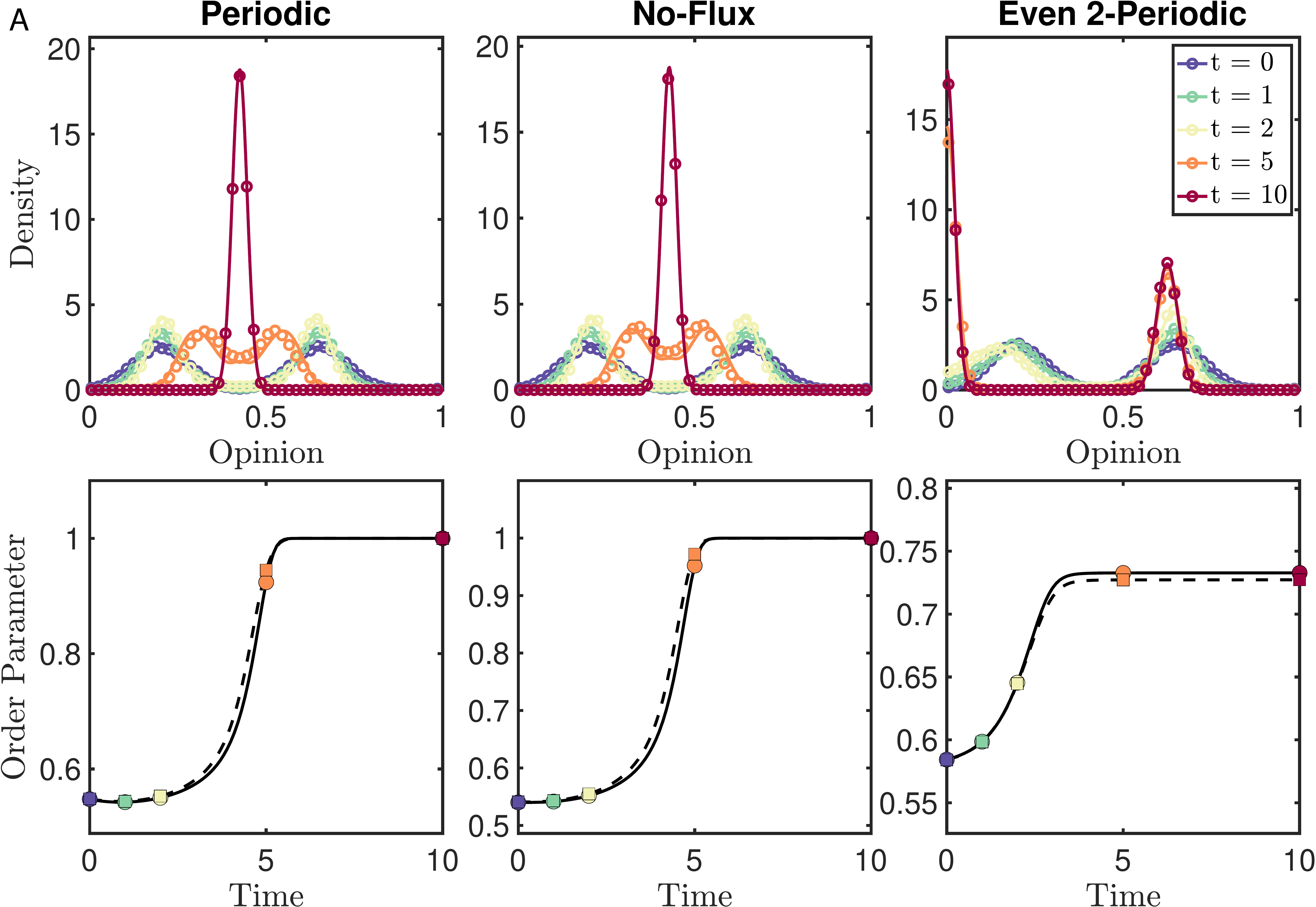}
\includegraphics[width = 0.5\textwidth]{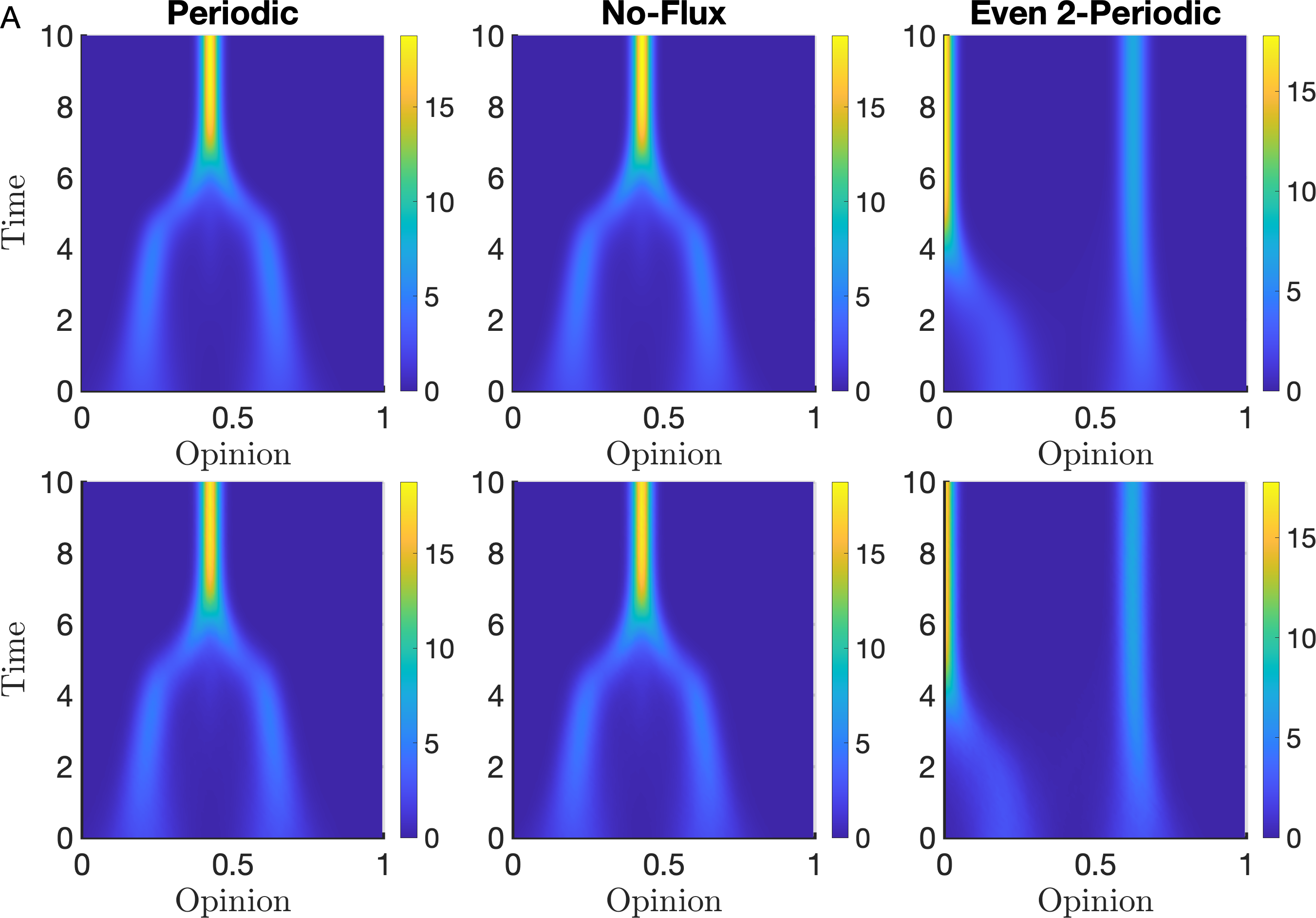}}\\[2mm]
\resizebox{\figwidth}{!}{
\includegraphics[width = 0.5\textwidth]{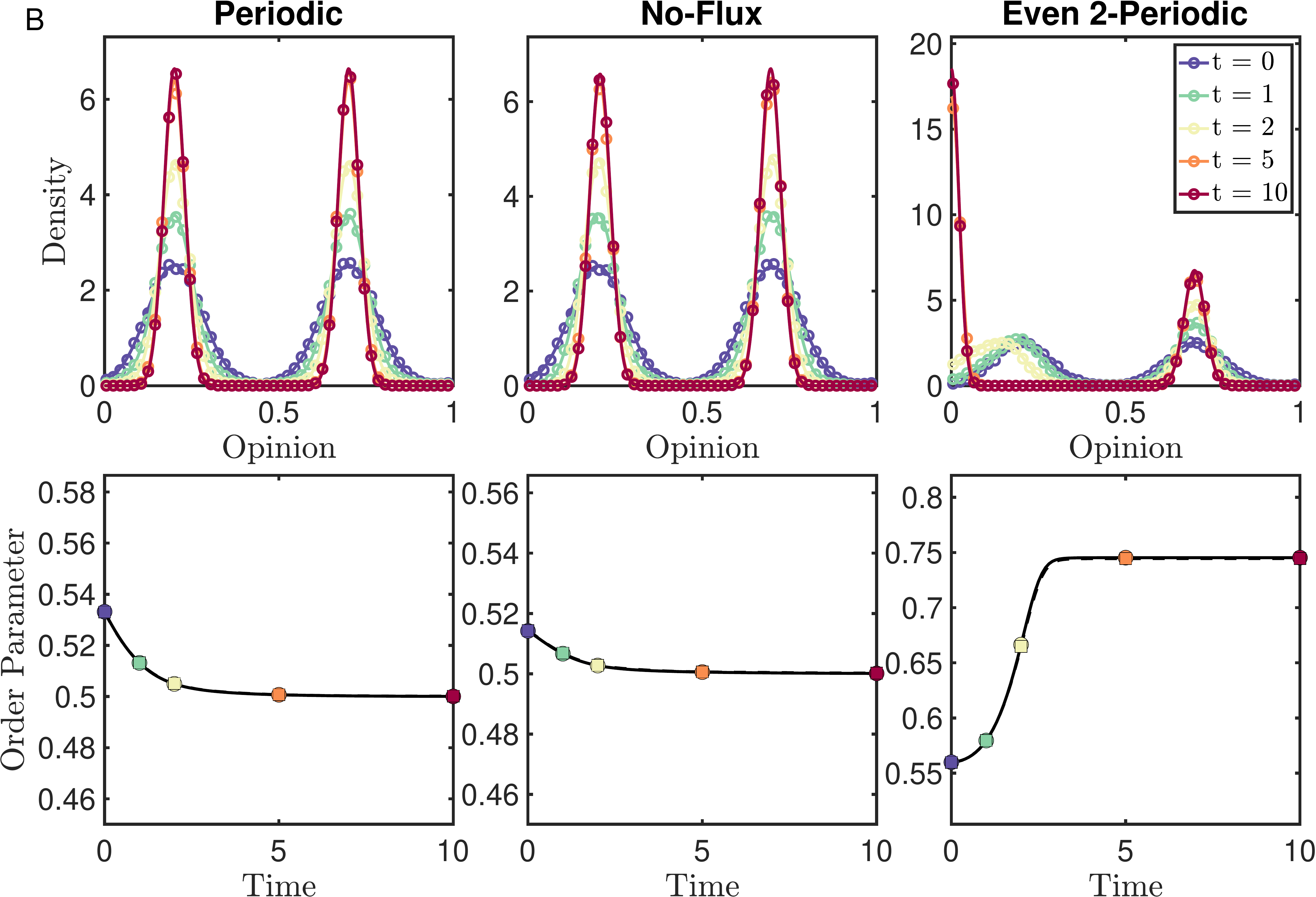}
\includegraphics[width = 0.5\textwidth]{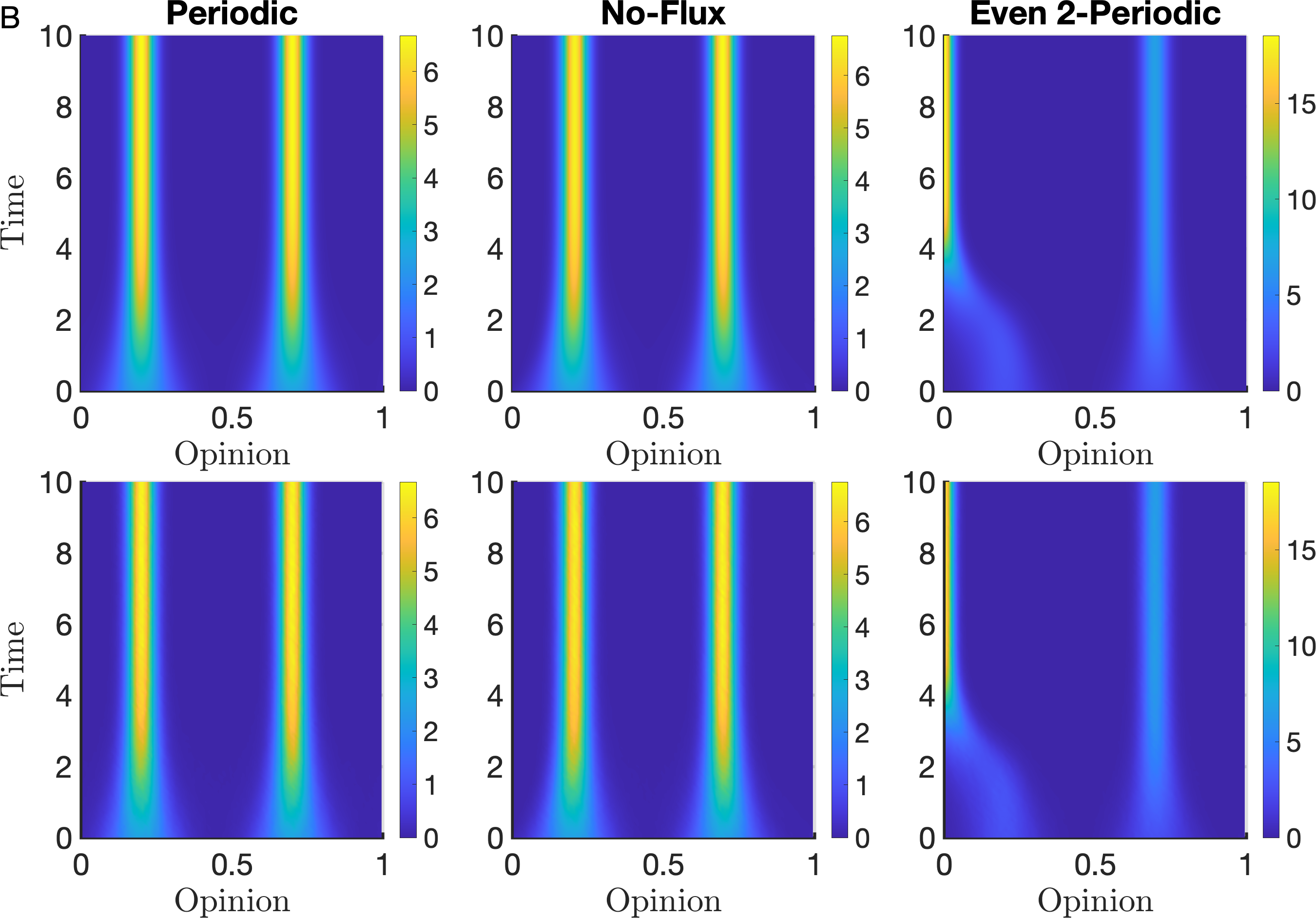}}\\[2mm]
\resizebox{\figwidth}{!}{
\includegraphics[width = 0.5\textwidth]{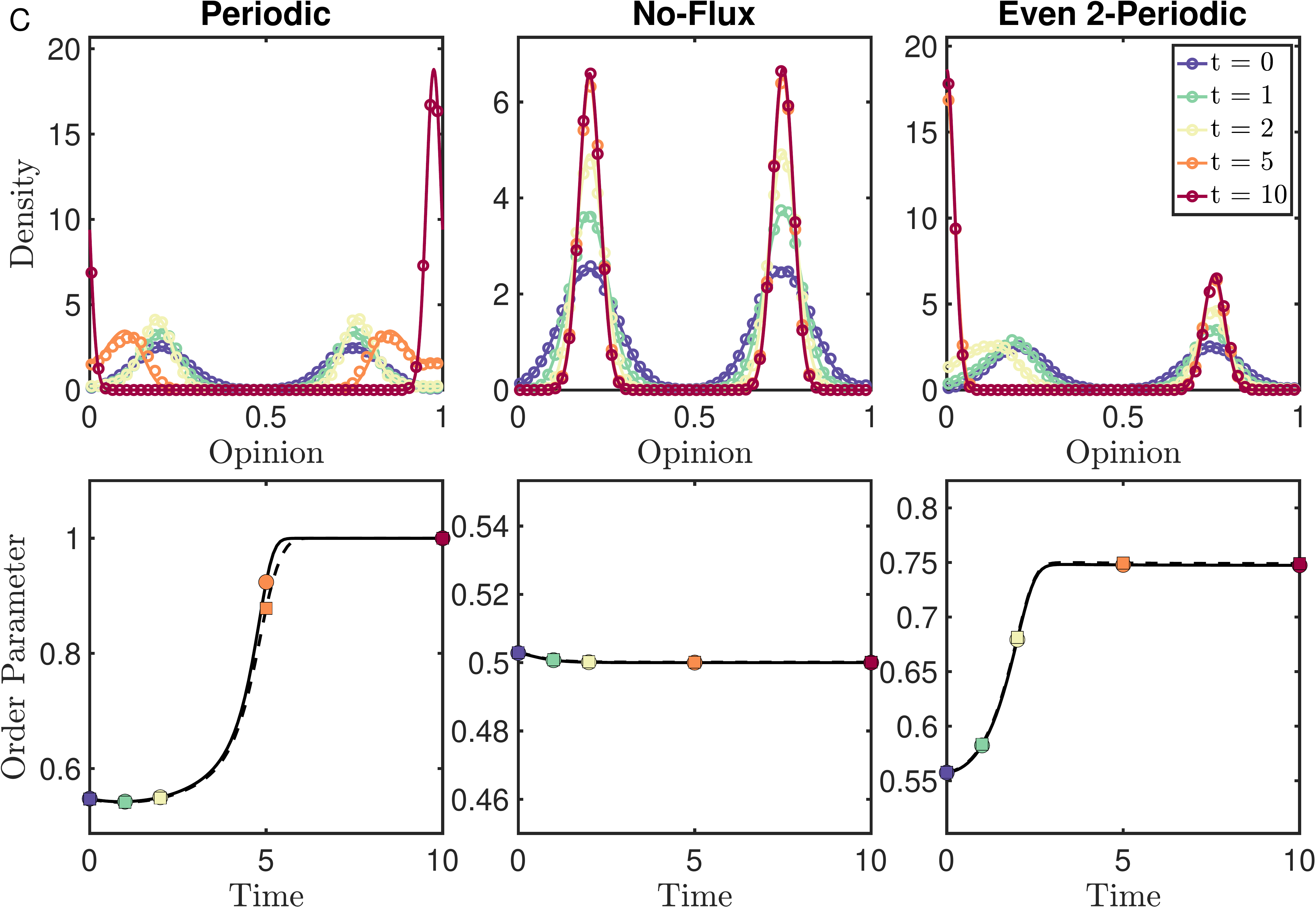}
\includegraphics[width = 0.5\textwidth]{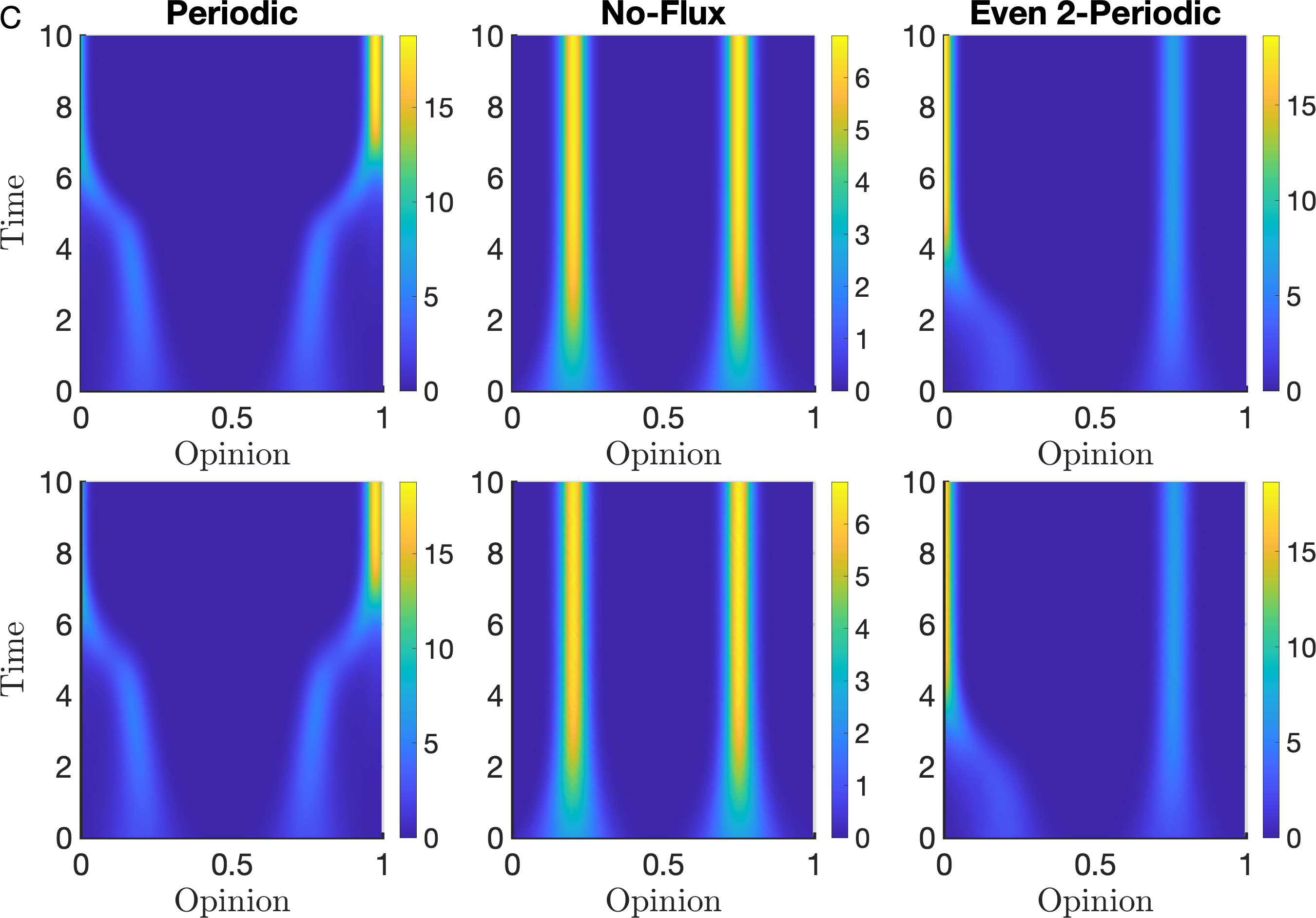}}
\caption{
As Figure~\ref{fig:Uniform_Small_Sigma_Snapshots} but for an initial condition
which is a linear combination of Gaussians.  
Labels correspond to those in the bottom right panel of Figure~\ref{fig:Gaussian},
for A ($R=0.3$, $y_{0,2} = 0.65$), B ($R=0.3$, $y_{0,2} = 0.7$), and C ($R=0.3$, $y_{0,2} = 0.75$).
Additionally, we show the solutions of the corresponding SDEs with $10^4$ particles.  Densities
shown by circles, order parameter by dashed lines and squares, and time-space plots in the
bottom row of each panel.
}
\label{fig:Two_Gaussians_Snapshots_Time}
\end{figure}

We focus on three particular
pairs of parameters $(y_{0,2},R)$, denoted A (0.65,0.3), B (0.7,0.3), and C (0.75,0.3) in 
Figures~\ref{fig:Gaussian} and~\ref{fig:Two_Gaussians_Snapshots_Time}.  In Figure~\ref{fig:Two_Gaussians_Snapshots_Time}
we plot the results up to $t=10$, which captures much of the interesting dynamics and
enables comparison to the SDE results for $10^4$ particles.
In all cases, the even 2-periodic boundary condition result is qualitatively different; the long-time
behaviour has two clusters, one close to zero and one close to the mean of the
second initial cluster.  The cause of this is the `mirror' density, which attracts the 
initial clusters at 0.2 and -0.2 together.  Note that the order parameter is close to 1, rather
than the usual value of $1/2$ denoting two clusters.
Increasing $y_{0,2}$ in the other two boundary conditions results in a switch from a single cluster 
for $y_{0,2}=0.65$ (A) to two clusters for $y_{0,2}=0.70$ (B).  The cause of this seems to be that increasing
the separation of the clusters for fixed $R$ causes the two clusters to sharpen, before
they can begin to coalesce, and they are then too well-separated to be drawn together.
This is related to the $2R$-conjecture regarding the separation of stable clusters~\cite{WLEC17}.
Finally, increasing $y_{0,2}$ to 0.7 (C) demonstrates an additional effect of periodic boundary
conditions. The long-term behaviour reverts
to a single cluster, this time centred near 1, rather than towards the centre of this interval.
The cause of this is the periodic nature of the domain -- there are two `distances' between
the initial clusters, and the shorter of the two now crosses the 0--1 point.  This causes the cluster
to form to the right of the initial cluster at 0.7, rather than to the left.  This is an issue for interpretation
in terms of extreme opinions, and suggests that if this is an aim of the model then the
no-flux boundary conditions are a more appropriate choice, both for stability and interpretability.

\section{Numerical Experiments: The Effect of Radicals} \label{s:Radicals}

In this section we will introduce a distribution of radicals, and investigate the 
sensitivity of the dynamics on this distribution.
Motivated by the observation
that the most interesting dynamics occur for small $\sigma$ and a uniform initial
distribution, we restrict to this regime here.

\subsection{Uniform Initial Condition and Gaussian Radicals}

We first consider a single Gaussian radical distribution of the form
\begin{equation}
	M \rho_r(y) = M Z^{-1} \exp\big( - C[d(y,y_0)]^2 \big),
	\label{eq:Gaussian_Radicals}
\end{equation}
where $Z$ is a normalisation constant, and $M$ determines
the mass of the radical population.
We now have five parameters in the system: $R$ and $\sigma$, as before,
and $y_0$, $C$, and M, which describe the mean, width and mass of the radicals, respectively.
To reduce the number of parameters, and to set the radical distribution to be relatively 
narrow, we fix $C = 800$.

In  Figure~\ref{fig:Uniform_Gaussian_Radicals} we fix two of the remaining parameters, and vary the other two:
$\sigma$--$R$ ($y_0 = 0.7$, $M=0.1$), 
$M$--$R$ ($\sigma = 0.02$, $y_0 = 0.7$), 
$y_0$--$\sigma$ ($R=0.1$, $M=0.1$), 
and
$y_0$--$R$ ($\sigma = 0.02$, $M = 0.1$).  The corresponding snapshots
are given in 
Figures~\ref{fig:Uniform_Gaussian_Radicals_R_sigma_Snapshots_Time}
--\ref{fig:Uniform_Gaussian_Radicals_y0_R_Snapshots_Time}.
The radical populations are shown in red (Figure~\ref{fig:Uniform_Gaussian_Radicals}),
or black (Figures~\ref{fig:Uniform_Gaussian_Radicals_R_sigma_Snapshots_Time}
--\ref{fig:Uniform_Gaussian_Radicals_y0_R_Snapshots_Time})
although due to their small size and strongly peaked normal distributions,
they are sometimes hard to distinguish.

\begin{figure}
\centering
\resizebox{\figwidth}{!}{
\includegraphics[width = 0.5\textwidth]{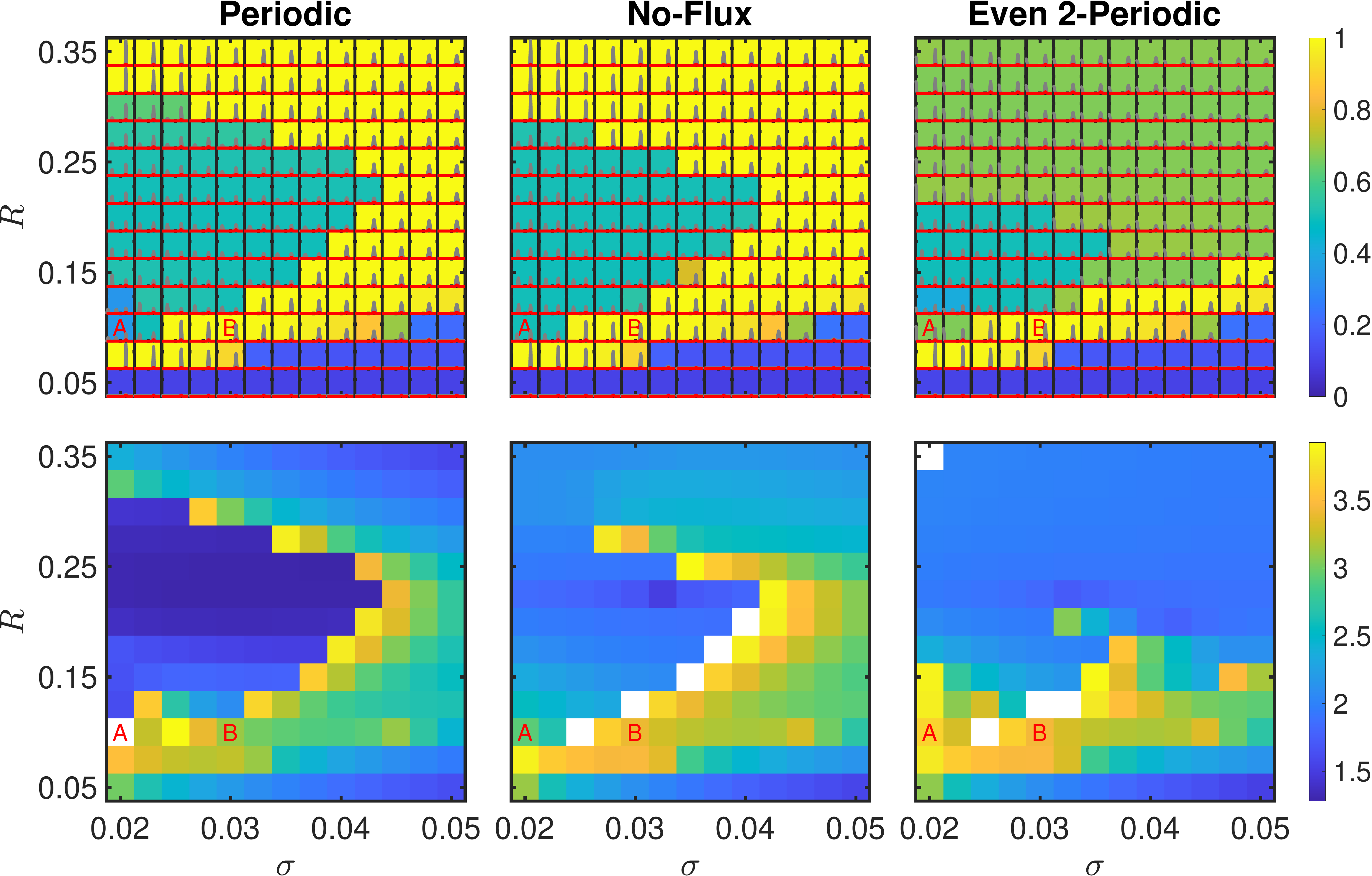}
\includegraphics[width = 0.5\textwidth]{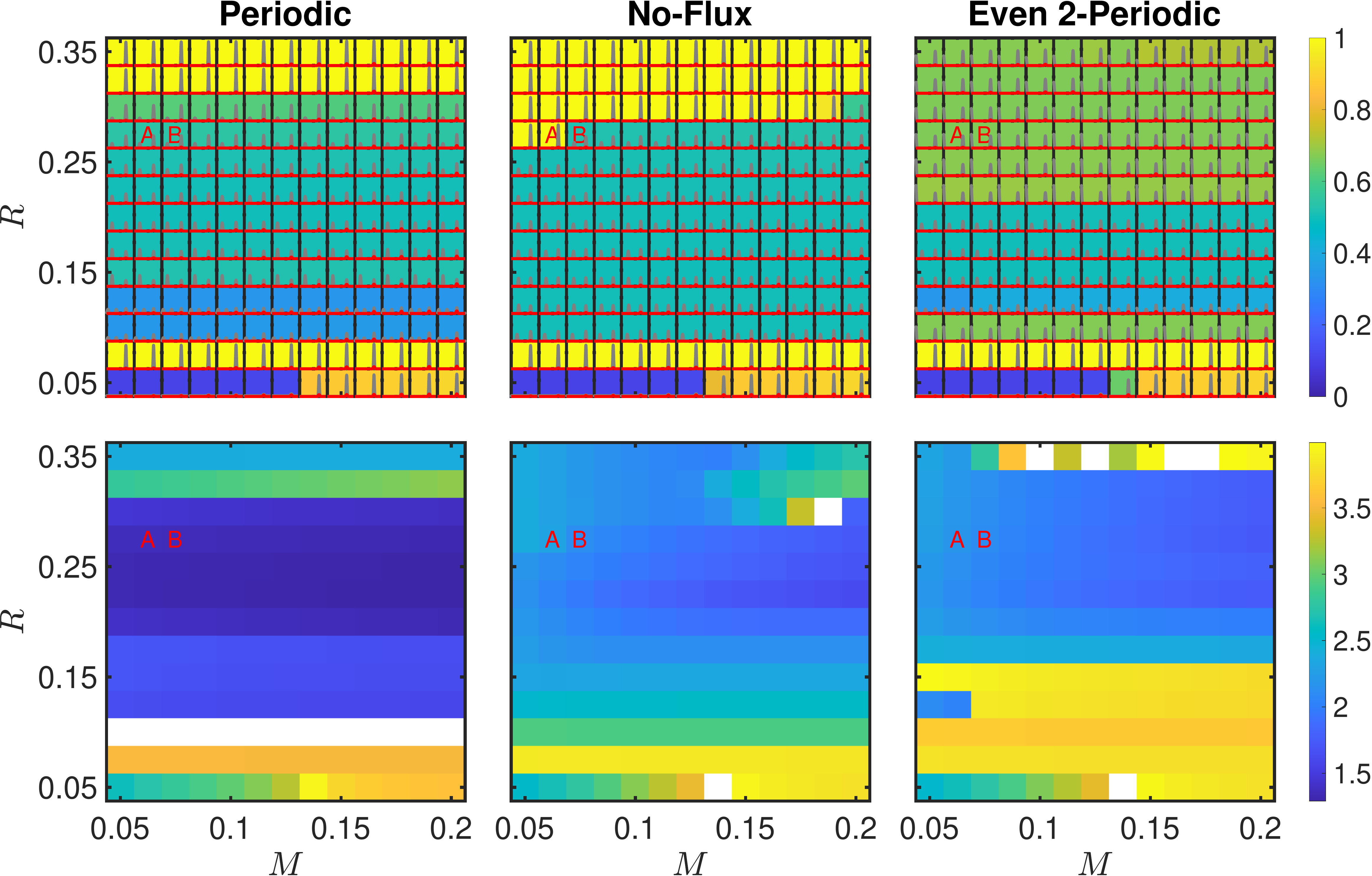}}\\[2mm]
\resizebox{\figwidth}{!}{
\includegraphics[width = 0.5\textwidth]{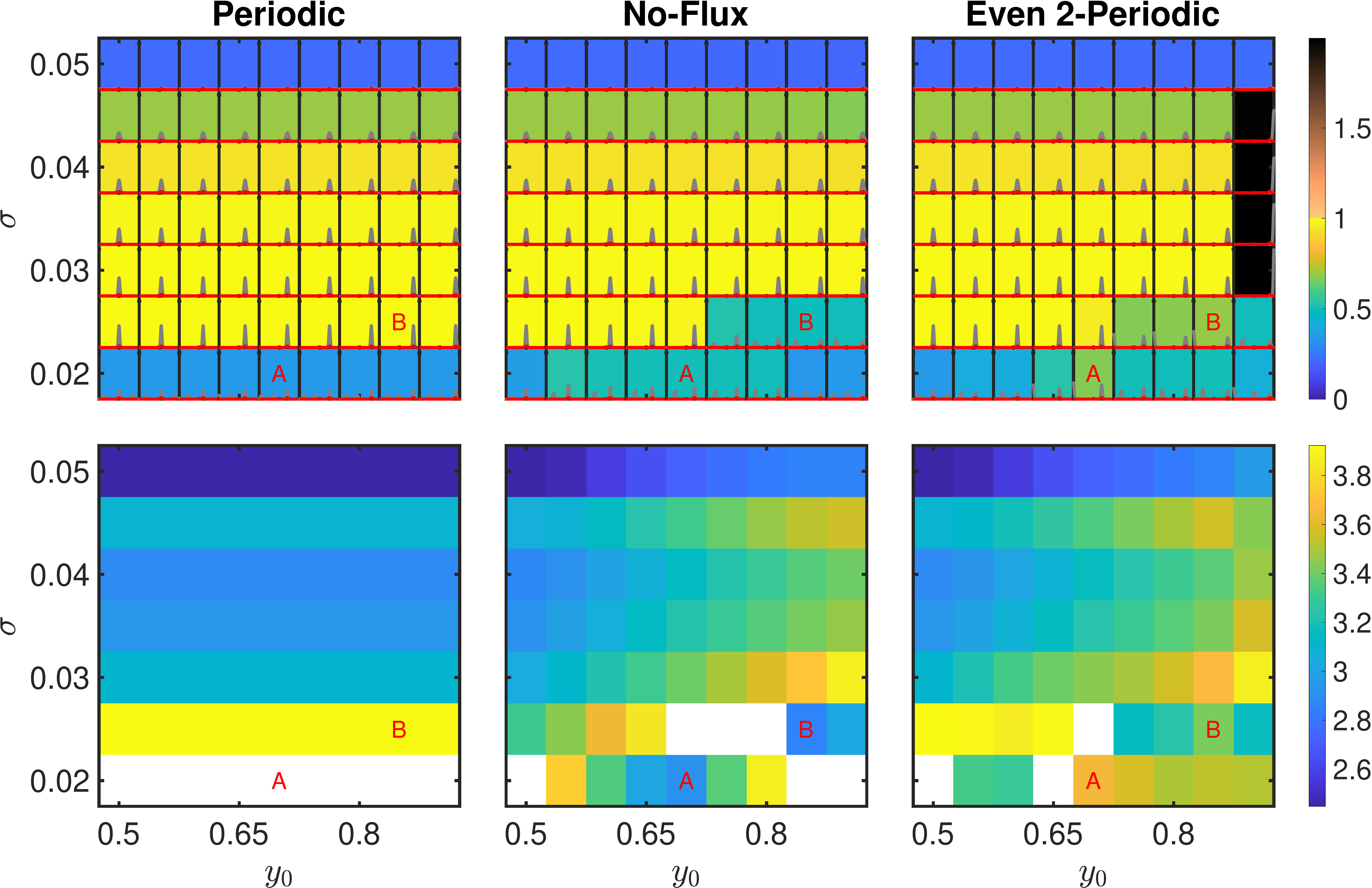}
\includegraphics[width = 0.5\textwidth]{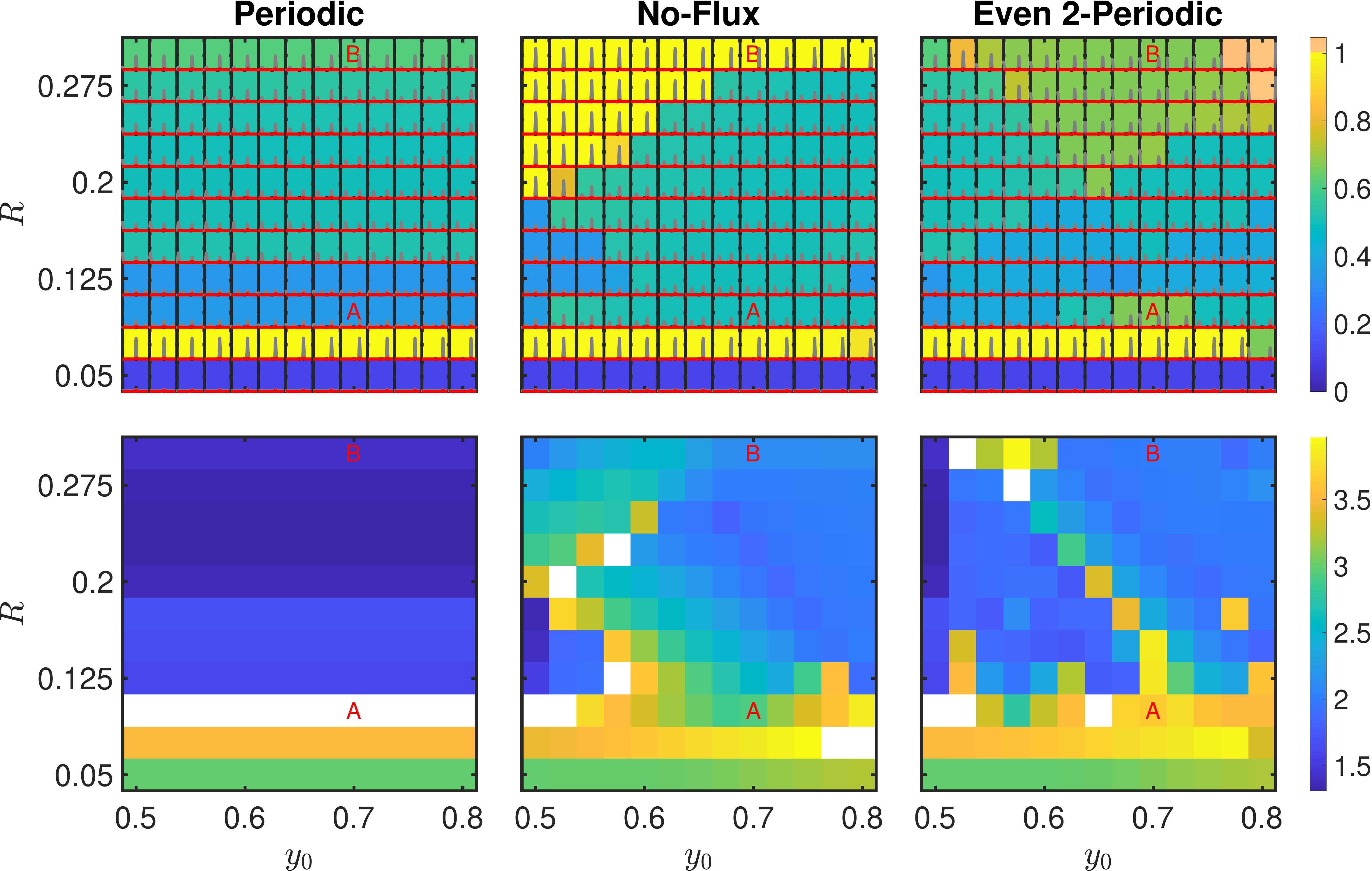}
}
\caption{As Figure~\ref{fig:Uniform} 
but for a uniform initial condition and a Gaussian radical distribution \eqref{eq:Gaussian_Radicals}. 
Here we vary: 
(top left) the strength of the noise ($\sigma$) and the 
size of the confidence interval ($R$); 
(top right) the mass of the radicals ($M$) and the confidence bound ($R$);
(bottom left) the mean position of the radical distribution ($y_0$) and the 
strength of the noise ($\sigma$);
(bottom right) the mean position of the radical distribution ($y_0$) and the 
confidence bound ($R$).  Radical distributions are shown in red.
}
\label{fig:Uniform_Gaussian_Radicals}
\end{figure}

\begin{figure}
\centering
\resizebox{\figwidth}{!}{
\includegraphics[width = 0.5\textwidth]{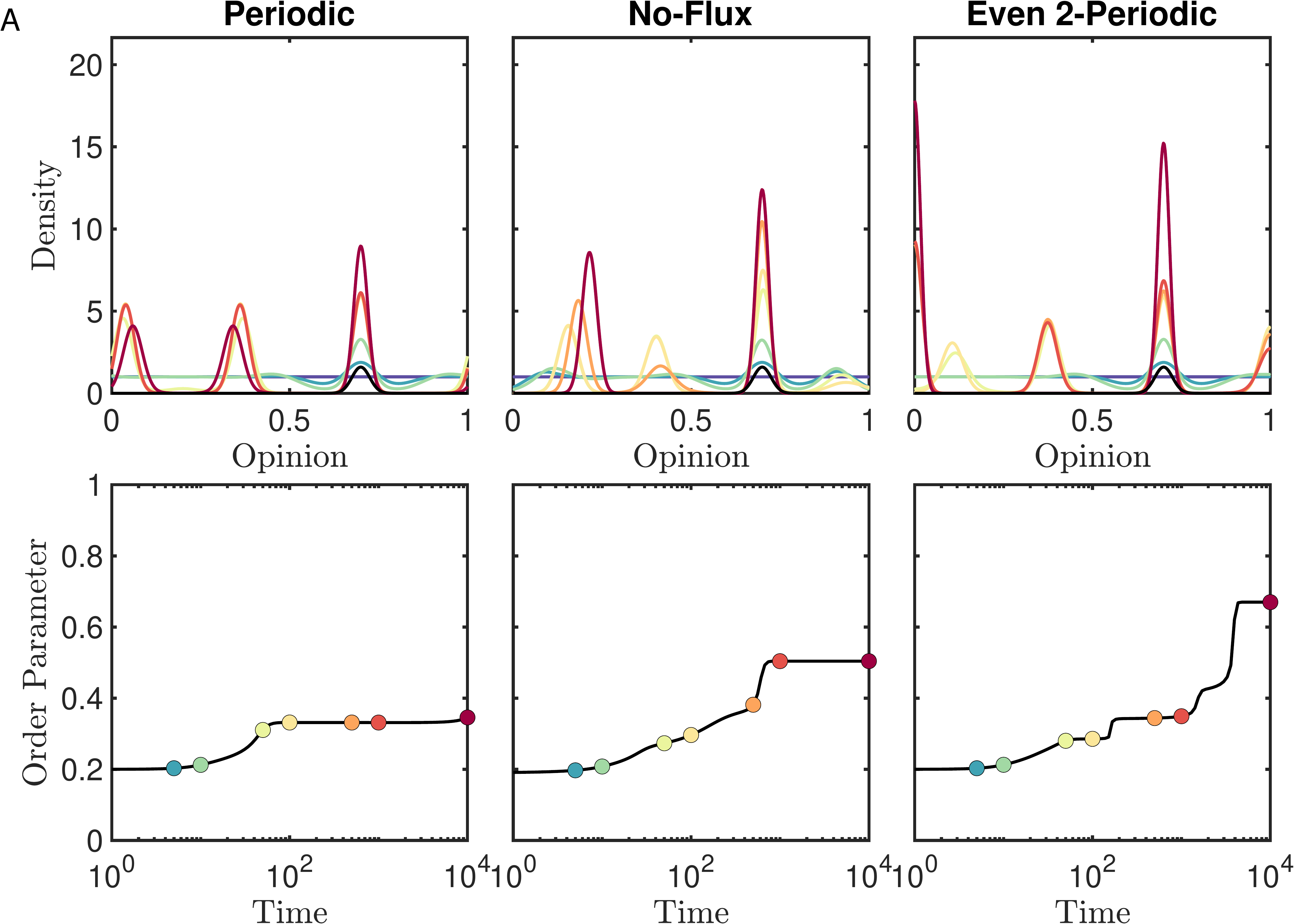}
\includegraphics[width = 0.5\textwidth]{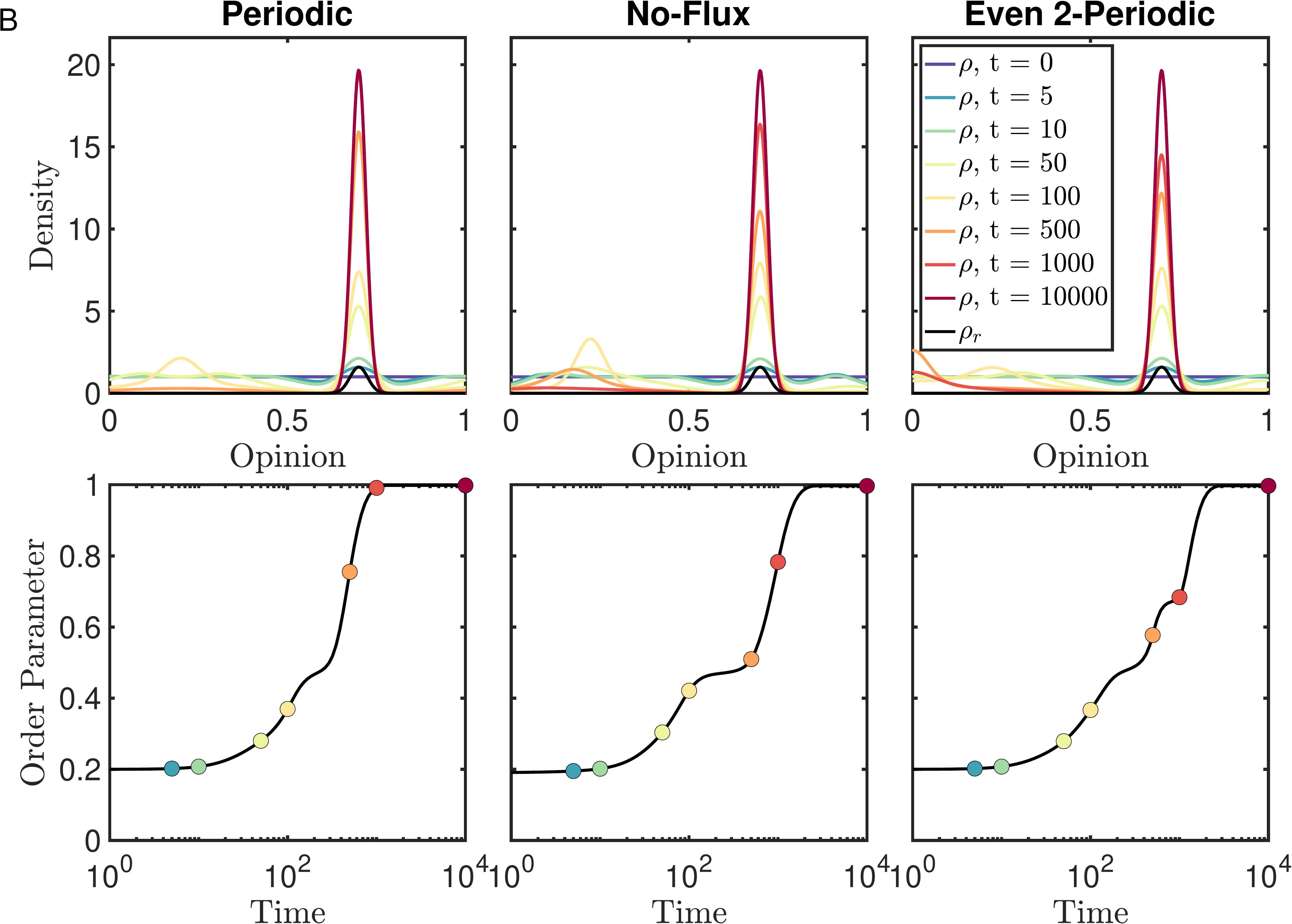}}\\[2mm]
\resizebox{\figwidth}{!}{
\includegraphics[width = 0.5\textwidth]{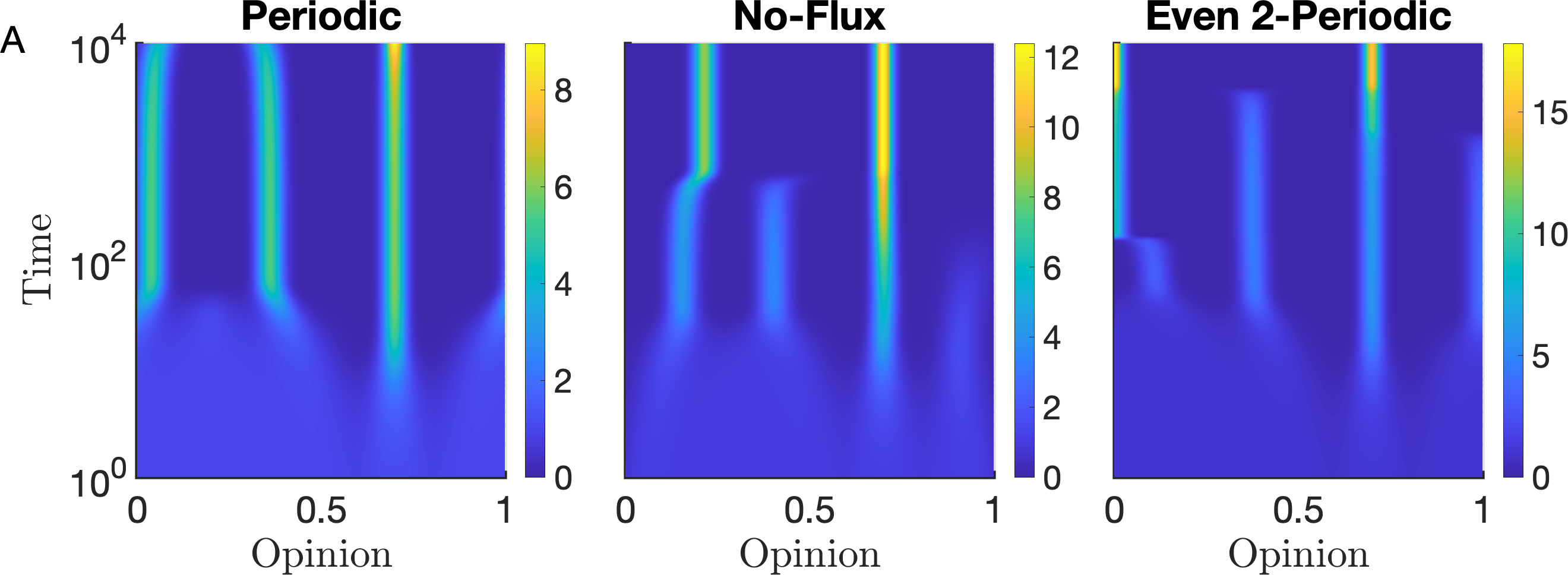}
\includegraphics[width = 0.5\textwidth]{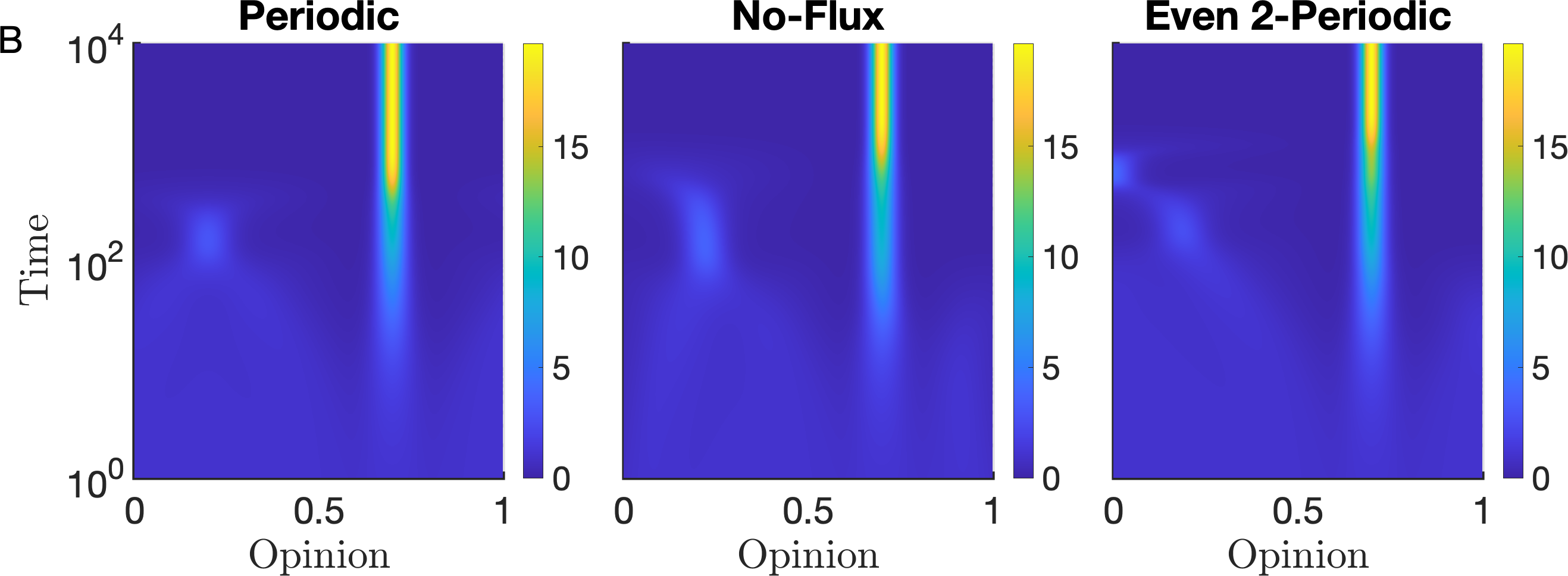}
}
\caption{
As Figure~\ref{fig:Uniform_Small_Sigma_Snapshots} 
but for a uniform initial condition
and a Gaussian radical distribution.
Labels correspond to parameter values in the top left panel of 
\ref{fig:Uniform_Gaussian_Radicals}, for A ($\sigma = 0.02$ and $R=0.1$) 
and B ($\sigma = 0.03$ and $R=0.1$).
}
\label{fig:Uniform_Gaussian_Radicals_R_sigma_Snapshots_Time}
\end{figure}

\subsubsection{$R$--$\sigma$ [Figures~\ref{fig:Uniform_Gaussian_Radicals} (top left)
and~\ref{fig:Uniform_Gaussian_Radicals_R_sigma_Snapshots_Time}]}

For small, fixed $R$, in all three boundary conditions we see a similar behaviour as
in the case of no radicals; small $\sigma$ results in a single cluster, 
which is now centred around the mean of the radical distribution, whilst increasing
$\sigma$ results in a uniform, or almost-uniform distribution.
This is perhaps what one would expect intuitively.
For larger $R$, the long-time dynamics is less intuitive.
For the periodic and no-flux boundary conditions, fixing $R$ and increasing $\sigma$
causes a transition from multiple clusters (two or three) to a single one; one cluster
is always centred around the radical distribution.
For the even 2-periodic case, there are three qualitatively
different long-time distributions.  For small sigma, two clusters emerge, as for
the other boundary conditions, and for larger $\sigma$, there is once again a single
cluster.  However, for intermediate $\sigma$ there is a different state with a cluster
centred around the radical distribution, and another narrow cluster at zero; once
again this additional state is due to the attractive nature of the mirror population.
As before, we observe much longer equilibration times for regions of phase
space on the border of different long-time distributions.

In Figure~\ref{fig:Uniform_Gaussian_Radicals_R_sigma_Snapshots_Time}
we focus on two pairs of parameters with ($R$,$\sigma$) equal to 
(0.1,0.02) [A] and (0.1,0,03) [B].
For A, the boundary condition has a strong effect on the dynamics.  
The periodic case rapidly forms three clusters, corresponding to an order 
parameter of around $1/3$.  As can be seen from the white region in 
in Figure~\ref{fig:Uniform_Gaussian_Radicals}
this simulation has not reached our definition of equilibrium, so it is possible
that this is not the final state.  
However, slightly increasing
$R$ appears to make the three-cluster state stable.
In the no-flux case, there are initially four clusters, two of which are
relatively weak, which eventually merge into two strong clusters, once centred
around the radicals, and one at the other side of the interval.  
As with the other boundary conditions, this suggests
that, with these parameter values, the radicals cause an initial division in the
population, but do not have enough influence to cause a single consensus to 
form.
Finally, in the even 2-periodic case, the dynamics is even more complex; this
is well demonstrated by the order parameter.
The four clusters at times around time $10^2$, look similar to those in the
no-flux case, and once again the final state has two clusters.
However, here the additional
clusters persist for much longer, and the second
cluster is near to zero, rather than centred at approximately 0.2.

For B, the final state for all three boundary conditions is relatively
similar, with a single, strong cluster centred around the radicals.  However,
once again, the dynamical path to this distribution depends on the boundary
conditions.  In all three cases, at times up to around $100$, a secondary
cluster on the left of the interval is visible.  In the periodic case this then rapidly
merges with the final cluster, whilst in the other two cases it is longer-lived,
first moving away from the final cluster, before eventually merging into it.
At time $500$ the secondary cluster in the no-flux case is 
located near zero, as in case A; here we suggest that a larger value of $\sigma$ 
increases the diffusion to a point at which the weaker clusters are dispersed
and can be captured by the strong cluster centred around the radicals.  
The clusters, especially the large one, are also slightly wider with the
increased value of $\sigma$, which may aid this coalescence.

\begin{figure}
\centering
\resizebox{\figwidth}{!}{
\includegraphics[height = 5.2cm]{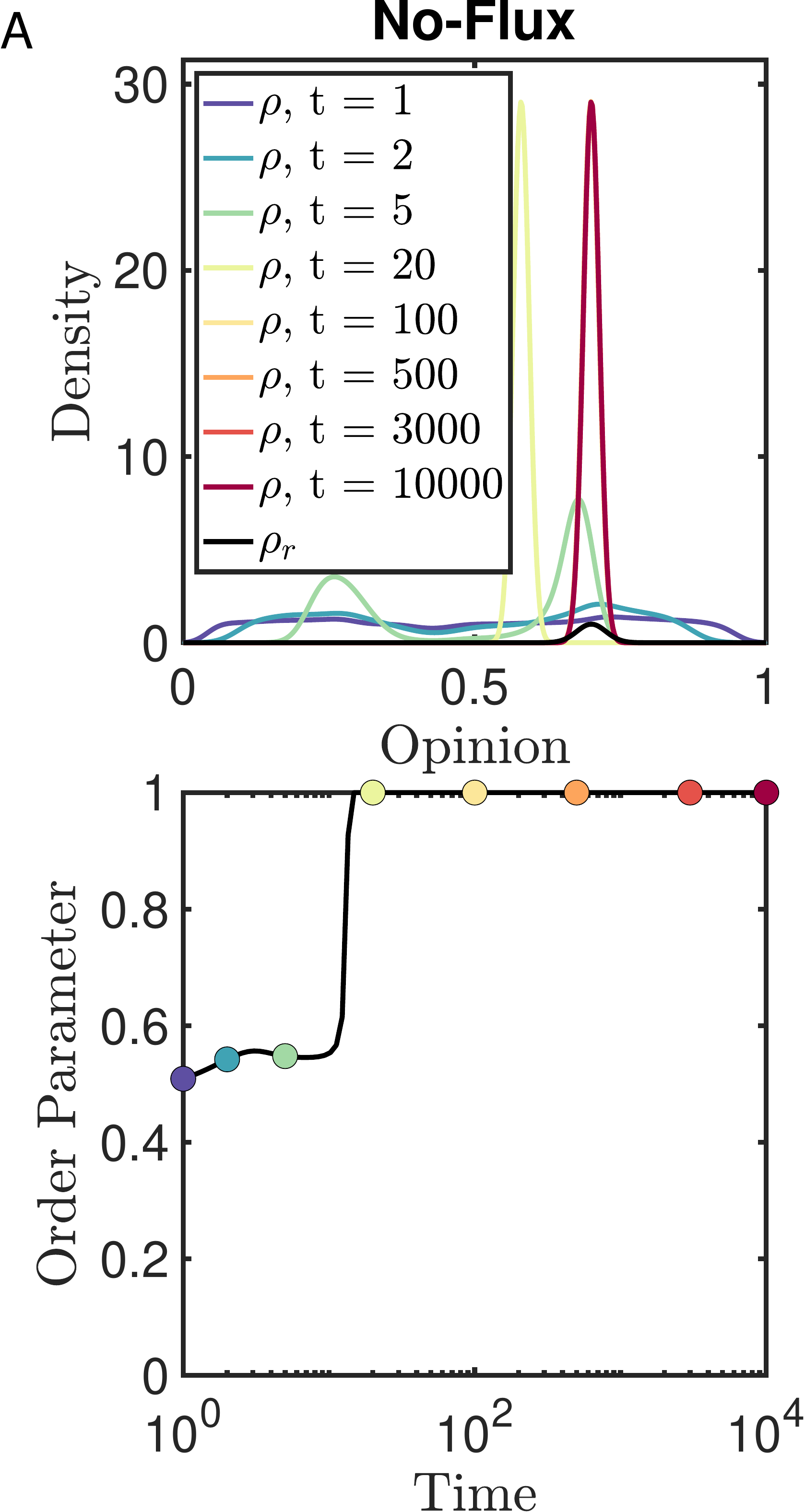}
\includegraphics[height = 5.2cm]{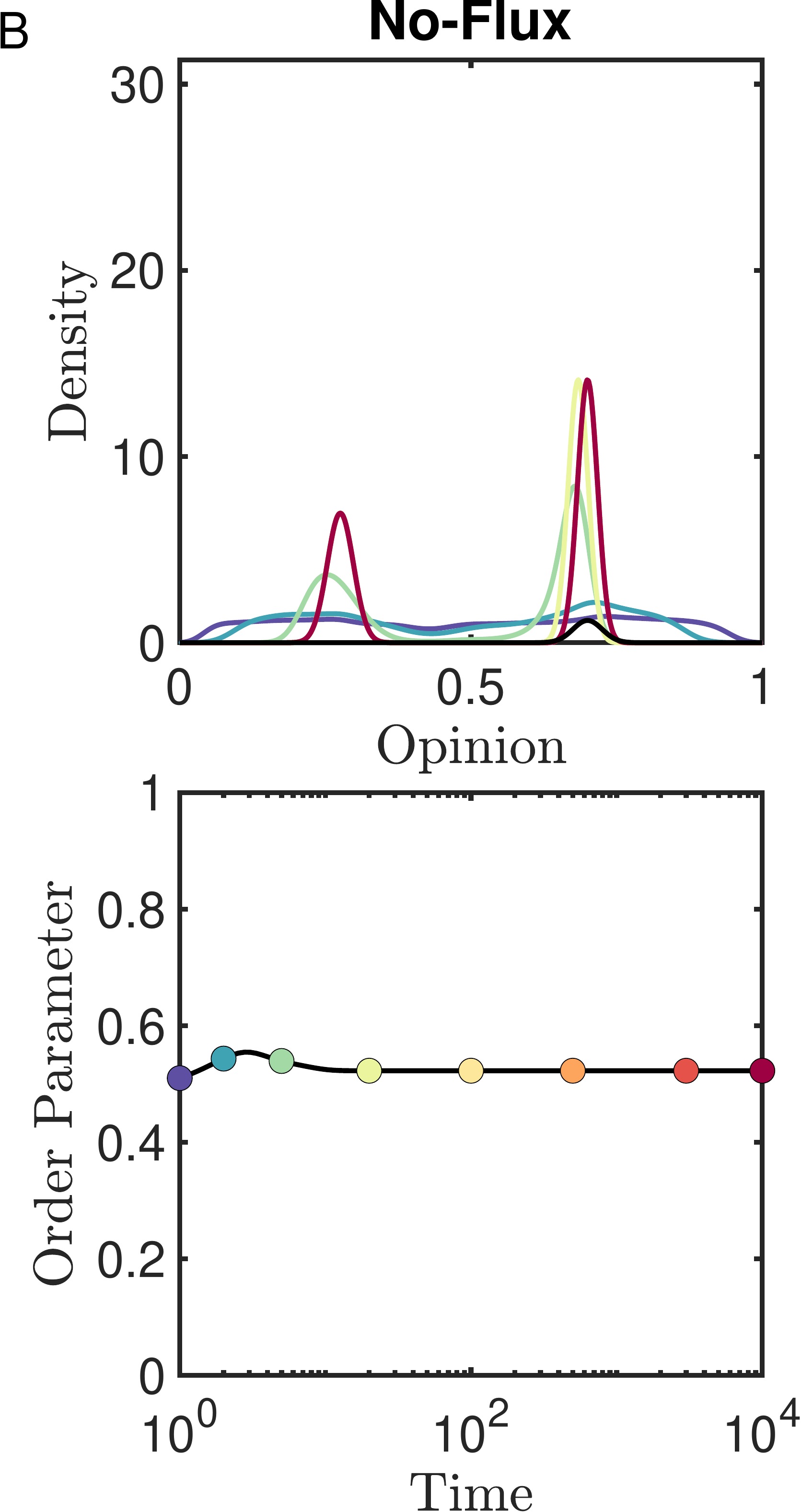}
\hspace*{1mm}
\vrule
\hspace*{1mm}
\includegraphics[height = 5.2cm]{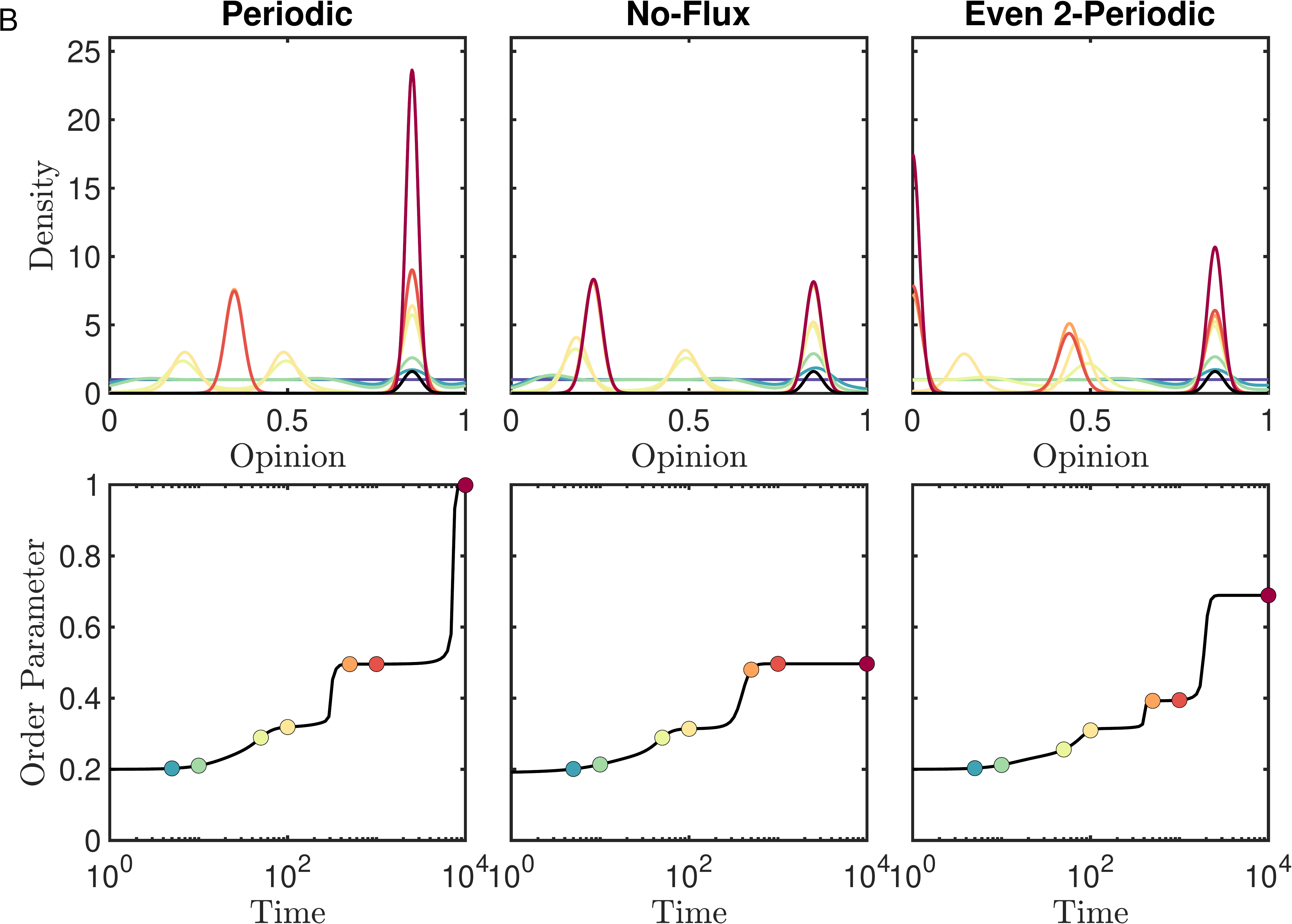}}\\[2mm]
\resizebox{\figwidth}{!}{
\includegraphics[height = 2.7cm]{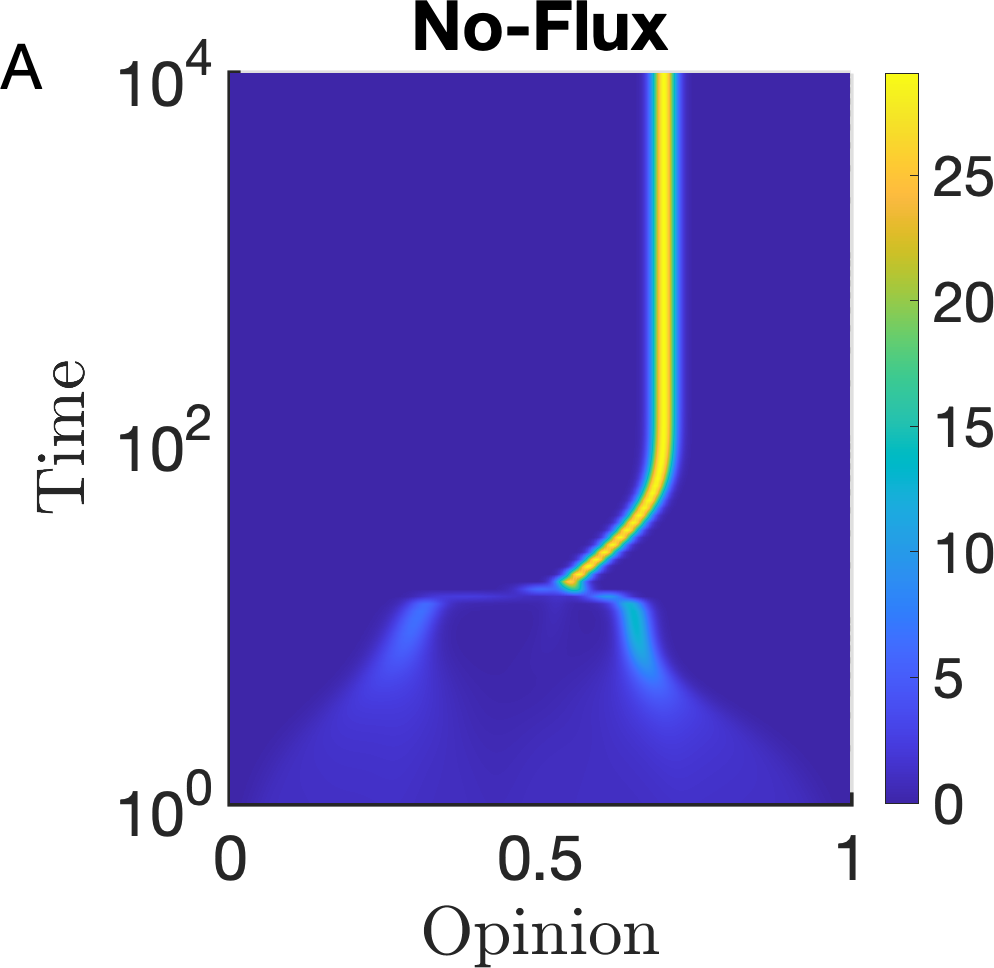}
\includegraphics[height = 2.7cm]{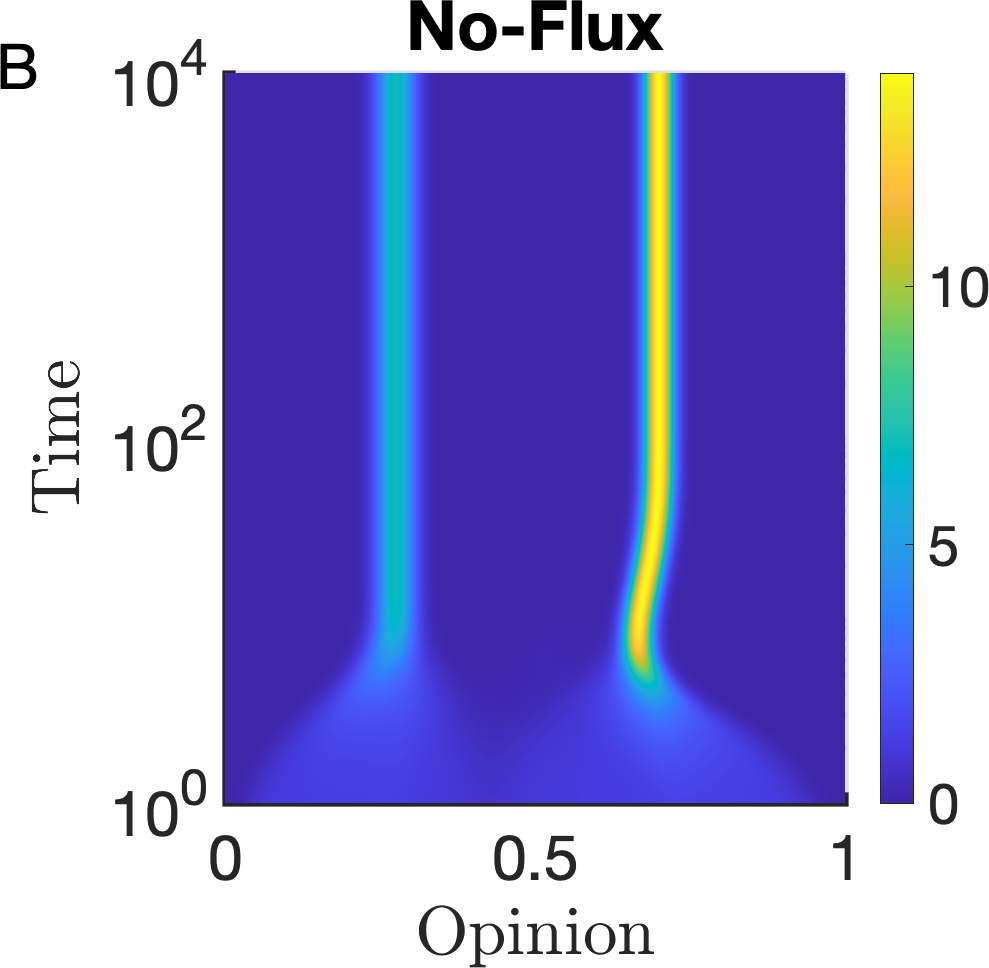} 
\hspace*{1mm}
\vrule
\hspace*{1mm}
\includegraphics[height = 2.7cm]{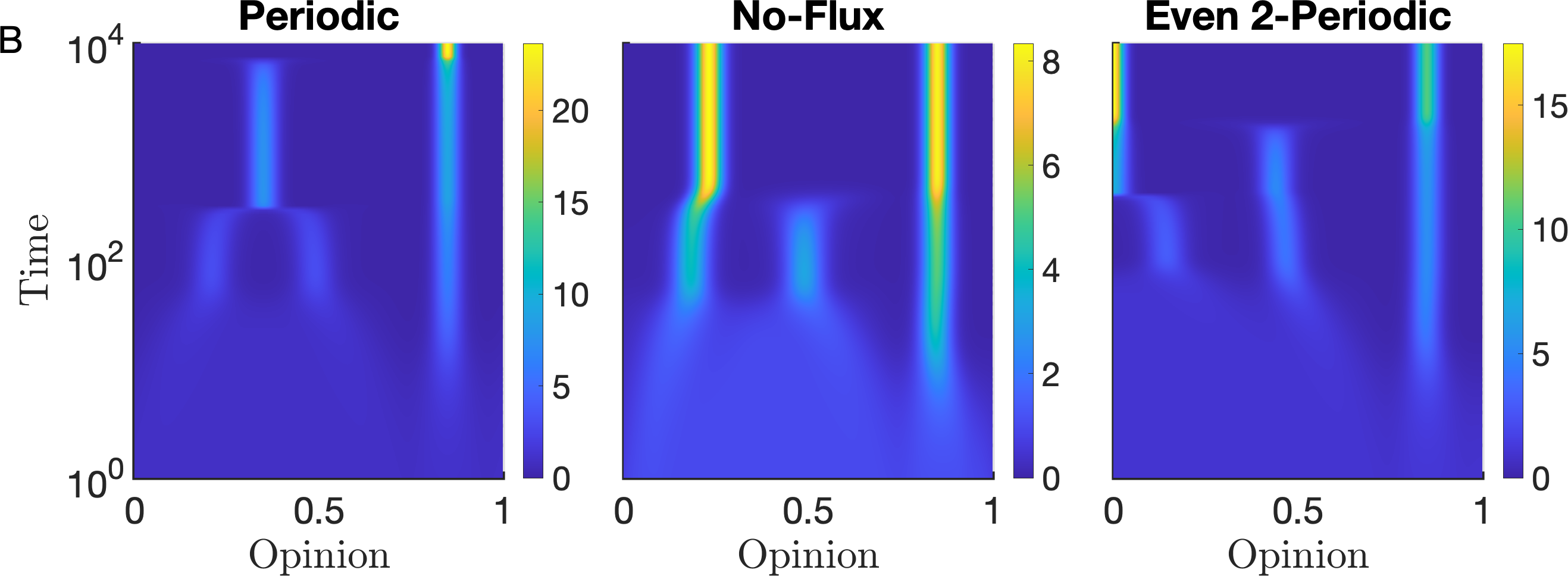}}
\caption
{
As Figure~\ref{fig:Uniform_Gaussian_Radicals_R_sigma_Snapshots_Time}
but for: (left) the labels in the top right panel of Figure~\ref{fig:Uniform_Gaussian_Radicals}
for A ($M=0.0625$ and $R=0.275$)  and B ($M=0.075$ and $R=0.275$);
(right) the labels in the bottom left panel of Figure~\ref{fig:Uniform_Gaussian_Radicals}
with B ($y_0 = 0.85$ and $\sigma = 0.025$).
}
\label{fig:Uniform_Gaussian_Radicals_M_R_y0_sigma_Snapshots_Time}
\end{figure}

\subsubsection{$M$--$R$ [Figures~\ref{fig:Uniform_Gaussian_Radicals} (top right)
and~\ref{fig:Uniform_Gaussian_Radicals_M_R_y0_sigma_Snapshots_Time} (left)]}

From Figure~\ref{fig:Uniform_Gaussian_Radicals}, it is clear that the dependence on $M$
is very weak in the two periodic cases, unless $R$ is very small.
For larger $R$, and no-flux boundary conditions, varying $M$ can produce unintuitive results.
Comparing 
A and B on the left of \ref{fig:Uniform_Gaussian_Radicals_M_R_y0_sigma_Snapshots_Time}
we see that increasing $M$ causes the long-time distribution to switch from a 
single cluster around the radical distribution (which is something one would expect), 
to a bimodal distribution, in which the increase in the number of
radicals has caused polarisation.
A possible explanation for this is that a larger radical cluster attracts the
nearby normal population more quickly, causing a split in the distribution
after short times; this can be seen in both cases A and B.  In case A,
the two clusters move towards each other (with the one near
the radicals actually moving away from the radical population), 
before forming a single cluster
away from the radical distribution, which then migrates towards the
radicals.
In case B, it appears that the radicals are now sufficiently strong to prevent
the initial cluster around them from moving towards the centre of the interval,
leaving the two polarised clusters.

\subsubsection{$y_0$--$\sigma$ [Figures~\ref{fig:Uniform_Gaussian_Radicals} (bottom left)
and~\ref{fig:Uniform_Gaussian_Radicals_M_R_y0_sigma_Snapshots_Time} (right)]}

As is clear from Figure~\ref{fig:Uniform_Gaussian_Radicals}, the results of the
periodic case are independent of $y_0$, up to a shift in the opinion axis.
In the other cases, the final order parameter and shape of the distribution are essentially independent of 
$y_0$ for all but small $\sigma$, or for large (or, by symmetry, small) $y_0$ when the no-flux and even 2-periodic BCs 
become influential.
As $y_0$ increases, there is a clear, corresponding shift in position of final-time
cluster, and, as before, increasing $\sigma$ leads to disorder.
Case A corresponds to Case A in Figure~\ref{fig:Uniform_Gaussian_Radicals_R_sigma_Snapshots_Time}.
Case B (see right of Figure~\ref{fig:Uniform_Gaussian_Radicals_M_R_y0_sigma_Snapshots_Time}) 
corresponds to slightly larger $\sigma$ and increased $y_0$, 
for which the dynamics is strongly-dependent on
the choice of boundary conditions.  The periodic case has a single cluster, 
whilst no-flux has two, which are almost the same size, and even 2-periodic
has two, but with one near zero and significantly taller.

\begin{figure}
\centering
\resizebox{\figwidth}{!}{
\includegraphics[width = 0.5\textwidth]{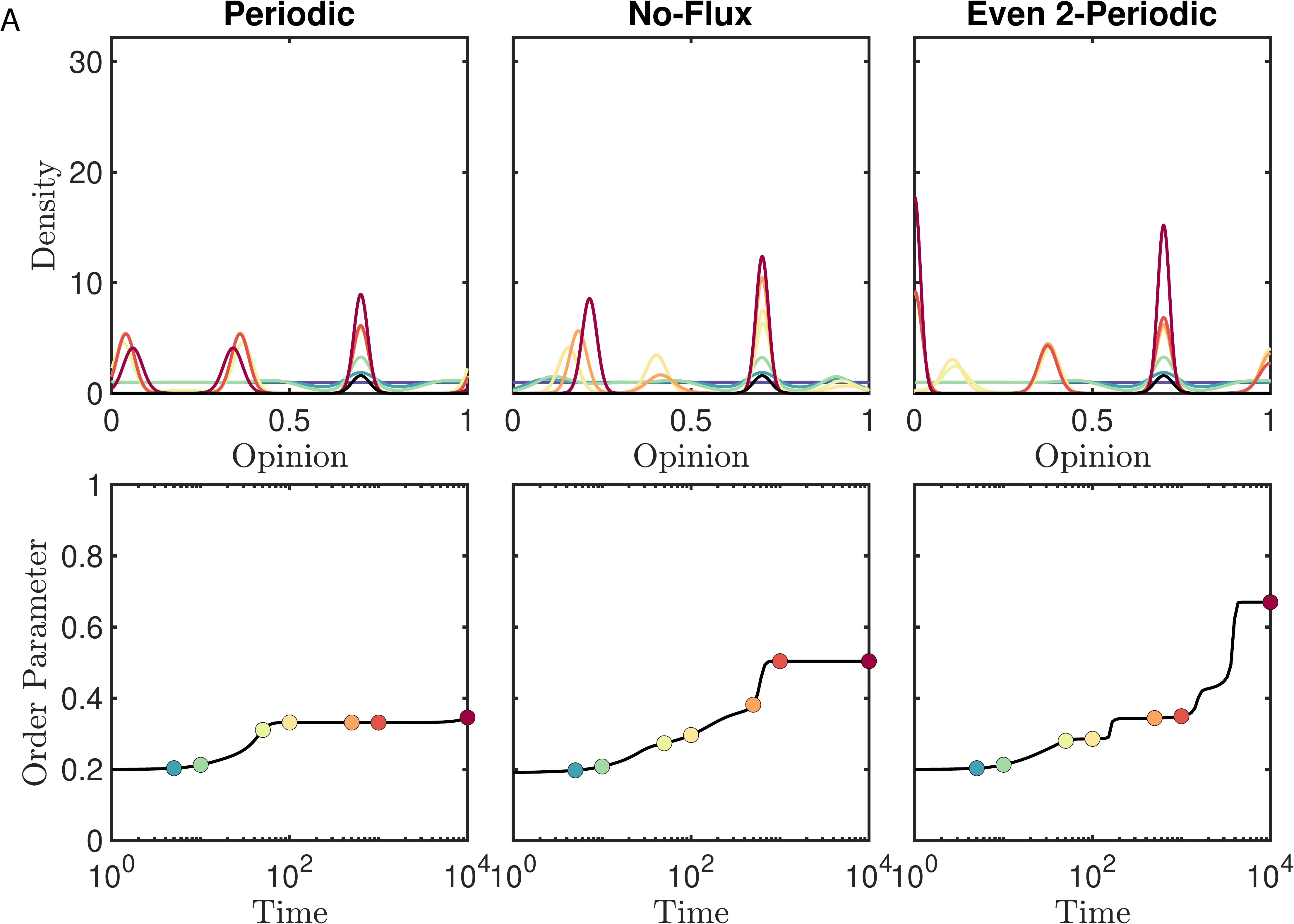}
\includegraphics[width = 0.5\textwidth]{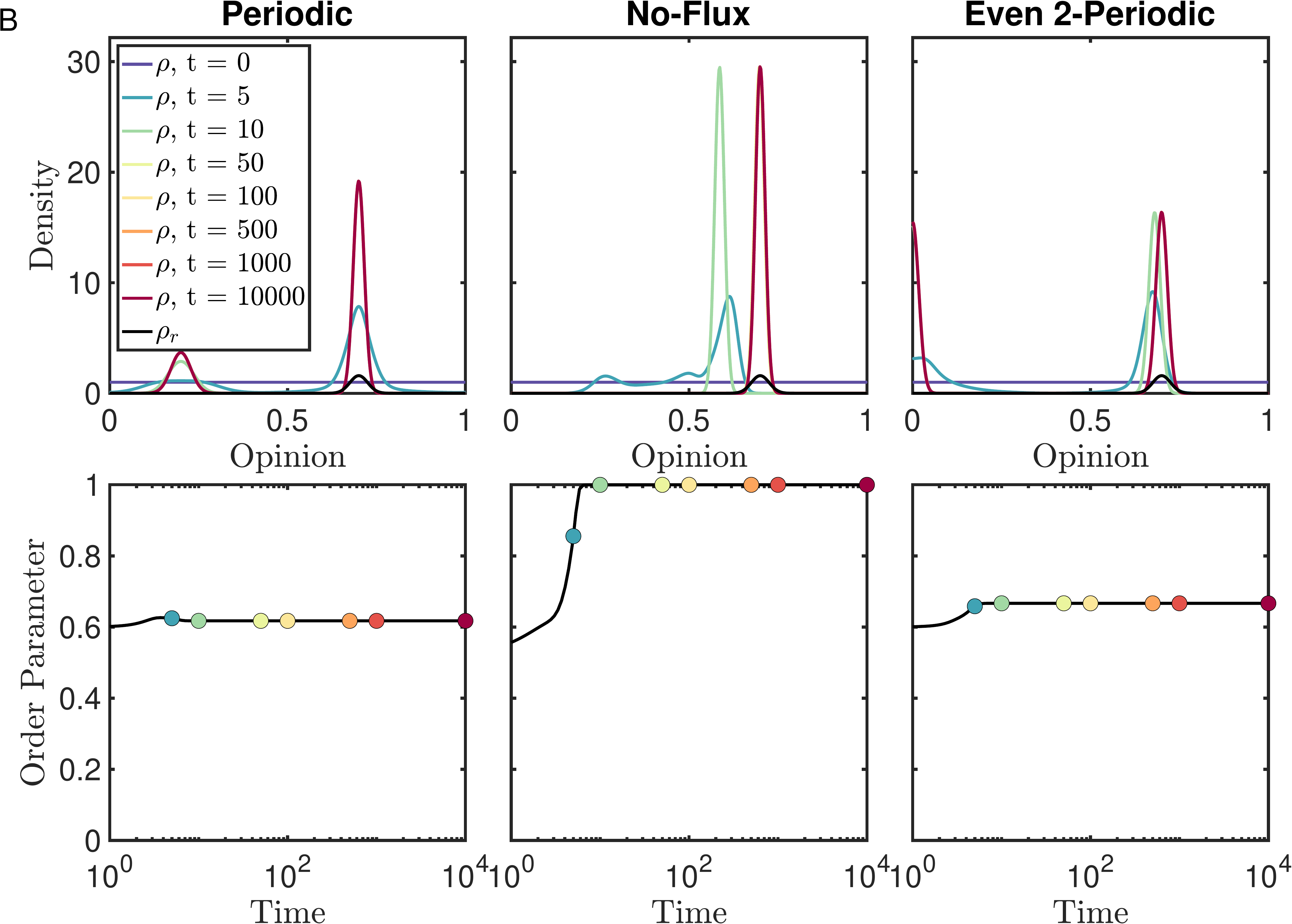}}\\[2mm]
\resizebox{\figwidth}{!}{
\includegraphics[width = 0.5\textwidth]{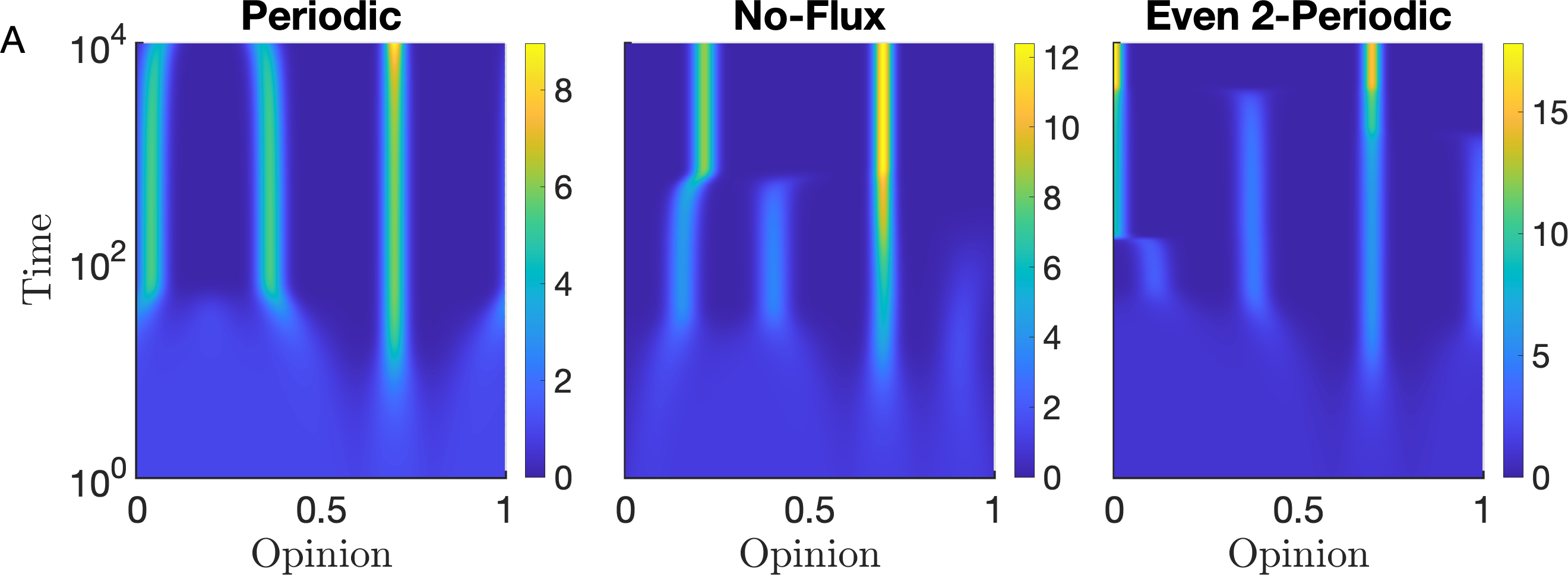}
\includegraphics[width = 0.5\textwidth]{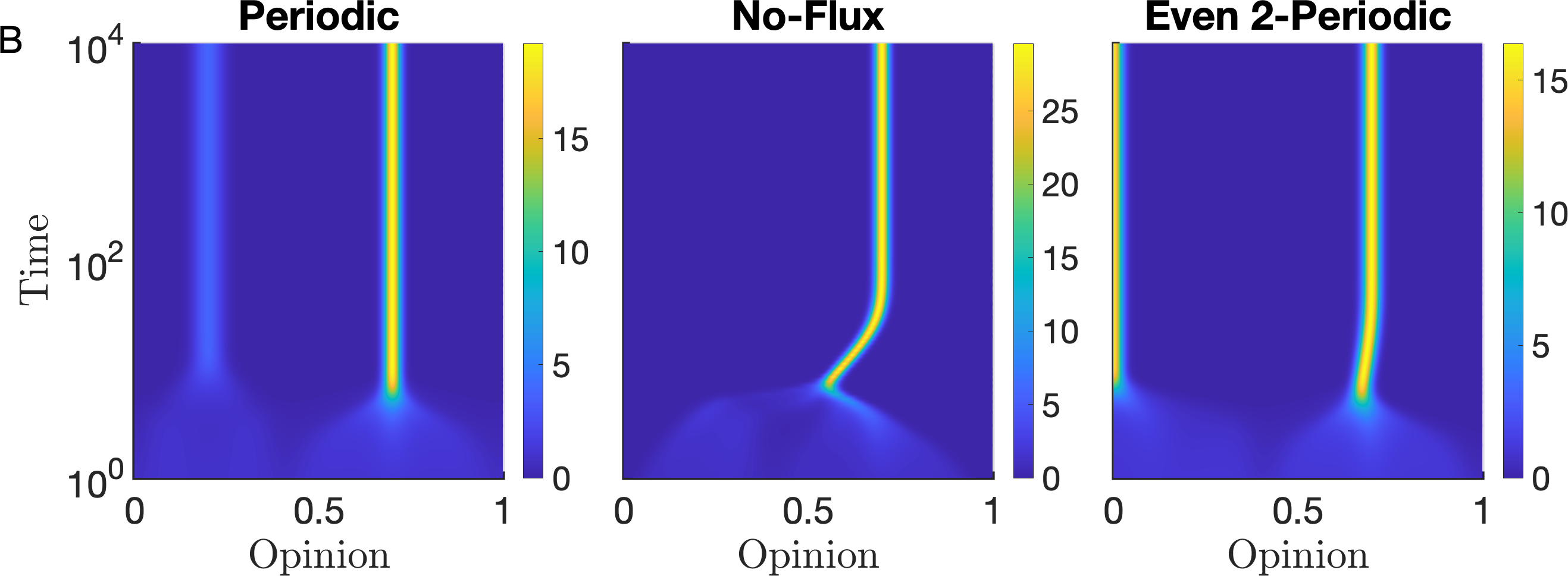}
}
\caption{As Figure~\ref{fig:Uniform_Gaussian_Radicals_R_sigma_Snapshots_Time}
but for the labels in the bottom right panel of Figure~\ref{fig:Uniform_Gaussian_Radicals},
with A ($y_0 = 0.7$ and $R = 0.1$),
and B ($y_0 = 0.7$ and $R = 0.3$).
}
\label{fig:Uniform_Gaussian_Radicals_y0_R_Snapshots_Time}
\end{figure}

\subsubsection{$y_0$--$R$ [Figures~\ref{fig:Uniform_Gaussian_Radicals} (bottom right)
and~\ref{fig:Uniform_Gaussian_Radicals_y0_R_Snapshots_Time}]}

%
In Figure~\ref{fig:Uniform_Gaussian_Radicals} we see that the different regimes 
of the no-flux and even 2-periodic cases
are much richer, with multiple parameter regions with qualitatively different
long-time dynamics.
As could be expected, increasing $R$, and hence the interaction range, leads to bigger differences between the 
results for different boundary conditions.
In Figure~\ref{fig:Uniform_Gaussian_Radicals_y0_R_Snapshots_Time}, we show some
representative dynamics, demonstrating the richness and complexity as the
parameters are varied.

\subsection{Uniform Initial Condition and Double-Gaussian Radicals}


As a second example we consider a double-Gaussian distribution of radicals
\begin{equation}
	M \rho_r(y) = M Z^{-1} \Big[
	\lambda \exp\big( - C[d(y,y_{0,1})]^2 \big) 
	+ (1-\lambda) \exp\big( - C[d(y,y_{0,2})]^2 \big) \Big],
	\label{eq:TwoGaussian_Radicals}
\end{equation}
where, once again, $Z$ is the normalisation constant for the term in square brackets, and $M$ is the mass
of the radical distribution. Here, $\lambda \in [0,1]$ is a parameter describing the relative masses of the
two Gaussians.  The physical interpretation here is that we have two competing groups of radicals. 
To simplify exposition, whilst demonstrating interesting effects, we first fix $C= 800$  (to ensure a 
concentrated radical distribution) and $M=0.1$.  
We consider two choices of $\lambda$, namely $\lambda=0.5$, which
corresponds to a completely even split of radicals, and $\lambda = 0.499$, which is a slight bias towards
one of the radical opinions.

\begin{figure}
\centering
\resizebox{\figwidth}{!}{
\includegraphics[width = 0.5\textwidth]{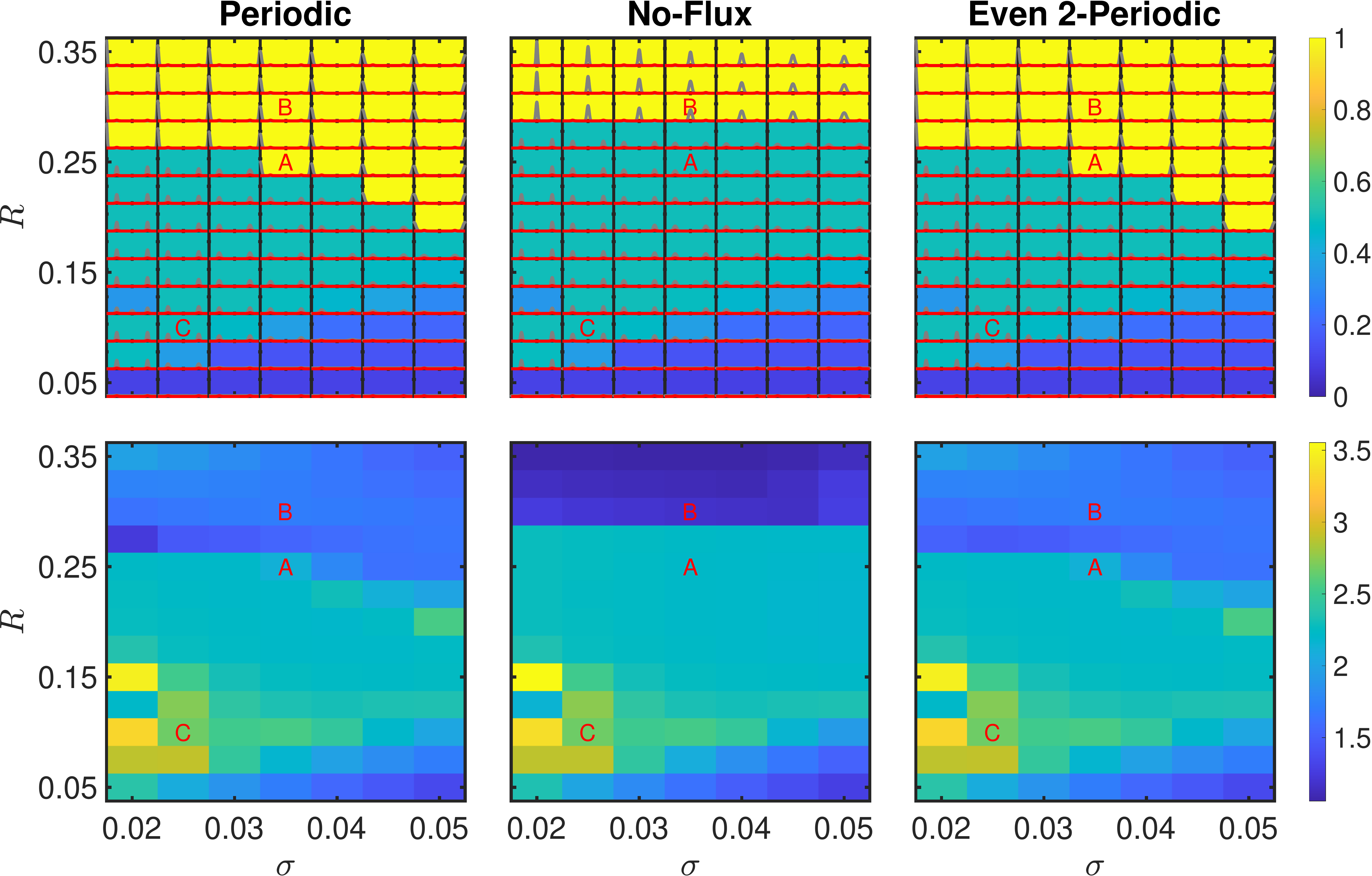}
\includegraphics[width = 0.5\textwidth]{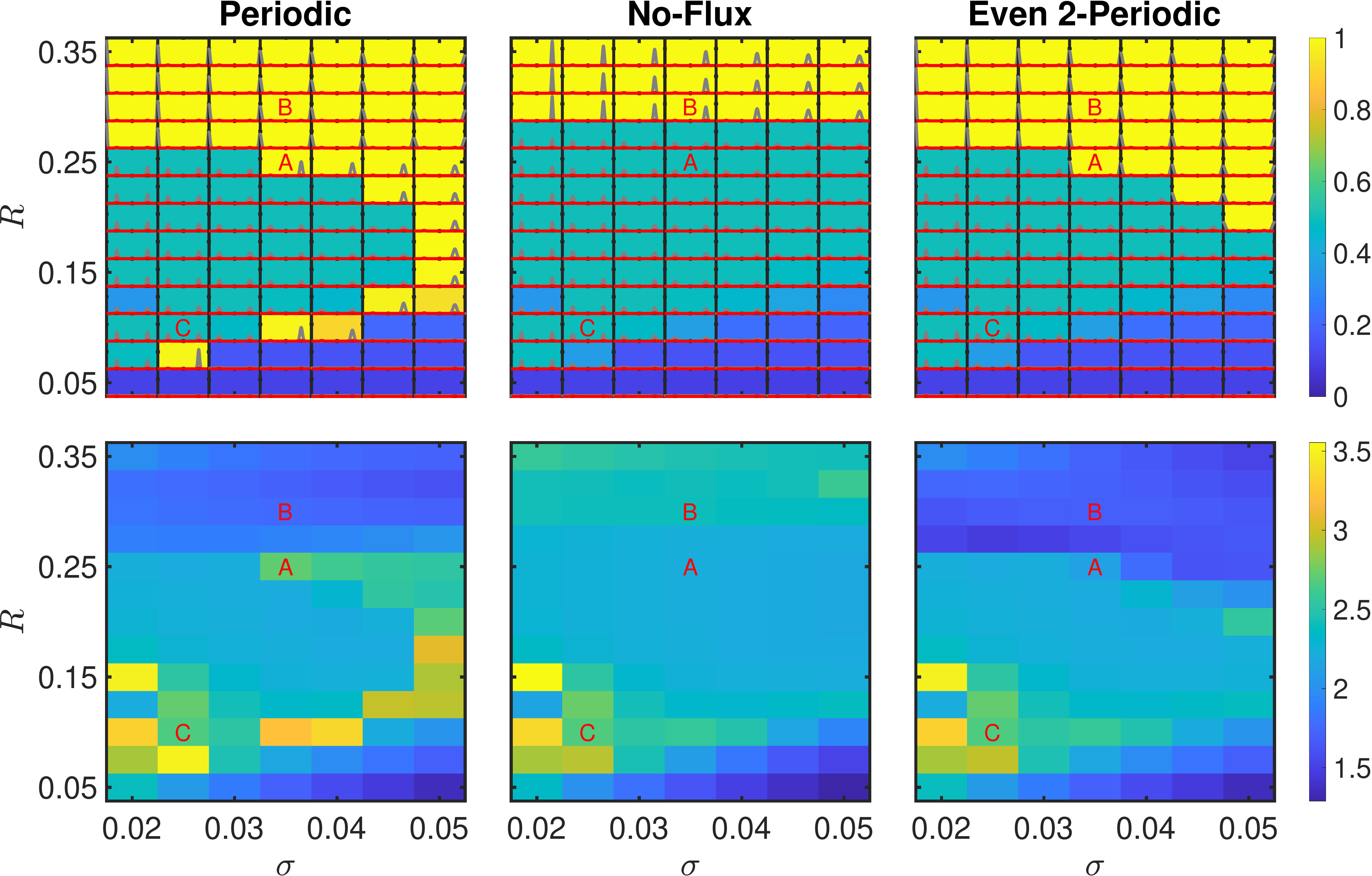}
}
\caption{As Figure~\ref{fig:Uniform} 
but for a uniform initial condition
and a double-Gaussian radical distribution \eqref{eq:TwoGaussian_Radicals} with $\lambda = 0.5$ (left)
and $\lambda=0.499$ (right). Here we vary
the strength of the noise ($\sigma$) and the 
confidence bound ($R$).
}
\label{fig:Uniform_TwoGaussians_Radicals}
\end{figure}

In Figure~\ref{fig:Uniform_TwoGaussians_Radicals}, we choose choose $y_{0,1} = 1-y_{0,2} = 0.2$,
and vary $R$ and $\sigma$ with $\lambda = 0.5$ (left) and $\lambda = 0.499$ (right).
By symmetry, the periodic and even 2-periodic cases are identical, for $\lambda = 0.5$.
We note similar trends to before: increasing $R$ favours the formation of a single cluster,
whilst increasing $\sigma$ tends to produce an almost uniform distribution.  For the periodic and
even 2-periodic cases, for moderate $R$, increasing $\sigma$ causes a transition from two clusters
to a single one; we interpret this as the noise becoming large enough to overcome the attraction of the radical distribution.  
This does not occur in the no-flux case, at least for the range of parameters studied here.
When a single cluster is formed, we see significant effects of the boundary conditions.
For no-flux, the cluster is either centred in the middle of the interval ($\lambda=0.5$),
or on top of the dominant radical cluster ($\lambda=0.499$).  In contrast, the periodic and 
even 2-periodic cases result in clusters
centred around 0/1, caused by the shortest
distance between the radical clusters being across 0/1, rather than through 0.5.
Physically, this means that two competing populations of radicals can result in either a moderate
cluster (for no-flux), or two extreme clusters (for periodic and even 2-periodic).  This is a significant effect
of the boundary conditions, and suggests issues when interpreting such results without giving careful
consideration to the modelling choices.  Similar sensitivities with respect to the choices of
$y_{0,1}$ and $y_{0,2}$ are demonstrated in Supplementary Material Section SM5.

In cases A ($\sigma = 0.035$ and $R = 0.25$) and B ($\sigma = 0.035$ and $R = 0.3$), 
we see that strong dependence on the choice of $\lambda$. In the periodic case
with parameters A, the short-time dynamics are similar for both values of $\lambda$, but the symmetry 
breaking with $\lambda = 0.499$ is very clear in the long-time dynamics.  For the no-flux case with
parameters B, two  initial clusters merge into a single cluster at the centre of the interval, with the asymmetry becoming
visible only at longer times when the cluster migrates to the right with $\lambda = 0.499$.  We highlight that these
are far from the only interesting and non-intuitive results from the model.
For parameters C ($\sigma = 0.025$ and $R = 0.1$)
we show only the $\lambda = 0.5$ periodic case, as the other cases are very similar.
Here we interpret the two radical clusters as competing political viewpoints.  Whilst the long-time
behaviour in all cases is a pair of equal clusters centred around the radicals, our interest here lies
in the intermediate dynamics, where there is a clear third peak centred around zero.
As discussed in~\cite{BS16} in the context of the Scottish independence referendum,
this is a common feature of such situations in which the population mostly polarises, but leaves
a number of `undecided' individuals, who only move to one of the popular opinions after long times.  It would
be interesting to study such situations further.

\begin{figure}
\centering
\resizebox{\figwidth}{!}{
\includegraphics[width = 2.6cm]{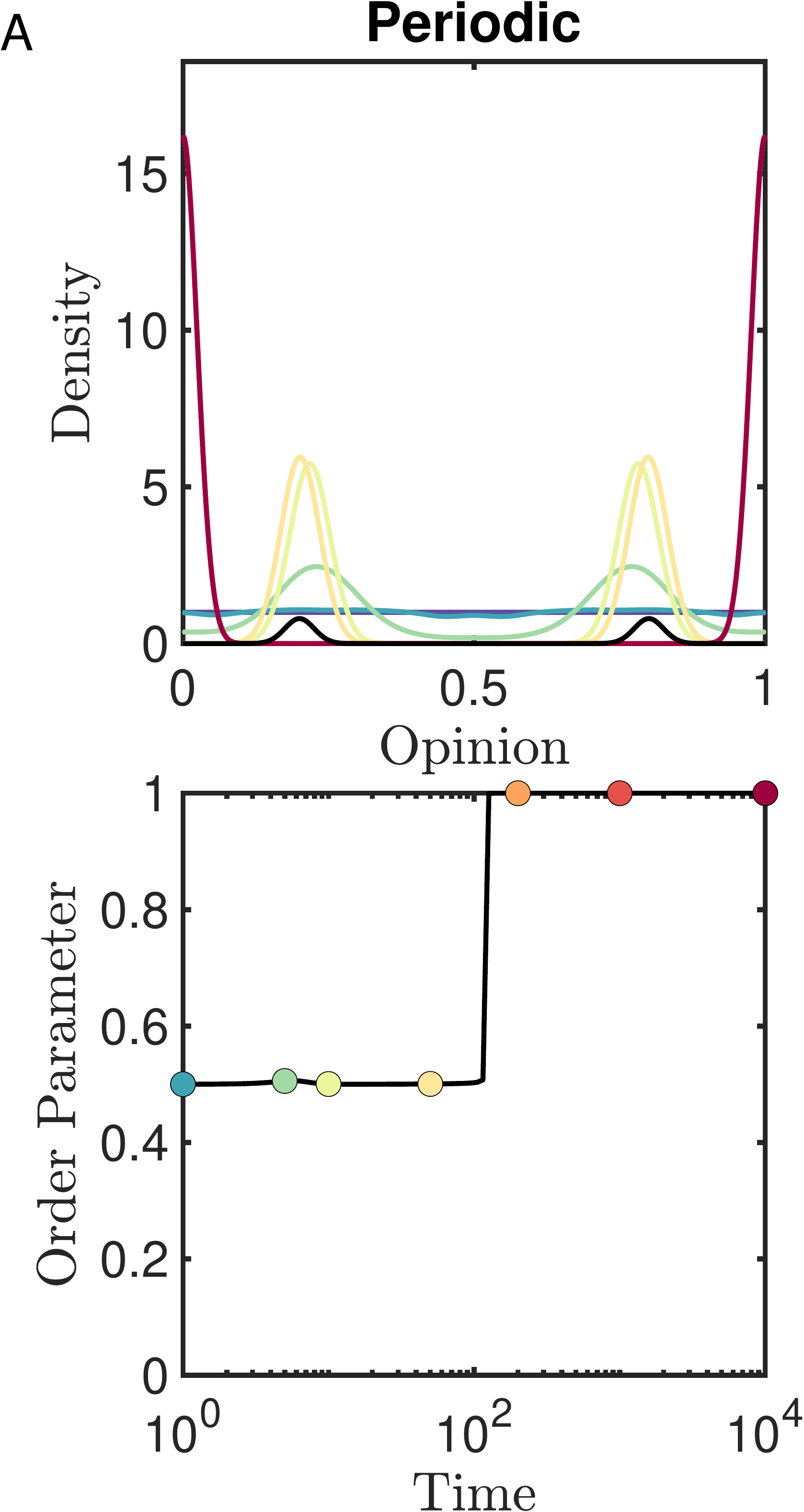}
\includegraphics[width = 2.6cm]{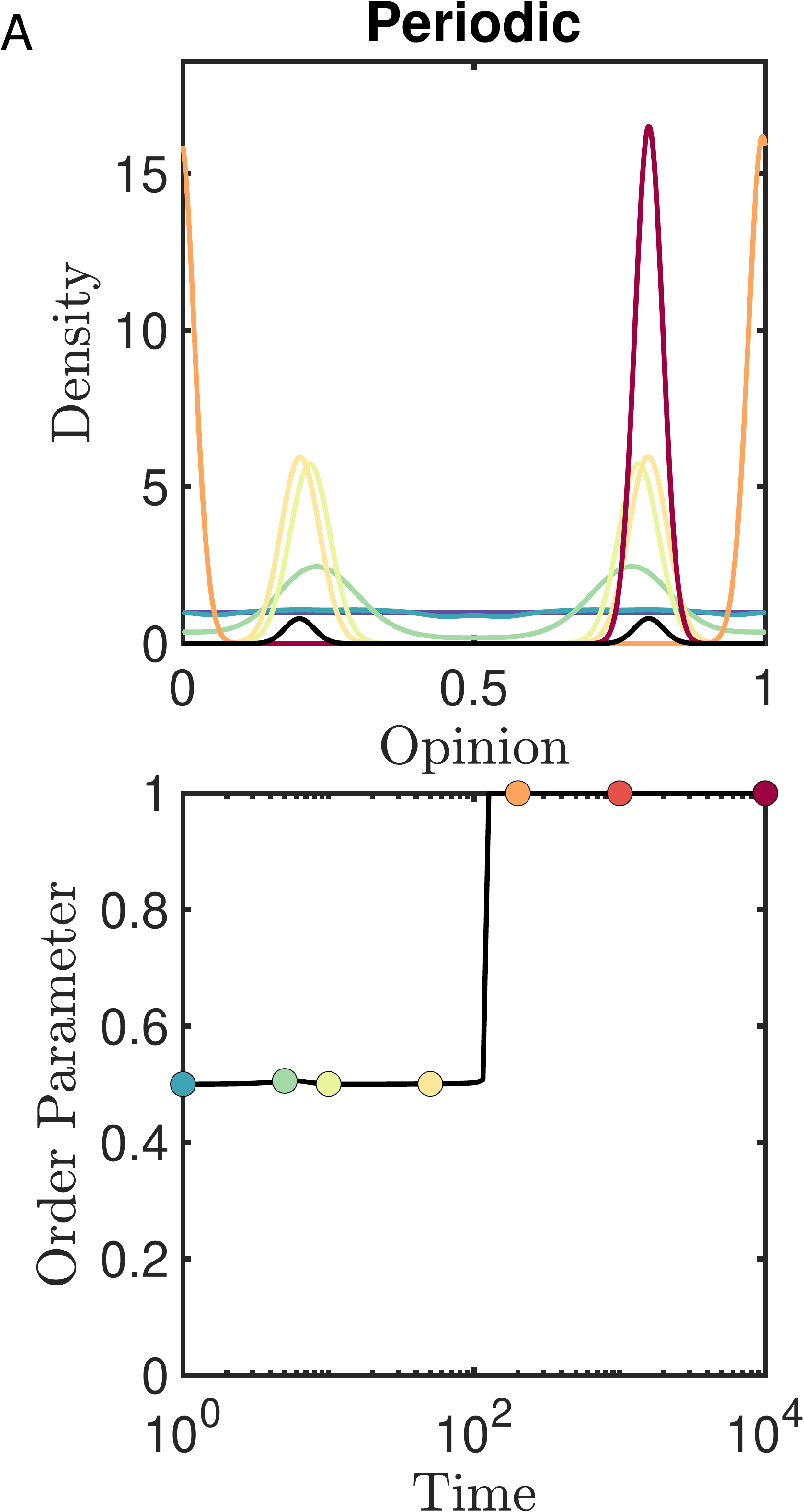}
\includegraphics[width = 2.6cm]{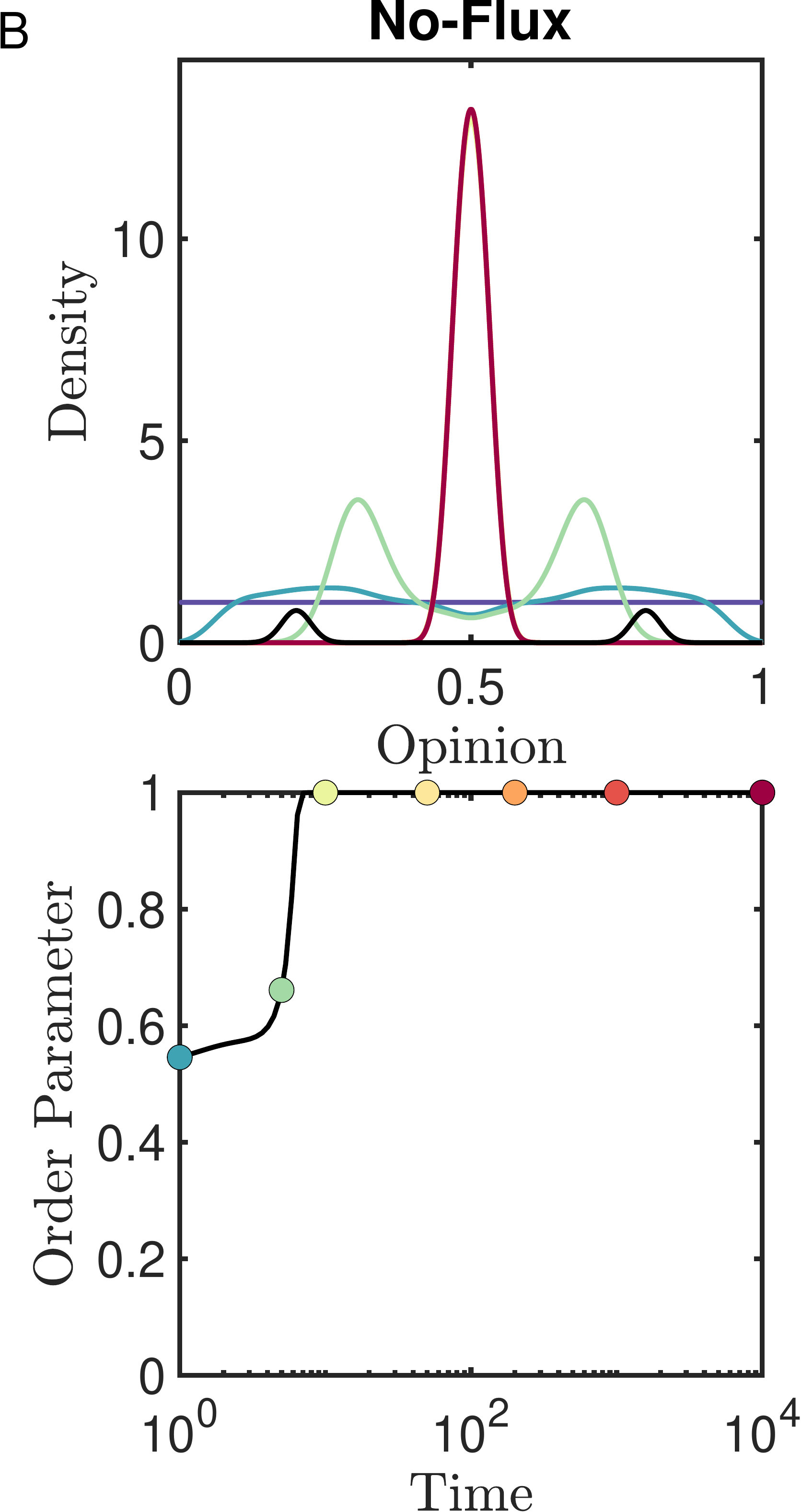}
\includegraphics[width = 2.6cm]{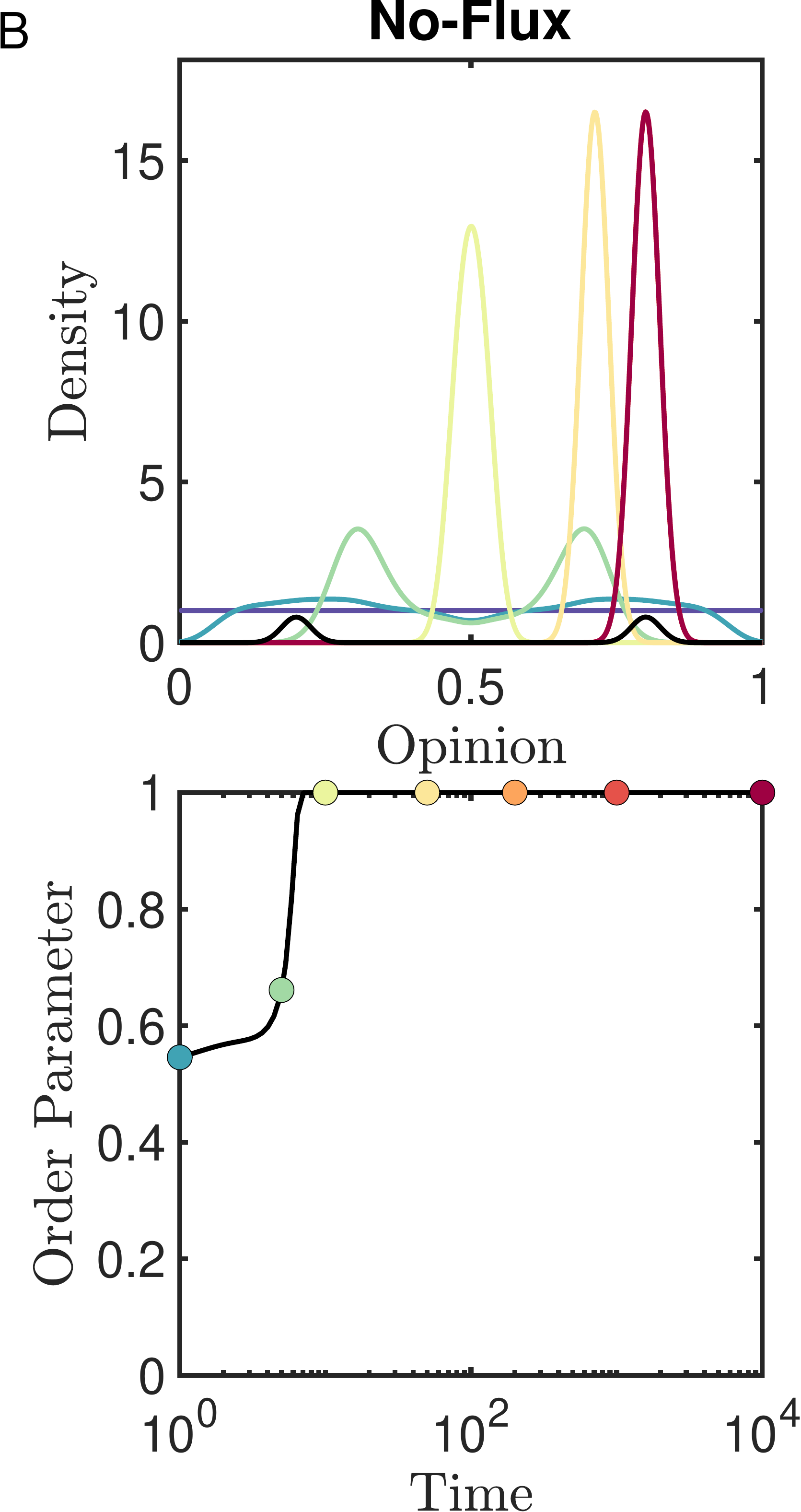}
\includegraphics[width = 2.6cm]{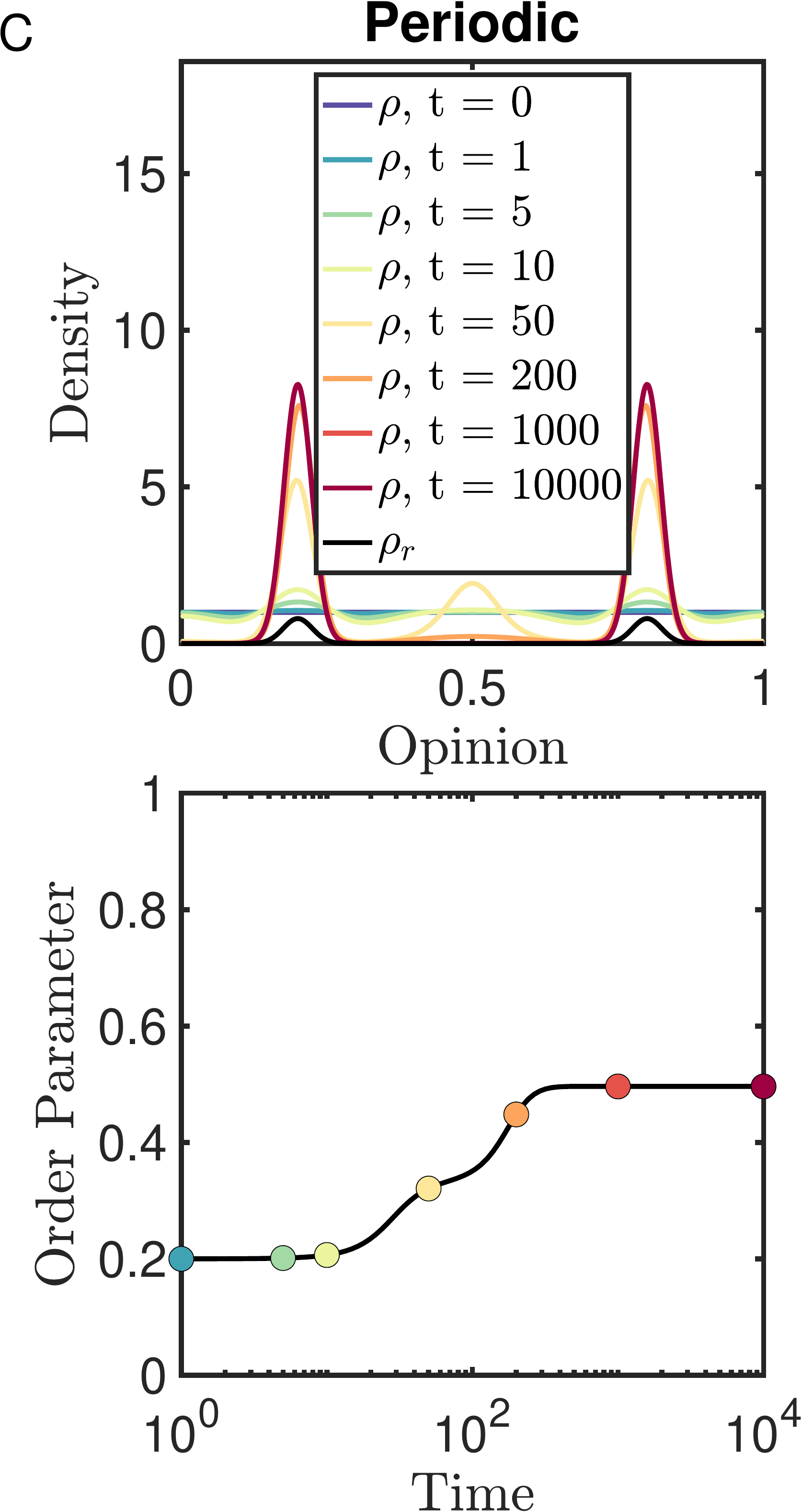}}\\
\resizebox{\figwidth}{!}{
\includegraphics[height = 2.55cm]{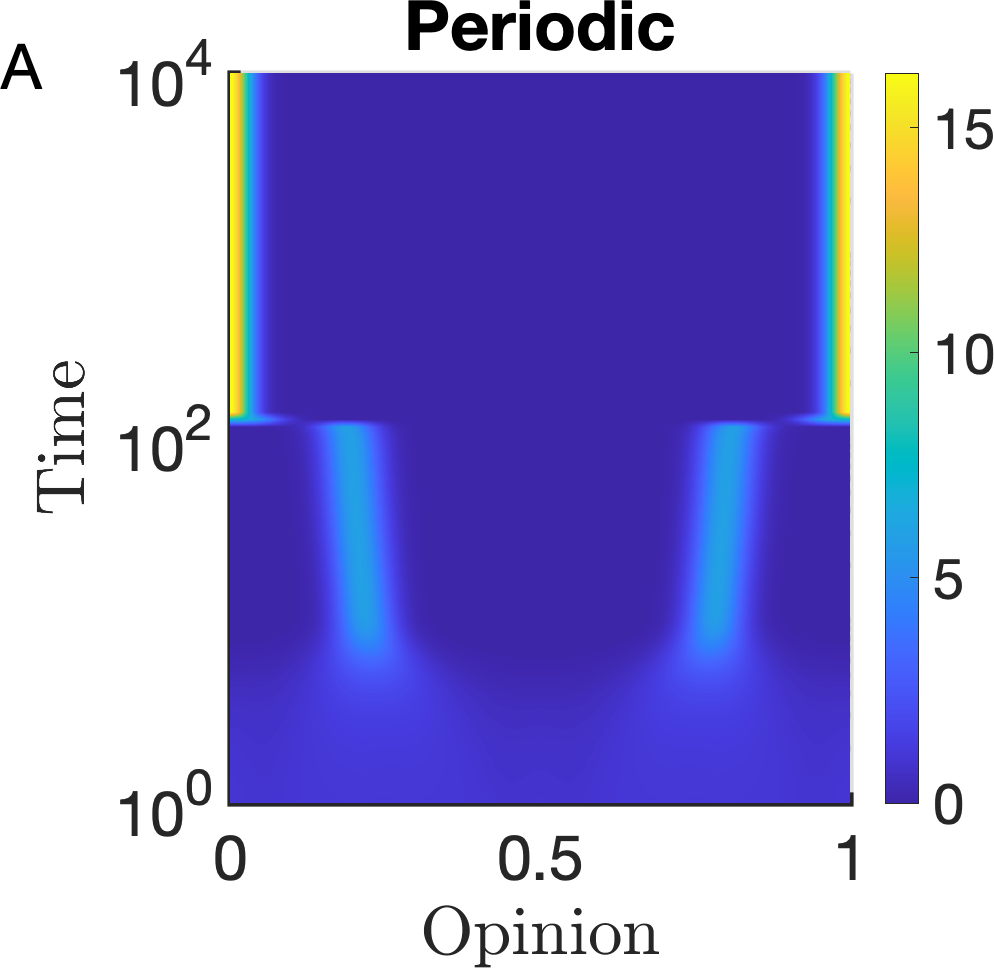} 
\includegraphics[height = 2.55cm]{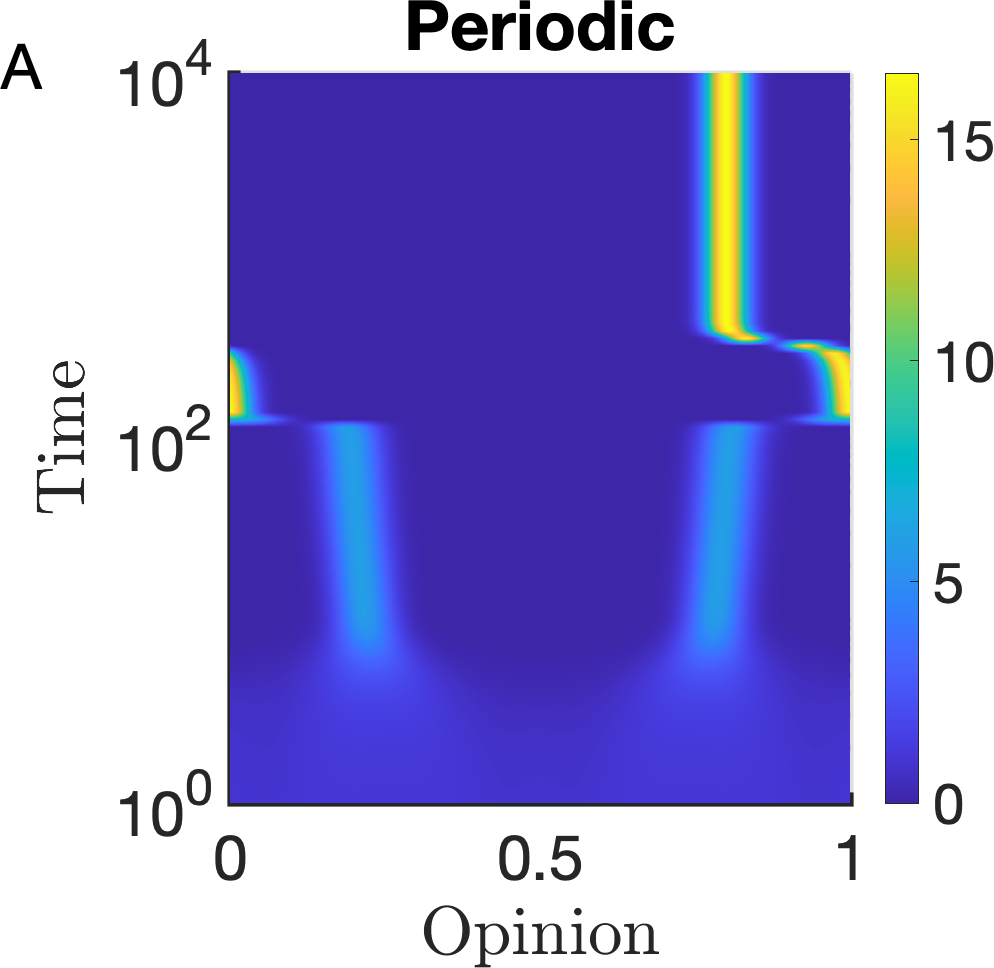} 
\includegraphics[height = 2.55cm]{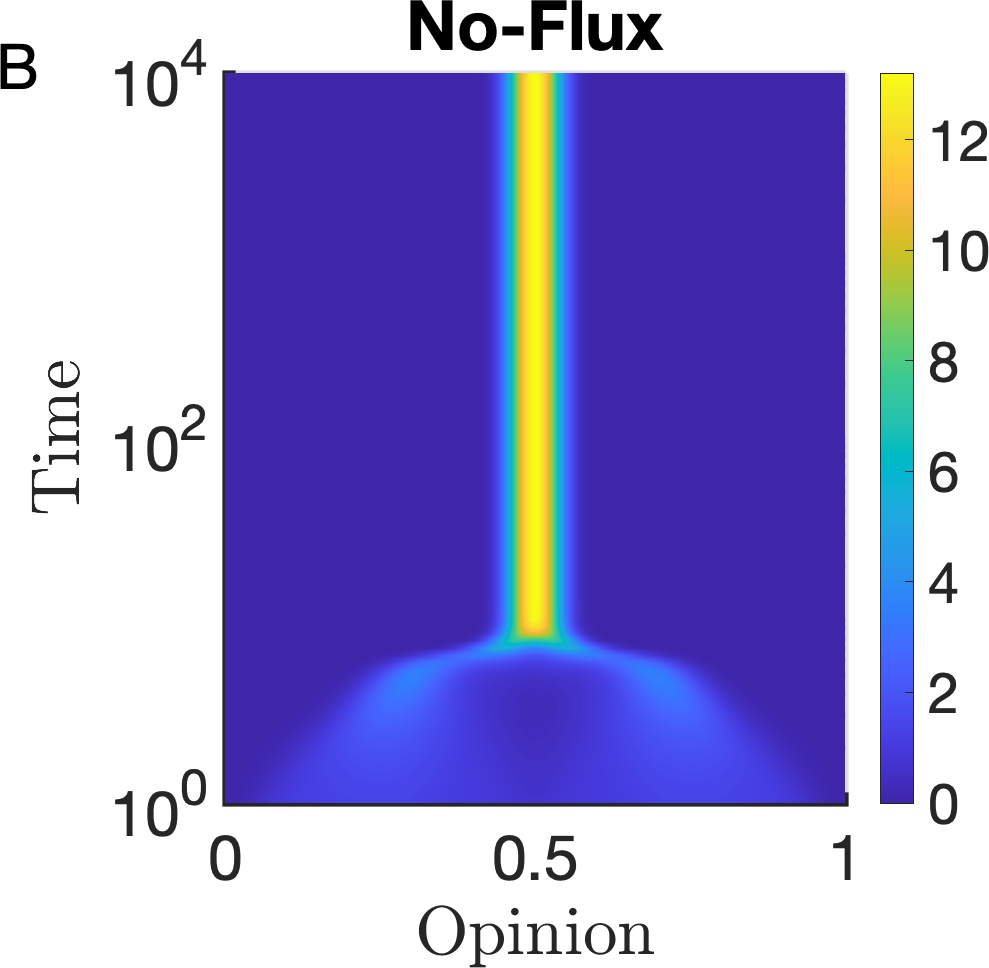}
\includegraphics[height = 2.55cm]{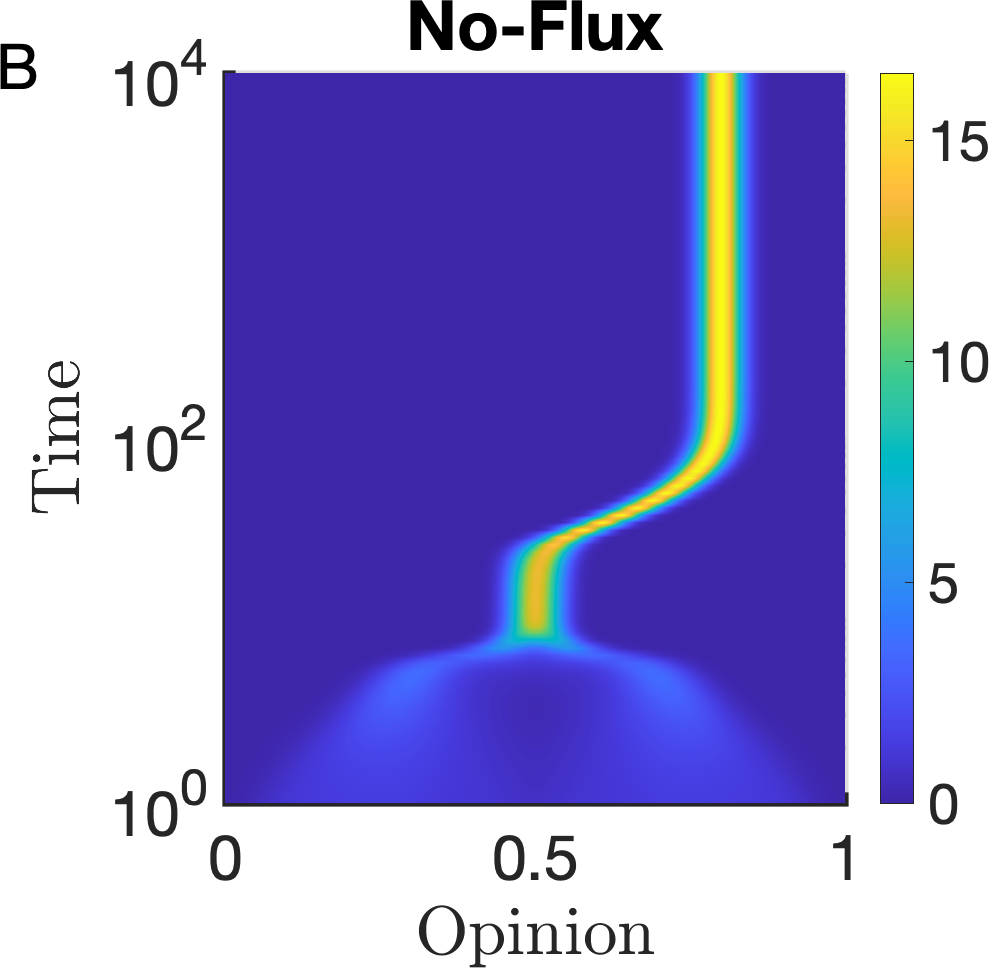}
\includegraphics[height = 2.55cm]{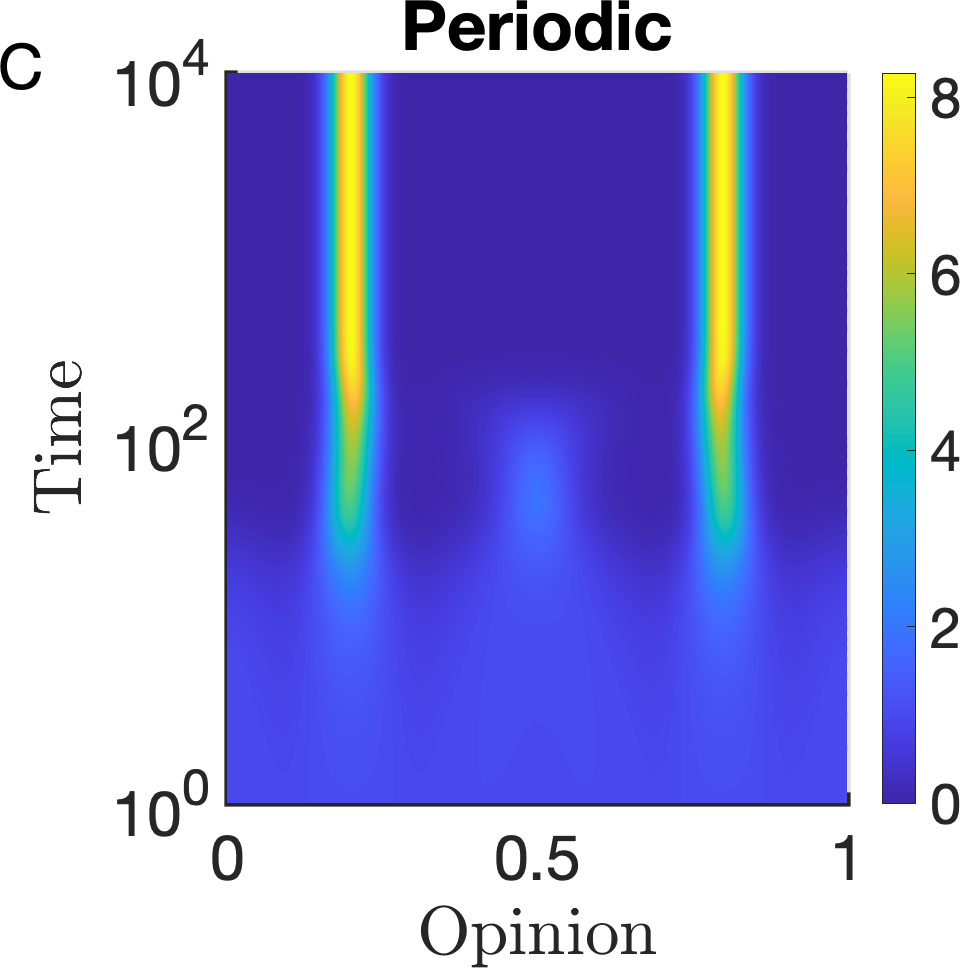}}
\caption
{As Figure~\ref{fig:Uniform_Small_Sigma_Snapshots}, but with a uniform initial condition
and double Gaussian radical distribution.
Labels correspond to parameter values
in Figure~\ref{fig:Uniform_TwoGaussians_Radicals}, with A ($\sigma = 0.035$ and $R = 0.25$).
B ($\sigma = 0.035$ and $R = 0.3$).
C ($\sigma = 0.025$ and $R = 0.1$).
}
\label{fig:Uniform_TwoGaussians_Radicals_Snapshots}
\end{figure}

\section{Conclusions and Outlook} \label{s:Conclusions}

We have demonstrated the significant effect of the choice of boundary condition in 
SDE and PDE bounded confidence models, as well as the sensitivity of such models to
small changes in various parameters, with and without the inclusion of radical distributions.
This work clearly demonstrates that no-flux boundary conditions are the correct choice from a 
modelling perspective, and also in terms of bridging the gap between agent-based~\cite{HK02,HK15}
and PDE models.
In particular, the no-flux choice most faithfully recreates the underlying dynamics of the original
deterministic models~\cite{HK02,HK15}.

There are many possible extensions, some of which have been studied in the literature for 
other models in opinion dynamics: confidence intervals which are
asymmetric~\cite{HK02}, heterogeneous~\cite{HO16,MB12,ZZTK16}, or time-dependent~\cite{MG10};
influence that is negative~\cite{PT18}, or which increases with separation~\cite{MT14};
stubbornness or inertia~\cite{PT18};
persuasiveness or supportiveness~\cite{HKS01};
multi-dimensional models~\cite{FLPR05, PLR06, NT12, BBCN13, EBTT13, ET15, M16};
more general models for the noise, for example coloured, multiplicative, or non-Gaussian (see ~\cite{GPV19}); 
slowly-varying radical distributions;
a rigorous, systematic derivation of the order parameter, perhaps using data-driven approaches.
We note that most of these can be incorporated in both our models and numerical schemes in relatively
straightforward ways, although there are open questions regarding SDEs with no-flux boundary conditions~\cite{LS84}.  

On the analytical side, there are interesting questions regarding rigorous analysis of the SDE model, as well as the rigorous derivation
of the PDE model,  For the no-flux case, the linear stability analysis of the stationary state(s) is a challenging problem, due to the nonlinear and nonlocal boundary conditions. Furthermore, the study of fluctuations around the mean field limit, in particular close to the phase transition will provide us with detailed information about the behaviour of the finite-agent system.  In the absence of radical groups and when the underlying network has fully connectivity (corresponding to the large $R$ limit) and for when $\sigma$ sufficiently small, a centred Gaussian is an approximate stationary state when we consider no-flux boundary conditions. This is similar to the case of periodic boundary conditions~\cite{WLEC17,GPY17}, and follows from the fact that localized stationary states do not feel the effect of the boundary conditions. The stability of such approximate localised states, corresponding to clusters, is a very interesting problem that we plan to address in future work. 

In terms of applications, we plan to use real-world data, particularly from referenda, to study the use of such models in practical settings, with a focus on the transient dynamics rather than the long-term behaviour.  There are also interesting questions regarding the use of such models when applied to advertising, or charismatic leaders~\cite{HK15}.  Another important open question is how to determine the modelling parameters $\sigma$ and $R$.  This is related to inference for McKean SDEs, and we hope to address this topic in future work.  We direct the interested reader to a study of estimating parameters in mean field models~\cite{SKPP21}, which could act as a starting point for such studies.  By estimating modelling parameters, such as the noise, it should be possible to develop diagnostic tools for predicting consensus formation, or lack thereof.

\section*{Funding}
GAP was supported by the EPSRC through grant numbers EP/P031587/1, EP/L024926/1, and EP/L020564/1. This research was funded in part by JPMorgan Chase \& Co. Any views or opinions expressed herein are solely those of the authors listed, and may differ from the views and opinions expressed by JPMorgan Chase \& Co. or its affiliates. This material is not a product of the Research Department of J.P. Morgan Securities LLC. This material does not constitute a solicitation
or offer in any jurisdiction.

\section*{Acknowledgments}
BDG is grateful to Andrew Archer, Valerio Restocchi, and David Sibley for helpful discussions.

\section{Data Availability}
The supplementary material is available at\\ \url{https://www.maths.ed.ac.uk/~bgoddard/files/OD-SupplementaryMaterial.pdf}

\bibliographystyle{plain}
\bibliography{OpinionDynamics}

\end{document}